\documentclass[12pt,sort&compress]{elsarticle}

\usepackage{bm}

\usepackage{mathrsfs}
\usepackage{amsmath}
\usepackage{amsfonts}
\usepackage{amsthm}
\usepackage{color}
\usepackage{xcolor}
\usepackage{siunitx}
\usepackage{bbm}

\usepackage{caption}
\usepackage{subcaption}

\usepackage{float}
\usepackage{url}
\usepackage{multicol}
\usepackage[top=1in,bottom=1in,left=1in,right=1in]{geometry}
\usepackage[pdfborder={0 0 0},colorlinks,allcolors=blue]{hyperref}
\usepackage{txfonts}
\usepackage{setspace}
\usepackage{enumitem}
\usepackage{lipsum}
\usepackage{siunitx,booktabs}
\usepackage{nth}
\usepackage{accents} 
\usepackage{arydshln}
\theoremstyle{definition}
\theoremstyle{definition}
\newtheorem{remark}{Remark}

\allowdisplaybreaks

\newcommand{\vertiii}[1]{{\left\vert\kern-0.2ex\left\vert\kern-0.2ex\left\vert #1 
    \right\vert\kern-0.2ex\right\vert\kern-0.2ex\right\vert}}

\newcommand{\vertii}[1]{{\left\vert\kern-0.2ex\left\vert #1 
    \right\vert\kern-0.2ex\right\vert}}

\usepackage{listings}
\definecolor{light-gray}{gray}{0.95}
\makeatletter
\def\ps@pprintTitle{%
   \let\@oddhead\@empty
   \let\@evenhead\@empty
   \def\@oddfoot{\reset@font\hfil\thepage\hfil}
   \let\@evenfoot\@oddfoot
}
\makeatother

\begin{document}
\begin{frontmatter}

\title{Atomistically-informed continuum modeling and isogeometric analysis of 2D materials over holey substrates}

\author[umn]{Moon-ki Choi\fnref{fn1}}
\author[ucsd]{Marco Pasetto\fnref{fn1}}
\author[unilu]{Zhaoxiang Shen\fnref{fn1}}
\author[umn]{Ellad B. Tadmor}
\author[ucsd]{David Kamensky\corref{cor1}}
\ead{dmkamensky@eng.ucsd.edu}
\fntext[fn1]{These three authors contributed equally to this work and are ordered alphabetically by surname.}
\cortext[cor1]{Corresponding author}

\address[ucsd]{Department of Mechanical and Aerospace Engineering, University of California, La Jolla, CA 92093, USA}
\address[unilu]{Department of Engineering, Faculty of Science, Technology and Medicine, University of Luxembourg, Esch-sur-Alzette 4365, Luxembourg}
\address[umn]{Department of Aerospace Engineering and Mechanics, University of Minnesota, Minneapolis, MN 55455, USA}


\begin{abstract}
This work develops, discretizes, and validates a continuum model of a molybdenum disulfide (MoS$_2$) monolayer interacting with a periodic holey silicon nitride (Si$_3$N$_4$) substrate via van der Waals (vdW) forces.  The MoS$_2$ layer is modeled as a geometrically nonlinear Kirchhoff--Love shell, and vdW forces are modeled by a Lennard-Jones potential, simplified using approximations for a smooth substrate topography.  The material parameters of the shell model are calibrated by comparing small-strain tensile and bending tests with atomistic simulations.  This model is efficiently discretized using isogeometric analysis (IGA) for the shell structure and a pseudo-time continuation method for energy minimization. The IGA shell model is validated against fully-atomistic calculations for several benchmark problems with different substrate geometries. The continuum simulations reproduce deflections, strains and curvatures predicted by atomistic simulations, which are known to strongly affect the electronic properties of MoS$_2$, with deviations well below the modeling errors suggested by differences between the widely-used reactive empirical bond order (REBO) and Stillinger--Weber (SW) interatomic potentials. Agreement with atomistic results depends on geometric nonlinearity in some cases, but a simple isotropic St.\,Venant--Kirchhoff model is found to be sufficient to represent material behavior.  We find that the IGA discretization of the continuum model has a much lower computational cost than atomistic simulations, and expect that it will enable efficient design space exploration in strain engineering applications. This is demonstrated by studying the dependence of strain and curvature in MoS$_2$ over a holey substrate as a function of the hole spacing on scales inaccessible to atomistic calculations. The results show an unexpected qualitative change in the deformation pattern below a critical hole separation.
\end{abstract}


\end{frontmatter}

\section{Introduction}\label{sec:introduction}

Two-dimensional (2D) materials are crystalline materials that are a single unit cell thick. The most famous such material is graphene \cite{Novoselov2004}, but recently many others have been synthesized with thousands predicted to exist \cite{mounet2018}. Some 2D materials are semiconductors, and thus potentially useful for development of new electronic components that could outperform the metal-oxide-silicon transistors ubiquitous in modern computing. An important example in this class are the transition metal dichalcogenides (TMDs), such as molybdenum disulfide (MoS$_2$) \cite{Wang2012}.  The band gap and other electronic properties of MoS$_2$ (and other 2D materials) can be tuned by applying mechanical strains \cite{TuningEpropertiesMoS2,Zib2012,BandMassMoS2,Castellanos-Gomez2013}.  This is referred to as strain engineering, and is an active field of research within nanoelectronics.  A recently-developed approach to apply strains to 2D materials without the need for sustained external loading is through van der Waals (vdW) interaction with an underlying substrate \cite{holey_zhang}.  This paper focuses on a single layer of MoS$_2$ (in the 2H phase) interacting with a silicon nitride (Si$_3$N$_4$) substrate perforated by a periodic pattern of holes, i.e., a ``holey substrate.'' A key aspect of this problem, described in \cite{holey_zhang}, is that the substrate is not flat, but rather has significant periodic topography introduced by the manufacturing process of the holes. This topography is a critical component in straining the MoS$_2$ layer and must be included in any modeling approach. Although this paper focuses on MoS$_2$, we anticipate that most of our conclusions will carry over to other instances of 2D materials interacting with patterned substrates.  

One approach to predicting strains in 2D materials, like MoS$_2$, is to perform molecular statics (MS) simulations that directly model atomic nuclei as discrete particles interacting via empirical interatomic potentials \cite{Momeni2020}. Indeed, this was the approach taken in \cite{holey_zhang}.  However, the computational cost of MS becomes prohibitive for all but the smallest physical systems.\footnote{Although the number of atoms in a 2D material system is far smaller than bulk systems, computational difficulties resulting from the large mismatch between intralayer forces and weak vdW interactions between layers and substrate lead to very slow convergence during energy minimization. This limits the size of problems that can be studied, especially since the benefits of parallelization are limited.}  The goal of the present work is to develop a continuum mechanical model that treats a 2D material as a thin shell structure governed by a partial differential equation (PDE) system, while still incorporating nonlocal vdW interactions with an underlying substrate.  This will allow efficient methods for numerical PDEs to replace costly MS simulations.   

Sauer and collaborators have previously pursued similar objectives with a focus on graphene.  In particular, they have proposed to use isogeometric analysis (IGA) \cite{HughesIGApaper,IGAbook} of Kirchhoff--Love shell theory \cite{Kiendl2011,paper_KLshell,KLshell-hyperelastic} in this context, developing and calibrating specialized constitutive models \cite{Shirazian2018,Ghaffari2018,Mokhalingam2020} and simulating vdW interactions with substrates through a Lennard-Jones (LJ) \cite{LJ1924} potential \cite{Ghaffari2018a,Ghaffari2018b}.  We follow a broadly similar approach of using IGA of Kirchhoff--Love shells to simulate 2D materials, but introduce the following novel contributions:
\begin{itemize}
    \item An extension of the approach to a 2D material that has thickness in the out-of-plane direction.
    \item An efficient approximation of the LJ potential, tailored to holey substrates.
    \item A robust energy minimization approach based on a damped dynamic analysis in pseudo-time.
    \item A detailed comparison with MS simulations of 2D material--holey substrate interaction, examining both accuracy and computational performance.
    \item An open-source implementation leveraging modern code generation techniques for numerical PDEs.
\end{itemize}
We believe that these contributions represent a significant step toward enabling efficient design space exploration in the field of strain engineering.  

The remainder of this paper is structured as follows. Section \ref{sec:problem-description} describes the molecular problem setup and its features. Section \ref{sec:continuum} introduces a continuum model of MoS$_2$ interacting with a holey substrate, and its discretization using IGA.  Section \ref{sec:ms} then describes the atomistic model it will be calibrated with (Section \ref{sec:calibration}) and validated against (Section \ref{sec:validation}). Section \ref{sec:application} demonstrates the ability of the isogeometric shell analysis to simulate larger systems than would be practical with atomistic methods. Section \ref{sec:conclusions} summarizes our findings and discusses potential future extensions of the work.

\section{Interaction between $\rm{MoS_2}$ and a holey $\rm{Si_3N_4}$ substrate}\label{sec:problem-description}

\begin{figure}[!htbp]\centering
\includegraphics[width=0.6\textwidth]{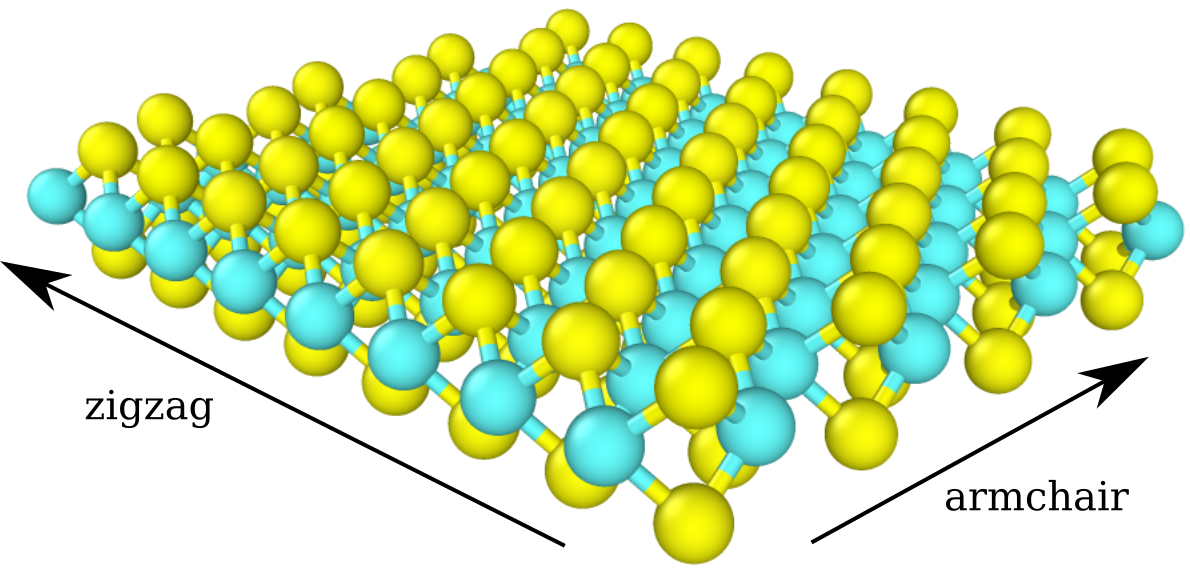}
\caption{Structure of a $\rm{MoS_2}$ monolayer, with Mo atoms in cyan and S atoms in yellow.  The vertical thickness of the layer (between upper and lower S nuclei) is approximately 0.3\,nm.  The material contains two distinct types of in-plane directions, referred to as ``armchair'' and ``zigzag'', and labelled above. (Due to structural symmetries, several armchair and zigzag directions exist, but only one of each is labelled for clarity.)} 
\label{fig:mos2_structure}
\end{figure}

\begin{figure}[!htbp]\centering
\includegraphics[width=0.6\textwidth]{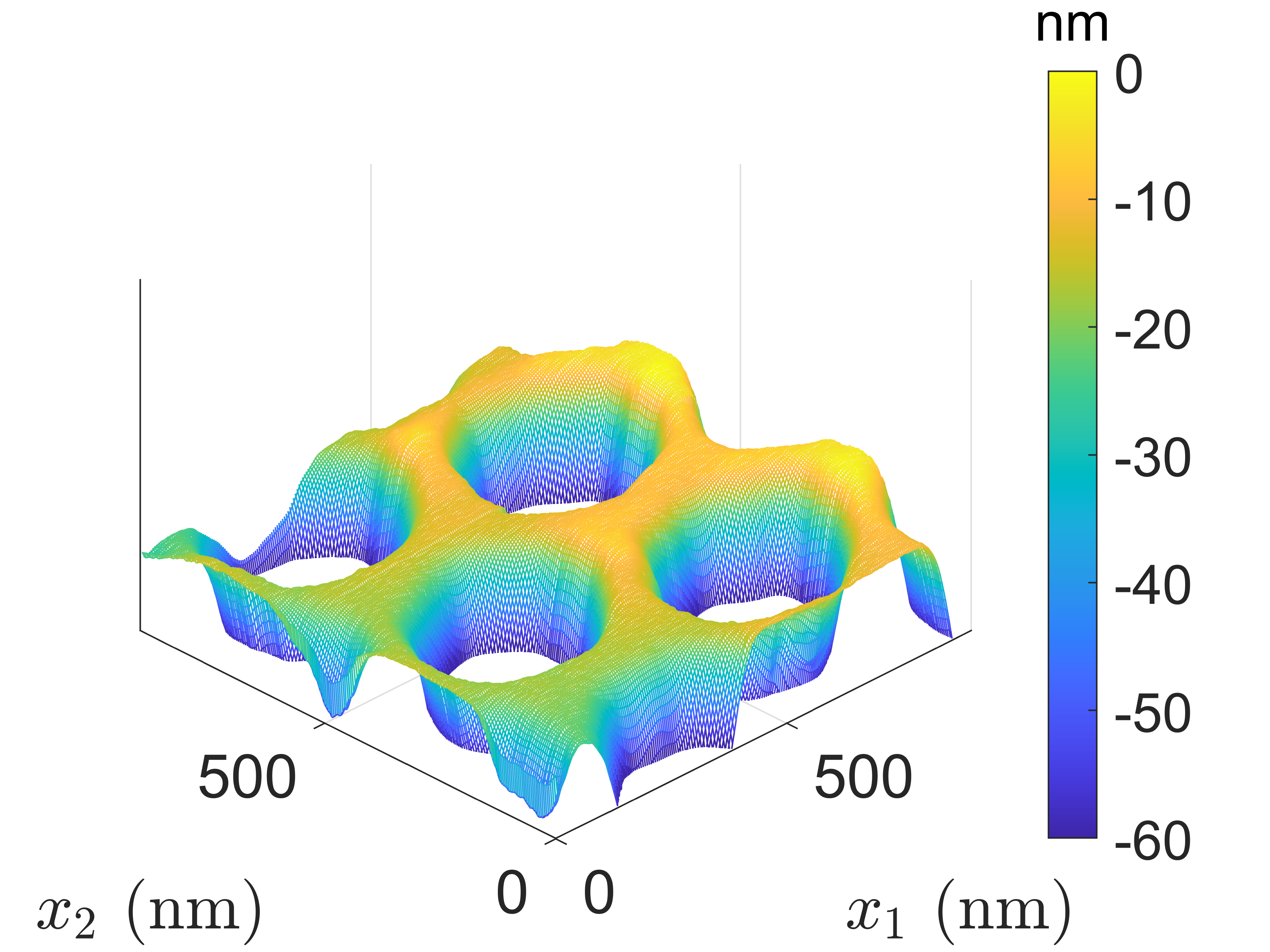}
\caption{Surface geometry of a holey $\rm{Si_3N_4}$ substrate, as measured by AFM for holes with nominal radius $R_0=100$\,nm and hole spacing of $L=400$\,nm.  N.b. the difference in overall scale from Figure \ref{fig:mos2_structure} and the difference in vertical and horizontal axis scaling. The AFM measurements show that the substrate exhibits a cratered structure with the holes separated by elevated ridges. (AFM data was provided to us by Yichao Zhang (University of Minnesota). Similar measurements are reported in \cite{holey_zhang}.)}
\label{fig:afm_substrate}
\end{figure}

As mentioned in Section \ref{sec:introduction}, the present study focuses on $\rm{MoS_2}$ (in the 2H phase) as a representative 2D material and $\rm{Si_3N_4}$ as a substrate material.  In addition to its electronic properties discussed in Section \ref{sec:introduction}, $\rm{MoS_2}$ is also interesting from a geometric standpoint, in that it consists of three sub-layers, with a central sub-layer of Mo atoms sandwiched between two sub-layers of S atoms, bonded together as shown in Figure \ref{fig:mos2_structure}.  The continuum model of Section \ref{sec:continuum} will need to account for this trilayer internal structure of $\rm{MoS_2}$ to accurately approximate a fully-atomistic model.  The choice of $\rm{Si_3N_4}$ as a substrate material is motivated by its mechanical and chemical stability, including resistance to solvents and acids \cite{si34_review_dwyer}.  There are  well-established techniques to apply photolithography \cite{pattern_ge,porous_si3n4_ge} and electron beam lithography \cite{phononic_si3n4_tambo} to $\rm{Si_3N_4}$, to manufacture holey substrates in a controlled manner, as needed for strain engineering applications.  A representative plot of the surface of holey substrate is shown in Figure \ref{fig:afm_substrate}, which is derived from atomic force microscopy (AFM) of a holey $\rm{Si_3N_4}$ substrate manufactured by Ted Pella, Inc. \cite{holey_zhang}. The AFM measurements show that the substrate has significant topography near the holes, which has a strong effect on the resulting strain in the MoS$_2$ layer. 

Note that the sub-nanometer distances between neighboring atoms in the $\rm{MoS_2}$ layer are much smaller than geometric features of the substrate, whose height varies smoothly over tens-to-hundreds of nanometers (e.g., see Figure \ref{fig:afm_substrate}).  Strains induced by interactions with the substrate are expected to vary at the latter scale, which is our motivation for considering continuum approximations of the $\rm{MoS_2}$ layer (Section \ref{sec:kl}).  The length scale in the LJ potential used to model layer--substrate interaction are also sub-nanometer, so, from the perspective of an atom close enough to the substrate to feel significant LJ forces, the substrate will appear approximately flat.  This fact will be leveraged to derive efficient approximations of the LJ potential in our continuum modeling framework (Section \ref{sec:geom-interact-approx}).

In this work, we follow the convention of letting the substrate occupy a half-space below some horizontal plane, such that the layer of 2D material rests horizontally on top of the substrate, hovering at some distance determined by the transition from attractive to repulsive layer--substrate interaction forces.  However, these designations of ``up'' and ``down'' are arbitrary, as gravity is negligible in nanomechanics, so the physics do not imply a preferred vertical direction. We describe geometry using a global Cartesian coordinate chart $(x_1, x_2, x_3)$ of physical space, such that $x_1$ and $x_2$ parameterize horizontal planes and $x_3$ parameterizes the vertical direction.

\section{Continuum modeling and discretization}\label{sec:continuum}
This section introduces our continuum modeling framework, where the MoS$_2$ layer is modeled as a Kirchhoff--Love shell (Section \ref{sec:kl}) and the substrate is assumed to have a sufficiently smooth surface that it is able to be locally approximated by a half-space behind a tangent plane (Section \ref{sec:tangent-plane}).  These assumptions make it possible to discretize the continuum model efficiently using IGA, taking full advantage of modern code generation functionality to obtain a high-performance implementation (Section \ref{sec:iga}).  
\subsection{Geometrically-nonlinear Kirchhoff--Love shell}\label{sec:kl}
2D materials typically exhibit both membrane and bending stiffness, along with a Poisson effect in the in-plane deformation.  We aim to capture these properties in a continuum model posed over a 2D manifold.  In this work, we assume isotropy with respect to in-plane directions (see Remark \ref{rem:isotropy} for justification of this assumption), in which case, the bending and membrane stiffness and Poisson's ratio of a given 2D material can be matched by calibrating a Kirchhoff--Love shell model.  The Kirchhoff--Love model is {\em formally} derived by considering an isotropic elastic material occupying a 3D region that is thin along one dimension, then using kinematic simplifications and a plane stress assumption to obtain a problem posed on the 2D midsurface manifold of this region.  In the case of a 2D material, the thin direction is only several atoms thick, and thus well below the length scale at which continuum modeling is valid.  However, we can nonetheless calibrate a Kirchhoff--Love model to match the bending stiffness, membrane stiffness, and Poisson's ratio of an atomistic model.  Thus, the 3D continuum model used to formally express the Kirchhoff--Love formulation does not correspond to any real bulk material.  For this reason, we refer to it as a ``formal 3D material.''

The reference-configuration midsurface of the Kirchhoff--Love shell is assumed to be a 2D manifold $\Gamma_0$.  The main kinematic assumption behind Kirchhoff--Love theory is that the in-plane part of the Green--Lagrange strain $\mathbf{E}$ at some point in the formal 3D material is given by
\cite{Kiendl2011}
\begin{equation}\label{eq:in-plane-gl-strain-defn}
    \mathbf{E}_{\text{2D}} = \bm{\varepsilon} + \xi_3\bm{\kappa}\text{ ,}
\end{equation}
where $\bm{\varepsilon}$ is the 2D membrane strain of the midsurface, $\bm{\kappa}$ is a 2D bending strain, and $\xi_3$ is an arc-length coordinate in the through-thickness direction, giving the signed distance from $\Gamma_0$ in the reference configuration.  
The membrane strain is defined as half the change in midsurface metric tensor, i.e.,
\begin{equation}
 \bm{\varepsilon} = \frac{1}{2}\left(\mathbf{g}-\mathbf{G}\right)\text{ ,}
\end{equation}
where $\mathbf{G}$ is the metric tensor on $\Gamma_0$, and $\mathbf{g}$ is the pullback of the deformed midsurface metric.
Bending strain is defined as minus the change in second fundamental form of the midsurface, i.e.,
\begin{equation}\label{eq:kappa-defn}
    \bm{\kappa} = \mathbf{B} - \mathbf{b}\text{ ,}
\end{equation}
where $\mathbf{B}$ is the second fundamental form of $\Gamma_0$ and $\mathbf{b}$ is the pullback of the second fundamental form of the deformed midsurface.
The out-of-plane Green--Lagrange strain $E_{33}$ is determined by imposing the plane stress condition $S_{33}=0$ on the second Piola--Kirchhoff stress $\mathbf{S}$, while transverse shear strains $E_{13}$ and $E_{23}$ are assumed to be zero.
We model the formal 3D material using the St.\,Venant--Kirchhoff constitutive relation,
\begin{equation}\label{eq:svk}
    \mathbf{S} = \left(\frac{E\nu}{(1+\nu)(1-2\nu)}\right)(\operatorname{tr}\mathbf{E})\mathbf{I} + 2\left(\frac{E}{2(1+\nu)}\right)\mathbf{E}\text{ ,}
\end{equation}
which we have stated here in full 3D form (without applying the aforementioned plane stress condition), where $E$ is the Young's modulus, and $\nu$ is the Poisson's ratio.  This is a generalization of isotropic linear elasticity, where the Cauchy stress and engineering strain are replaced by large-deformation counterparts.  It is formally hyperelastic, with energy density
\begin{equation}
\psi = \frac{1}{2}\mathbf{S}:\mathbf{E}    
\end{equation}
per unit volume in a stress-free reference configuration.  The St.\,Venant--Kirchhoff model is sometimes said not to be a ``true'' hyperelastic model, because it exhibits instabilities under strong compression \cite[Section 7.3, Examples 7.2 and 7.3]{Tadmor:CMT}.  However, these are not of concern in the present application, where a 2D material would buckle under compressive loading long before the material instability is encountered.  

The total elastic energy of a thin shell with reference-configuration 2D midsurface manifold $\Gamma_0$ and thickness $h_\text{th}$ is then
\begin{equation}\label{eq:w-int}
    U_\text{int} = \int_{\Gamma_0}\left(\int_{-h_\text{th}/2}^{+h_\text{th}/2}\psi\,d\xi_3\right)\,d\Gamma_0\text{ ,}
\end{equation}
where $\xi_3$ is an arc-length parameter in the through-thickness direction, giving signed distance from $\Gamma_0$ in the reference configuration.  The Kirchhoff--Love kinematic assumptions can be used to express $\mathbf{E}$ at any value of $\xi_3$ in terms of only a midsurface displacement field $\mathbf{u}:\Gamma_0\to\mathbb{R}^3$ and its derivatives, so that $\int d\xi_3$ can be computed analytically for each point in $\Gamma_0$ and $U_\text{int}$ can be understood as a functional mapping $\mathbf{u}$ to total elastic energy.  

The precise formulation used here is identical to that explained in detail by \cite[Section 3.2]{Kiendl2011}.  What is relevant to the present study is the existence of the energy functional \eqref{eq:w-int}, which depends on three free parameters: $E$, $\nu$, and $h_\text{th}$.  These parameters can be calibrated to obtain arbitrary bending stiffness, membrane stiffness, and Poisson's ratio.  N.b. that the thickness $h_\text{th}$ is {\em not} measured directly from molecular geometry, but is considered a free parameter in the calibration, following our interpretation of the 3D elastic material as a formal device rather than a real substance.  The details of this calibration will be explained in Section \ref{sec:calibration} (after first also introducing the atomistic models (Section \ref{sec:ms}) that it will be calibrated to approximate).  

\subsection{Approximate interaction with holey substrate}\label{sec:tangent-plane}
Our model of interactions between the 2D material and substrate involves two levels of approximation.  The first (Section \ref{sec:continuum-interact-approx}) is an application of continuum assumptions to convert sums over pairs of atoms into integrals.  The second (Section \ref{sec:geom-interact-approx}) is a series of geometric approximations to simplify calculations in the specific context of the holey substrate application.  

\subsubsection{Continuum approximation of Lennard-Jones interaction}\label{sec:continuum-interact-approx}
We assume that the Si$_3$N$_4$ substrate occupies the region
\begin{equation}\label{eq:omegasub}
    \Omega_\text{sub} = \left\{\mathbf{x}\in\mathbb{R}^3~:~x_3 < z_\text{sub}\left(\xi_1\left(x_1,x_2\right),\xi_2\left(x_1,x_2\right)\right)\right\}\text{ ,}
\end{equation}
where $\{\xi_\alpha\}_{\alpha=1}^2$ is a curvilinear coordinate chart of $\mathbb{R}^2$ and $z_\text{sub}:\mathbb{R}^2\to\mathbb{R}$ is a smooth (i.e., at least $C^1$) single-valued function defining the surface of the substrate.

The shell midsurface in the reference configuration is then assumed to be 
\begin{equation}
    \Gamma_0 = \left\{\mathbf{x}\in\mathbb{R}^3~:~x_3 = z_0\right\}\text{ ,}
\end{equation}
where $z_0\in\mathbb{R}$ is a constant height such that $\Gamma_0$ is strictly above the substrate (i.e., $z_\text{sub}(\bm{\xi}) < z_0~\forall\bm{\xi}\in\mathbb{R}^2$).  We may therefore use the coordinate chart $\{\xi_\alpha\}$ to parameterize $\Gamma_0$ in addition to the substrate height.  

The interaction of atoms of types A and B at positions $\mathbf{x}_\text{A}$ and $\mathbf{x}_\text{B}$ is modeled by a pairwise potential energy of the Lennard-Jones form:
\begin{equation}
\phi_{\text{A--B}}(r) = 4\epsilon_{\text{A--B}}\left(\left(\frac{\sigma_{\text{A--B}}}{r}\right)^{12} - \left(\frac{\sigma_\text{A--B}}{r}\right)^6\right)\text{ ,}
\end{equation}
where $r = \vert\mathbf{x}_\text{A} - \mathbf{x}_\text{B}\vert$, $\epsilon_{\text{A--B}}$ is an energy scale, and $\sigma_{\text{A--B}}$ is a length scale.  In the continuum model, the summation over such interactions is approximated by integration, weighted by the number density of atoms per unit volume in the substrate and per unit area in the 2D material \cite{zhang2017}. The interaction of one 2D material atom of type A with all substrate atoms of type B is given by
\begin{equation}\label{eq:Omega_sub-int}
    \psi_\text{A--B}(\mathbf{x}_\text{A}) = \int_{\Omega_\text{sub}}\rho_\text{B}\phi_{\text{A--B}}(r)\,d\mathbf{x}_\text{B}\text{ ,}
\end{equation}
where $\rho_\text{B}$ is the number density per unit volume of atoms of type B in the substrate, and the total interaction energy of atoms of types A and B is
\begin{equation}\label{eq:w-a--b}
    U_\text{A--B} = \int_{\Gamma_0}\rho_\text{A}\psi_{\text{A--B}}\left(\mathbf{x}_\text{A}(\bm{\xi})\right)\,d\Gamma_0,
\end{equation}
where $\rho_\text{A}$ is a number density per unit {\em area} of atoms in the 2D material, and $\mathbf{x}_\text{A}(\bm{\xi})$ is the deformed position of an atom of type A at parametric coordinates $\bm{\xi}$ in $\Gamma_0$.  

In the case of MoS$_2$, atoms of type Mo are taken to lie directly on the shell midsurface $\Gamma_0$ in the reference configuration, so $\mathbf{x}_\text{Mo}(\bm{\xi}) = \mathbf{X}(\bm{\xi}) + \mathbf{u}(\mathbf{X}(\bm{\xi}))$, where $\mathbf{X}:\mathbb{R}^2\to\mathbb{R}^3$ gives the reference-configuration positions in $\Gamma_0$ of parametric points.  S atoms, on the other hand, fall into two sub-layers, each offset from the midsurface by a distance $d_\text{Mo--S}$ in either direction, so that
\begin{equation}\label{eq:x_S_pm}
    \mathbf{x}^\pm_\text{S} = \mathbf{x}_\text{Mo} \pm d_\text{Mo--S}\mathbf{n}\text{ ,}
\end{equation}
where $\mathbf{n}$ is the unit normal to the deformed shell structure and the choice of $\pm$ depends on which sub-layer is considered.  We take the value of $d_\text{Mo--S}$ to be 0.1595\,nm, consistently with the geometry of the undeformed $\rm{MoS_2}$ monolayer (Figure \ref{fig:mos2_structure}). This distance is considered to be fixed and independent of the midsurface deformation, which follows by analogy from the Kirchhoff--Love kinematic assumptions used to define the deformation of continuum material away from the midsurface.

For an MoS$_2$ layer interacting with an Si$_3$N$_4$ substrate, the total combined elastic and Lennard-Jones energy of the shell structure in our continuum model is therefore
\begin{equation}\label{eq:total-U}
    U = U_\text{int} + U_\text{Mo--Si} + U_\text{Mo--N} + U^+_\text{S--Si} + U^-_\text{S--Si} + U^+_\text{S--N} + U^-_\text{S--N}\text{ ,}
\end{equation}
where superscript $+$ and $-$ for S--substrate interactions distinguish between the upper and lower sulfur sub-layers in the MoS$_2$ layer (cf. \eqref{eq:x_S_pm}).  The pairwise Lennard-Jones parameters and densities of different atom types are given in Tables \ref{tab:lj-params} and \ref{tab:densities}.  In this work, we seek a minimum of this combined energy functional $U$, by solving the problem:  Find $\mathbf{u}\in\mathcal{V}$ such that
\begin{equation}\label{eq:var-problem}
    D_\mathbf{v}U = 0\quad\forall\mathbf{v}\in\mathcal{V}\text{ ,}
\end{equation}
where $D_\mathbf{v}$ is a Gateaux derivative with respect to the shell midsurface displacement $\mathbf{u}$, in the direction of an arbitrary test function $\mathbf{v}$ in a displacement function space $\mathcal{V}$.  Although solving the problem \eqref{eq:var-problem} is technically only a necessary condition for a minimum of $U$, solutions in this application are typically local (if not global) minima, even if $U$ is not guaranteed to satisfy a property like convexity to ensure that \eqref{eq:var-problem} is a sufficient condition.
\begin{table}[!htb]\centering
\begin{tabular}{c|c|c|c|c}
A--B &  Mo--Si & Mo--N & S--Si & S--N\\
\hline
$\epsilon_\text{A--B}$ (eV) & 0.038120 & 0.077892 & 0.001799 & 0.003672\\
\hline
$\sigma_\text{A--B}$ (nm) & 0.33022 & 0.30261 & 0.37106 & 0.34345
\end{tabular}
\caption{Lennard-Jones parameters for interactions between an MoS$_2$ layer and an Si$_3$N$_4$ substrate \cite{Heinz2013,Halgren1992}.  Geometric and arithmetic means are used to determine $\epsilon$ and $\sigma$ for cross-species interactions, respectively \cite{holey_zhang}.}
\label{tab:lj-params}
\end{table}

\begin{table}[!htb]\centering
\begin{tabular}{c|c}
$\rho_\text{S}^+$ (nm$^{-2}$) & 11.35 \\
\hline
$\rho_\text{Mo}$ (nm$^{-2}$) & 11.35 \\
\hline
$\rho_\text{S}^-$ (nm$^{-2}$) & 11.35 \\
\hline
$\rho_\text{Si}$ (nm$^{-3}$) & 41.8552 \\
\hline
$\rho_\text{N}$ (nm$^{-3}$) & 55.8070
\end{tabular}
\caption{Areal and volumetric number densities for different atoms in MoS$_2$ and Si$_3$N$_4$.  (N.b. the difference in units.)}
\label{tab:densities}
\end{table}

\subsubsection{Geometric approximations}\label{sec:geom-interact-approx}
\begin{figure}[!htbp]\centering
\includegraphics[width=0.7\textwidth]{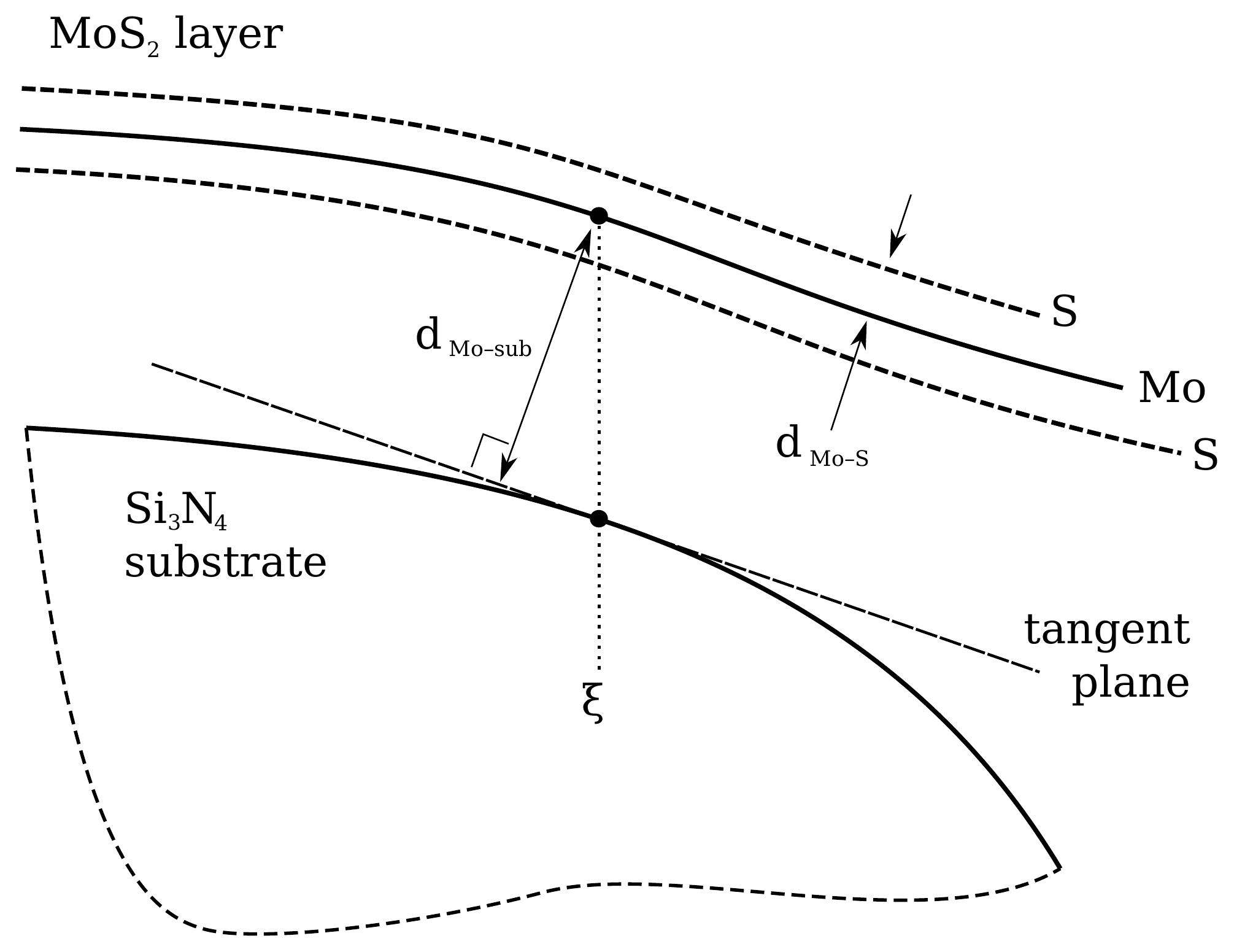}
\caption{Illustration of the geometric approximation made in calculating Lennard-Jones forces from the substrate.}
\label{fig:tangent-plane-method-notation}
\end{figure}

We now introduce geometric approximations to simplify the calculation of $\psi_\text{A--B}$, eliminating the challenging integral over $\Omega_\text{sub}$.  Our main approximation, similar to the approach taken in \cite{zhang2017,zhang2018}, is to assume that for the purpose of computing interactions of Mo and S atoms with parametric coordinates $\bm{\xi}$, the substrate can be approximated as the region below a tangent plane to the surface $z_\text{sub}$, at the same coordinates $\bm{\xi}$. This allows for approximation of \eqref{eq:Omega_sub-int} by an analytical solution to the half-space integral:
\begin{equation}\label{eq:psi-a--b-approx}
    \psi_\text{A--B}(\mathbf{x}_\text{A}(\bm{\xi}))\approx \frac{6}{5}\pi\epsilon_\text{A--B}\rho_\text{B}\sigma_\text{A--B}^2\left(\frac{5\sigma_\text{A--B}^4}{9d_\text{A--sub}^3} - \frac{2\sigma_\text{A--B}^{10}}{27d_{\text{A--sub}}^9}\right)\text{ ,}
\end{equation}
where $d_\text{A--sub}$ is the distance of $\mathbf{x}_\text{A}$ from the tangent plane, viz.,
\begin{equation}\label{eq:d-A-sub}
    d_\text{A--sub} = \left(\mathbf{x}_\text{A}(\bm{\xi}) - \mathbf{x}_\text{sub}(\bm{\xi})\right)\cdot\mathbf{n}_\text{sub}(\bm{\xi})\text{ ,}
\end{equation}
in which $\mathbf{x}_\text{sub}(\bm{\xi}) = \left(x^1(\bm{\xi}),x^2(\bm{\xi}),z_\text{sub}(\bm{\xi})\right)^T$ (i.e., the point on the substrate surface ``directly below'' $\mathbf{x}_\text{Mo}(\bm{\xi})$), and $\mathbf{n}_\text{sub}(\bm{\xi})$ is the outward unit normal to $\Omega_\text{sub}$ at parametric coordinates $\bm{\xi}$.
This approximation is visualized in Figure \ref{fig:tangent-plane-method-notation}.

Our geometric approximations are rooted in the geometric features of our target system.  First, the local use of a half-space to approximate the substrate is based on the assumption that the substrate surface's maximum absolute principal curvature is small relative to the inverse of the Lennard-Jones length scale, i.e.,
\begin{equation}
    k_\text{max} = \max\{\vert k_1\vert,\vert k_2\vert\} \ll \sigma_{\text{A--B}}^{-1}\text{ ,}
\end{equation}
where $k_1$ and $k_2$ are the substrate surface's principal curvatures.  Defining the approximate substrate half-space using a tangent plane at the point $\mathbf{x}_\text{sub}(\bm{\xi})$ instead of the geometrically closest substrate point to $\mathbf{x}_A(\bm{\xi})$ also depends on this low-curvature assumption; in the limit that $k_\text{max} = 0$, there would of course be no distinction between choosing $\mathbf{x}_\text{sub}(\bm{\xi})$ and closest-point projection.  

\subsection{Isogeometric discretization using FEniCS and tIGAr}\label{sec:iga}
For material points in the shell's formal 3D continuum with $\left\vert\xi_3\right\vert > 0$, the expression for $\mathbf{E}$ involves {\em second} derivatives of $\mathbf{u}$.  Thus, the functional $U_\text{int}$ is only well-defined for displacement fields with square-integrable second derivatives, i.e., $\mathbf{u}\in (H^2(\Gamma_0))^3$.  In piecewise-polynomial approximation spaces, this corresponds to requiring $C^1$ continuity.  In \cite{zhang2018}, a subdivision finite element approach was adopted to achieve the required continuity. This work uses a $C^1$ quadratic B-spline \cite{PiegTil97} space defined on a periodic cell to approximate each component of the shell structure's displacement field $\mathbf{u}$, leading to the isogeometric rotation-free Kirchhoff--Love shell discretization first introduced in \cite{paper_KLshell}.  This formulation has seen extensive use across a variety of application domains \cite{Herrema2019,Morganti:2015fy,Ding2021,Zhao2022}, and has been demonstrated to have much greater per-degree-of-freedom accuracy than classical finite element analysis of thin shells (e.g., \cite[Tables 1 and 2]{Morganti:2015fy} or \cite[Figure 8]{Herrema2019}).  

We assume that the substrate height $z_\text{sub}$ is in the same B-spline space, and compute the integral over $\Gamma_0$ in \eqref{eq:w-a--b} using Gaussian quadrature over knot spans in the B-spline parameter space.  The approximation \eqref{eq:psi-a--b-approx} gives us a closed form approximation of the inner integral over $\Omega_\text{sub}$, involving only evaluations of (derivatives of) $\mathbf{u}$ and $z_\text{sub}$ at a single point $\bm{\xi}$ in the B-spline parameter space.  (First derivatives of $\mathbf{u}$ with respect to $\bm{\xi}$ are needed for the deformed-configuration surface normal $\mathbf{n}$ (cf. \cite[(3.26)]{Kiendl2011}) in the formula \eqref{eq:x_S_pm} for S sub-layer positions; first derivatives of $z_\text{sub}$ are likewise needed for the substrate normal vector $\mathbf{n}_\text{sub}$ in the formula \eqref{eq:d-A-sub} for the layer--tangent distance used in \eqref{eq:psi-a--b-approx}.)  This makes our model amenable to automated implementation using modern code generation techniques.  

In particular, we use the finite element automation software FEniCS \cite{Fenicsbook}, in conjunction with its extension for IGA, tIGAr \cite{tIGAr}.  The canonical FEniCS workflow is to specify variational forms in the high-level Python-based Unified Form Language (UFL) \cite{UFL}, then automatically compile \cite{compilerFenics} these forms into high-performance numerical kernels executed elementwise within an assembly loop, using the finite element library DOLFIN \cite{Logg:2010}.  The auxiliary library tIGAr extends this to IGA by applying the concept of Lagrange extraction \cite{Schillinger2016} to reuse finite element assembly procedures.  

For the results in this paper, we use an open-source implementation of our approximate continuum model available at \cite{snoin-code}.  This implementation leverages the existing library ShNAPr \cite{shnapr-code} (with verification results documented in \cite{shnapr}) to provide a UFL definition of $U_\text{int}$ for a Kirchhoff--Love shell.  We then add UFL definitions of $U_\text{A--B}$ for all relevant pairs of A and B to obtain \eqref{eq:total-U}.  The Gateaux derivative used to define the variational problem \eqref{eq:var-problem} is obtained automatically, in symbolic form, using computer algebra functionality in UFL, as is the Jacobian form needed to perform each step of Newton iteration (i.e.: Find $\Delta\mathbf{u}\in\mathcal{V}$ such that $D_{\Delta\mathbf{u}}D_\mathbf{v}U = -D_\mathbf{v}U~,~\forall\mathbf{v}\in\mathcal{V}$)\footnote{In practice, we apply Newton iteration in each step of an implicit pseudo-time stepping method, rather than directly to the static energy minimization; see Section \ref{sec:pseudotime}.}.  The resulting variational forms are quite algebraically complex, but the resulting compiled code is automatically optimized using The Smart Form Compiler (TSFC) \cite{Homolya2018} to maintain efficiency.

\subsection{Pseudo-time solution method}\label{sec:pseudotime}
Due to the nonlinearity of the Lennard-Jones potential, the static energy minimization problem \eqref{eq:var-problem} is difficult to solve directly with a fixed point iteration like Newton's method, even when augmented with standard line search strategies.  We find that a more robust solution procedure is to instead solve a dynamic problem with inertial and damping terms whose steady solution satisfies \eqref{eq:var-problem}.  In particular, we consider the following unsteady problem:  Find $\mathbf{u}\in\mathcal{V}$ such that $\forall \mathbf{v}\in\mathcal{V}$,
\begin{equation}
    \int_{\Gamma_0}\left(\rho_\text{fict}\ddot{\mathbf{u}} + c_\text{fict}\dot{\mathbf{u}}\right)\cdot\mathbf{v}\,d\Gamma_0 + D_\mathbf{v}U = 0\text{ ,}
\end{equation}
where $\ddot{\mathbf{u}}$ and $\dot{\mathbf{u}}$ are the shell structure's midsurface acceleration and velocity, $\rho_\text{fict}$ is a fictitious areal mass density, and $c_\text{fict}$ is a fictitious areal damping coefficient.  As this problem approaches a steady state and pseudo-time derivatives of $\mathbf{u}$ go to zero, it reduces to our target problem, \eqref{eq:var-problem}.  Because time accuracy of this fictitious unsteady trajectory is of no concern, we use the backward Euler time integration method with a time step size that is doubled whenever a time step's nonlinear solve converges and halved otherwise.  The values of $\rho_\text{fict}$ and $c_\text{fict}$ are chosen purely to expedite the convergence to a steady configuration, without regard to the actual inertia or dissipative mechanisms of the physical system.  In the present study, we focus on a specific choice of 2D and substrate materials, and simply tuned the parameters $\rho_\text{fict}$ and $c_\text{fict}$ empirically, finding that the values given in Table \ref{tab:pt-params} provide robust and efficient convergence in the cases considered.

\begin{table}[!htb]\centering
\begin{tabular}{c|c}
$\rho_\text{fict}$ (kg\,nm$^{-2}$) & $10^3$ \\
\hline
$c_\text{fict}$ (kg\,s$^{-1}$\,nm$^{-2}$) & 200 \\
\end{tabular}
\caption{Parameters of the pseudo-time solution method, empirically selected for MoS$_2$ interacting with Si$_3$N$_4$.}
\label{tab:pt-params}
\end{table}

\section{Atomistic modeling and discretization}\label{sec:ms}
To calibrate and validate the continuum model of Section \ref{sec:continuum}, we also directly simulate the atomistic system using MS.  

\subsection{Atomistic model of $\rm{MoS_2}$}\label{sec:ms-model}

In our atomistic models, the bonding between atoms in the $\rm{MoS_2}$ monolayer is modeled through empirical interatomic potentials.  Two different models are considered to explore the effect of the potential on the results and thereby obtain a sense of the uncertainty in the predictions.  First is the second-generation reactive empirical bond-order (REBO) potential \cite{Brenner2002} combined with Lennard-Jones interactions \cite{Liang2009}, with parameters for $\rm{MoS_2}$ obtained by \cite{REBO_stewart,REBO_source}.  Second is the Stillinger--Weber (SW) potential \cite{Stillinger1985}, as adapted to $\rm{MoS_2}$ by \cite{SW_jiang,SW_source}.  These interatomic potentials are implemented in the molecular dynamics/statics software LAMMPS \cite{LAMMPS_plimpton}.

\subsection{Modeling and discretization of the substrate potential}\label{sec:ms-approx-lj}

Due to the prohibitive cost of directly using MS to model the 3D substrate, the continuum model \eqref{eq:Omega_sub-int} is also used in the atomistic simulations, i.e., the substrate (but not the MoS$_2$ monolayer) is considered as a continuum.  However, the MS calculations {\em do not} make any of the geometric approximations of Section \ref{sec:geom-interact-approx}.  The integral over $\Omega_\text{sub}$ for a fixed point $\mathbf{x}_\text{A}$ is discretized directly, using the adaptive {\tt integral3()} quadrature function from {\sc Matlab}. Performing this quadrature for every atomic position $\mathbf{x}_\text{A}$ at every step of an MS energy minimization would still be impractically expensive, so the values of $\psi_\text{A--B}$ are instead precomputed for $\mathbf{x}_\text{A}$ at each point of a fine structured grid covering all anticipated deformed configurations of the $\rm{MoS_2}$ layer.  When evaluating the total energy in LAMMPS during the MS energy minimization process, the values of $\psi_\text{A--B}$ at actual atomic positions are obtained by interpolating the precomputed values with a $C^2$ cubic spline, ensuring $C^1$ force continuity \cite[Supporting Information]{holey_zhang}.  The details of the grids used for interpolation vary depending on the substrate geometry, and will be discussed alongside specific problems in Section \ref{sec:validation}.

\subsection{Initialization and energy minimization}\label{sec:ms-e-min}
To perform MS analyses, we seek a minimum of the total energy, which is a sum over the energies of the atoms comprising the $\rm{MoS_2}$ layer due to interactions with each other, as modeled by the interatomic potential (Section \ref{sec:ms-model}), and to their interactions with the substrate (Section \ref{sec:ms-approx-lj}).  This minimization is performed with LAMMPS, using the conjugate gradient (CG) method and the fast inertial relaxation engine (FIRE) solver \cite{FIRE_gu} as described below.  The nominal equilibrium positions of Mo and S atoms in an isolated $\rm{MoS_2}$ layer (with no substrate interaction) may differ slightly from the true equilibrium positions for a given choice of interatomic potential (e.g., REBO or SW).  This difference would effectively manifest as a pre-strain of the 2D material if the nominal atomic positions were used directly.  To avoid introducing such a pre-strain, we first minimize energy in the absence of a substrate, to obtain an initial reference configuration of the $\rm{MoS_2}$.  This reference configuration is then used as the initial condition for subsequent minimization in different boundary-value problems.

\subsection{Postprocessing of strain and curvature}\label{sec:ms-strain}
For both strain engineering applications and comparison with a continuum shell model, we need to be able to estimate local strains in the MS system.  Because our Kirchhoff--Love shell model is based on the Green--Lagrange strain tensor $\mathbf{E}$ and the mechanics of very thin shells are dominated by membrane strains (as opposed to bending strains), we estimate the in-plane part of $\mathbf{E}$ for the MS results for comparison with the shell analysis.  In continuum mechanics, $\mathbf{E}$ is defined by the property that
\begin{equation}
    \mathbf{U}\cdot\mathbf{E}\mathbf{V} = \frac{1}{2}\left(\mathbf{u}\cdot\mathbf{v} - \mathbf{U}\cdot\mathbf{V}\right)\text{ ,}
\end{equation}
where $\mathbf{U}$ and $\mathbf{V}$ are two vectors in the tangent space of the reference configuration, and $\mathbf{u} = \mathbf{F}\mathbf{U}$ and $\mathbf{v}=\mathbf{F}\mathbf{V}$ are their pushforwards by the deformation gradient $\mathbf{F}$.  The in-plane strain has three unique components, so, given at least three linearly-independent in-plane material directions $\{\mathbf{D}_i\}_{i=1}^{\geq 3}$ with pushforwards $\{\mathbf{d}_i\}$, one can obtain the in-plane components $E_{11}$, $E_{12}$, and $E_{22}$ by solving
\begin{equation}
    \mathbf{D}_i\cdot\mathbf{E}\mathbf{D}_i = \frac{1}{2}\left(\vert\mathbf{d}_i\vert^2 - \vert\mathbf{D}_i\vert^2\right)\quad,\quad i=1,2,\cdots,\geq 3\text{ ,}
\end{equation}
possibly in a least-squares sense if more than three directions are used.  For an $\rm{MoS_2}$ monolayer, we estimate $\mathbf{E}$ at each Mo atom by taking $\{\mathbf{D}_i\}$ as the six displacement vectors to neighboring Mo atoms, and $\{\vert\mathbf{d}_i\vert^2\}$ as their deformed lengths.  

\begin{remark}
Additional equations can be obtained by taking $\mathbf{U}=\mathbf{D}_i$ and $\mathbf{V}=\mathbf{D}_j$ for $i\neq j$, which reduces the minimum number of directions needed to two, but our system of interest already provides more than three natural choices of $\mathbf{D}_i$ corresponding to neighboring Mo atoms, so we consider only changes in length.  
\end{remark}

This procedure gives an effective Green--Lagrange strain $\mathbf{E}_\text{Mo}$ of the central Mo sub-layer.  If the full $\rm{MoS_2}$ layer followed Kirchhoff--Love kinematics, we would expect this to coincide with the shell theory's in-plane Green--Lagrange strain at its midsurface, $\xi_3 = 0$.  However, one should recall that the Kirchhoff--Love kinematics are only assumed to hold for the fictitious formal 3D material, as a device to obtain effective membrane and bending stiffness.  We observe from MS calculations that bending results in strain of the Mo sub-layer.\footnote{This is likely due to the effect of bending on the energies of the top and bottom S layers. In the reference state, both layers have the same energy, but during bending, they will differ, since one layer will be in compression and the other in tension, changing bond lengths and angles.}
The influence of bending on the Mo sub-layer strain must be symmetric with respect to the direction of bending (based on the symmetry of the atomic structure) and smooth (based on the functional form of interatomic potentials).  We therefore propose to estimate the strain in the Mo layer from continuum results as
\begin{equation}
    \mathbf{E}^\text{CM}_\text{Mo} = \bm{\varepsilon} + C_\text{bend}\left(\vert\kappa_1\vert^2\bm{\Lambda}_1\otimes\bm{\Lambda}_1 + \vert\kappa_2\vert^2\bm{\Lambda}_2\otimes\bm{\Lambda}_2\right)\text{ ,}\label{eq:e-mo-from-cont}
\end{equation}
where $\{\kappa_i\}$ are the eigenvalues of the bending strain $\bm{\kappa}$ (defined by \eqref{eq:kappa-defn}),\footnote{Eigenvalues and eigenvectors are technically of the associated tensor $\mathbf{G}^{-1}\bm{\kappa}$, but, in practice, we use a local orthonormal basis, where the distinction is moot.} $\{\bm{\Lambda}_i\}$ are the associated eigenvectors, $C_\text{bend}$ is a parameter with dimensions of length squared, to be calibrated from MS calculations, and $C_\text{bend}\vert\cdot\vert^2$ can be interpreted as a truncated Taylor expansion of some more complicated smooth even function.

Another deformation measure that we would like to compare between atomistic and continuum results is the bending strain $\bm{\kappa}$.  Because the reference configuration is flat, $\mathbf{B}=\mathbf{0}$, and this is simply the second fundamental form of the deformed configuration (up to sign and pullback), and its absolute eigenvalues (as used in \eqref{eq:e-mo-from-cont}) are the absolute principal curvatures of the deformed configuration.  To estimate this quantity from the MS results, we treat the Mo atoms as points on the midsurface.  As discussed earlier in the context of membrane strains, the identification of the Mo sub-layer with the Kirchhoff--Love shell's midsurface should not be considered exact, but it is effective for estimating the bending strain.  To estimate the second fundamental form $\mathbf{b}$ from the discrete set of points provided by the positions of Mo nuclei, we use the following relation:
\begin{equation}\label{eq:2nd-fund-form}
    \mathbf{b}(\mathbf{u},\mathbf{v}) = -(\nabla_\mathbf{u}\mathbf{n})\cdot\mathbf{v}\text{ ,}
\end{equation}
where $\mathbf{u}$ and $\mathbf{v}$ are tangent vectors to the surface and $\mathbf{n}$ is the unit normal vector to the surface.  We first estimate a normal vector for each Mo atom, by performing a principal component analysis of its position and of the positions of its six neighboring Mo atoms.  We then obtain a set of algebraic equations, by using pairs of vectors between neighboring Mo atoms as $\mathbf{u}$ and $\mathbf{v}$ in \eqref{eq:2nd-fund-form} and approximating the directional derivative $\nabla_\mathbf{u}\mathbf{n}$ by a finite difference between normal vectors at neighboring Mo atoms.  This set of equations is solved in a least-squares sense for the three unique components of a matrix representation of $\mathbf{b}$, in a procedure analogous to how we solved for the components of $\mathbf{E}_\text{Mo}$. 

\section{Calibration of continuum model parameters}\label{sec:calibration}
To obtain the effective values of $E$, $\nu$, and $h_\text{th}$ for the Kirchhoff--Love shell model of Section \ref{sec:kl}, we calibrate it to match the small-strain behavior of an MS model for two classical structural mechanics problems: constrained uniaxial tension of a membrane and bending of a doubly-clamped plate.  We then use a pure bending test to calibrate the parameter $C_\text{bend}$, used to estimate the effect of continuum bending strain on the Green--Lagrange strain of the Mo sub-layer (cf. \eqref{eq:e-mo-from-cont}).  

\subsection{Constrained uniaxial tension}\label{sec:tensile}
\begin{figure}[H] 
\begin{center}
\subfloat[Reference configuration.]{\includegraphics[trim = 0mm 0mm 0mm 0mm, clip=true,width=0.49\textwidth]{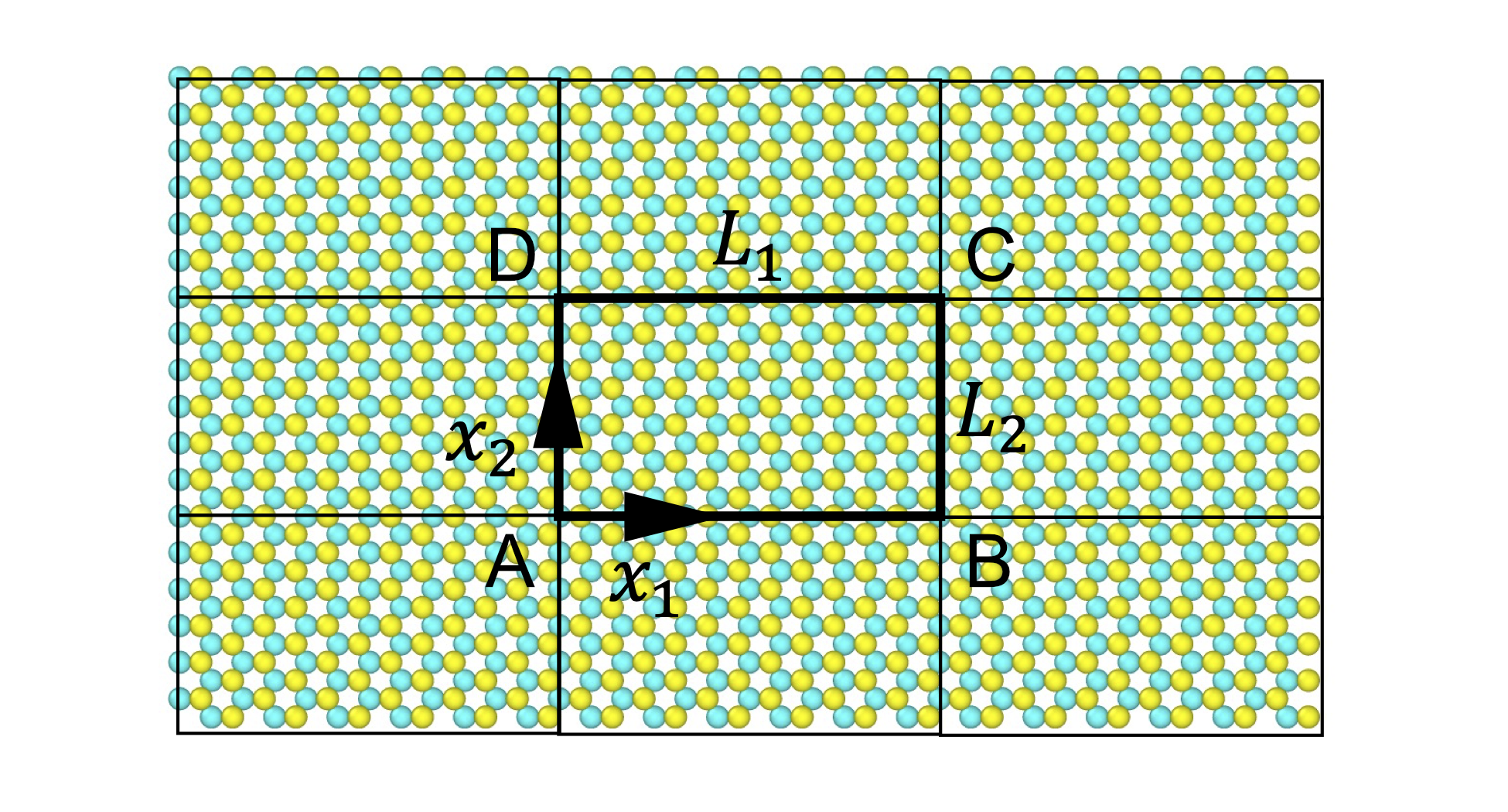}\label{MS_tensile_undeformed}}\hfill
\subfloat[Deformed configuration.]{\includegraphics[trim = 0mm 0mm 0mm 0mm, clip=true,width=0.49\textwidth]{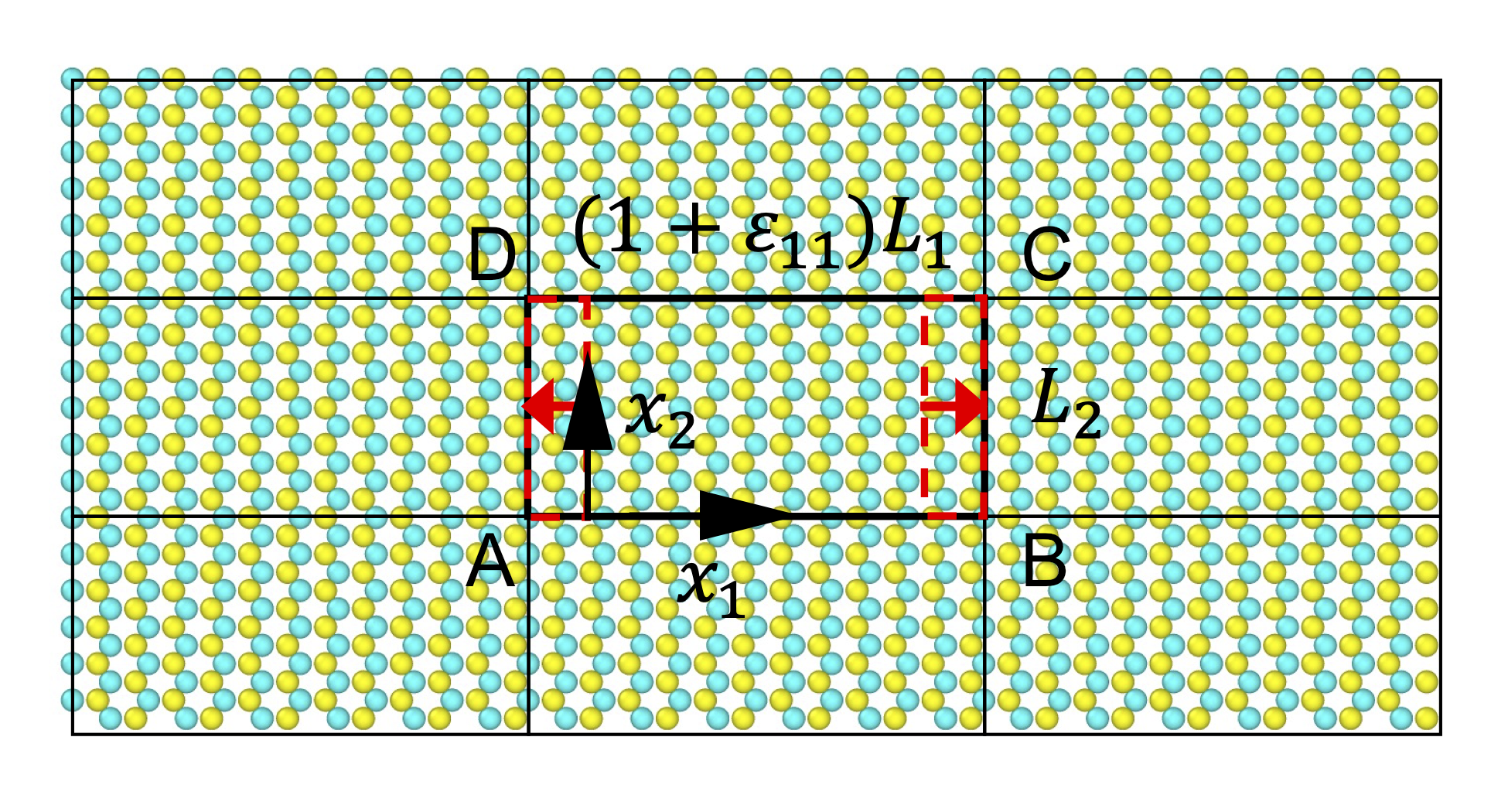}\label{MS_tensile_deformed2}}\hfill  
\caption{The monolayer $\rm{MoS_2}$ for the uniaxial tensile test.}
\label{fig:MS_tensile}
\end{center}
\end{figure}

Uniaxial tension is applied to a monolayer of MoS$_2$ that is prevented from contracting in the perpendicular direction. The geometric domain ABCD of the monoloayer in its reference and deformed configuration is shown in Figure \ref{fig:MS_tensile}, together with first periodic images.
The monolayer domain ABCD, i.e., the simulation box, has a size  of $3.2 \times 1.8\ \rm{nm}^2$. Periodic boundary conditions are applied along both the 1- and 2-directions. The monolayer is oriented with the armchair direction (see Figure~\ref{fig:mos2_structure}) along the $x_1$ axis.\footnote{See Remark~\ref{rem:isotropy} on the isotropy of the response and insensitivity to the monolayer's orientation.} Before performing the uniaxial tensile test, the energy of the $\rm{MoS_2}$ monolayer is minimized to obtain the equilibrium reference configuration, allowing the size of the simulation box to change along the 1- and 2-directions. The minimization is performed using the CG method and FIRE, successively, until the change in force is less than $10^{-9}$ eV/\text{\AA}. After the energy minimization, a small positive strain $\varepsilon_{11}$ is applied by increasing the length of the simulation box along the $1-$direction ($AB$ and $CD$). The length of the simulation box along the 2-direction ($AD$ and $BC$) is fixed to maintain $\varepsilon_{22} = 0$. FIRE is used to minimize the energy of the strained structure. 

During the minimization, all atoms are allowed to move in any direction, since the $x_1$ component of the Mo-S bond in the MoS$_2$ monolayer is not constant along the armchair direction (different $x_1$ components of Mo-S bonds will lead to different Mo-S distances when the atoms are remapped in the $x_1$ direction within the strained box).
We then measure the normal stress $\sigma_{11}$ and $\sigma_{22}$ of the minimized structure in the unit of force per length. 
Assuming a condition of plane stress ($\sigma_{33} = 0$), we have
\begin{equation}\label{eq:nu}
    \nu = \frac{\sigma_{22}}{\sigma_{11}},
\end{equation}
entirely in terms of quantities measured from the MS system. In addition, we have the following formula relating $E$ and $h_\text{th}$:
\begin{equation}\label{eq:E-h_th-tensile}
    \frac{E\varepsilon_{11}}{1-\nu^2} = \frac{\sigma_{11}}{h_\text{th}}\text{ .}
\end{equation}
We obtain a second equation for $E$ and $h_\text{th}$ from the bending test described next.  

\subsection{Doubly-clamped plate bending}\label{sec:bending}
\begin{figure} [ht]
\centering
\vspace{3pt}
\setlength{\abovecaptionskip}{1pt}   
\setlength{\belowcaptionskip}{0pt}   
\includegraphics[trim = 0mm 0mm 0mm 0mm, clip=true,width=0.9\textwidth]{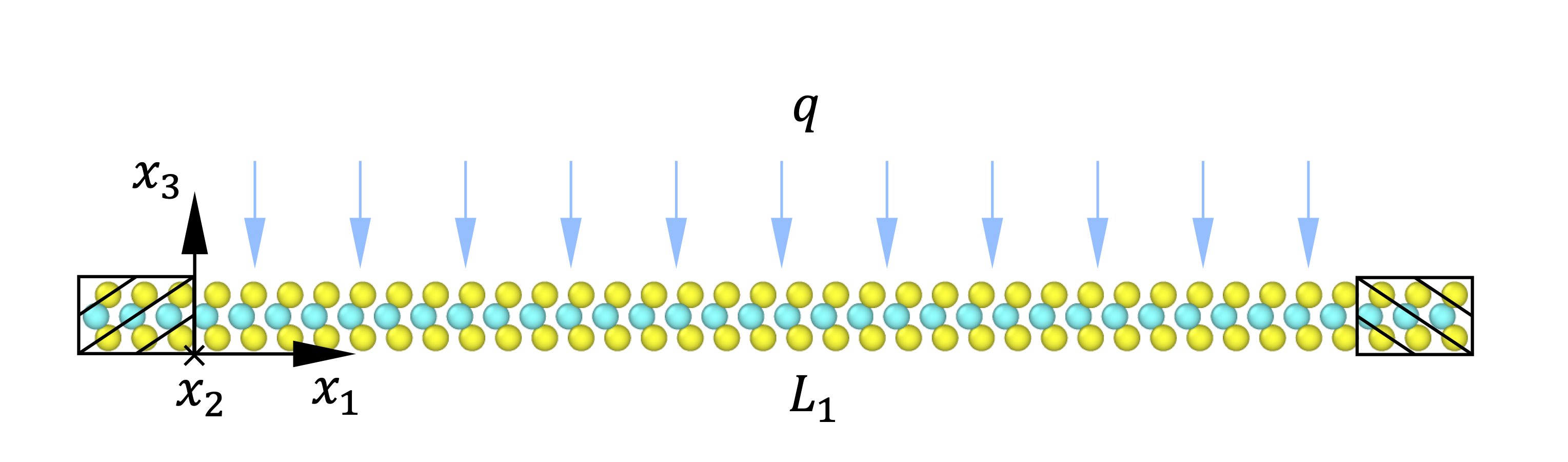}
\caption{Side view of a doubly-clamped $\rm{MoS_2}$ monolayer subjected to a uniform distributed load $q$ per unit area. The length of the ribbon in the 1-direction is $L_1$ and its width in the 2-direction is $L_2$.}
\label{fig:MS_bending_doubleclamped}
\end{figure}

An MoS$_2$ plate is clamped at both ends and subjected to a downward distributed load per unit area $q$. The geometry and loading of the plate are shown in Figure \ref{fig:MS_bending_doubleclamped}.

The monolayer has a size of $9.7 \times 2.7\ {\rm nm}^2$ with an additional 0.1 {\rm nm} on each side in the 1-direction employed for imposition of boundary conditions. Periodic boundary conditions are applied along the 1- and 2-directions of the extended domain during the relaxation phase. As in the uniaxial tension case, the plate is first relaxed to obtain the equilibrium reference configuration. After the relaxation, the atoms within 1.0\,nm of the right and left edges are constrained along $x_1$ and $x_3$ directions to represent a doubly-clamped boundary condition, and a periodic boundary condition is applied along the 2-direction. To apply a uniform distributed load on the monolayer, each Mo and S atom (except for those constrained near the edges) is subjected to an equal force in the negative 3-direction, such that the total force divided by the unconstrained area ($L_1L_2$) of the beam equals $q$.
For small loads, the beam theory solution for the deflection at the center of a doubly-clamped plate subjected to a distributed load $q$ is %
\begin{equation}
    w_\text{max} = \frac{L_1^4 L_2 q}{384 E^* I}\text{ ,}
    \label{eq:w_max}
\end{equation}
where $E^* = {E}/{(1-\nu^2)}$ and $I=h_\text{th}^3L_2/12$. Extracting $w_\text{max}$ from the MS simulation, we solve the system \eqref{eq:E-h_th-tensile} and \eqref{eq:w_max} for the remaining parameters $E$ and $h_\text{th}$.
\subsection{Pure bending}\label{sec:pure_bending}
\begin{figure} [ht]
\centering
\vspace{3pt}
\setlength{\abovecaptionskip}{1pt}   
\setlength{\belowcaptionskip}{0pt}   
\includegraphics[trim = 0mm 0mm 0mm 0mm, clip=true,width=0.8\textwidth]{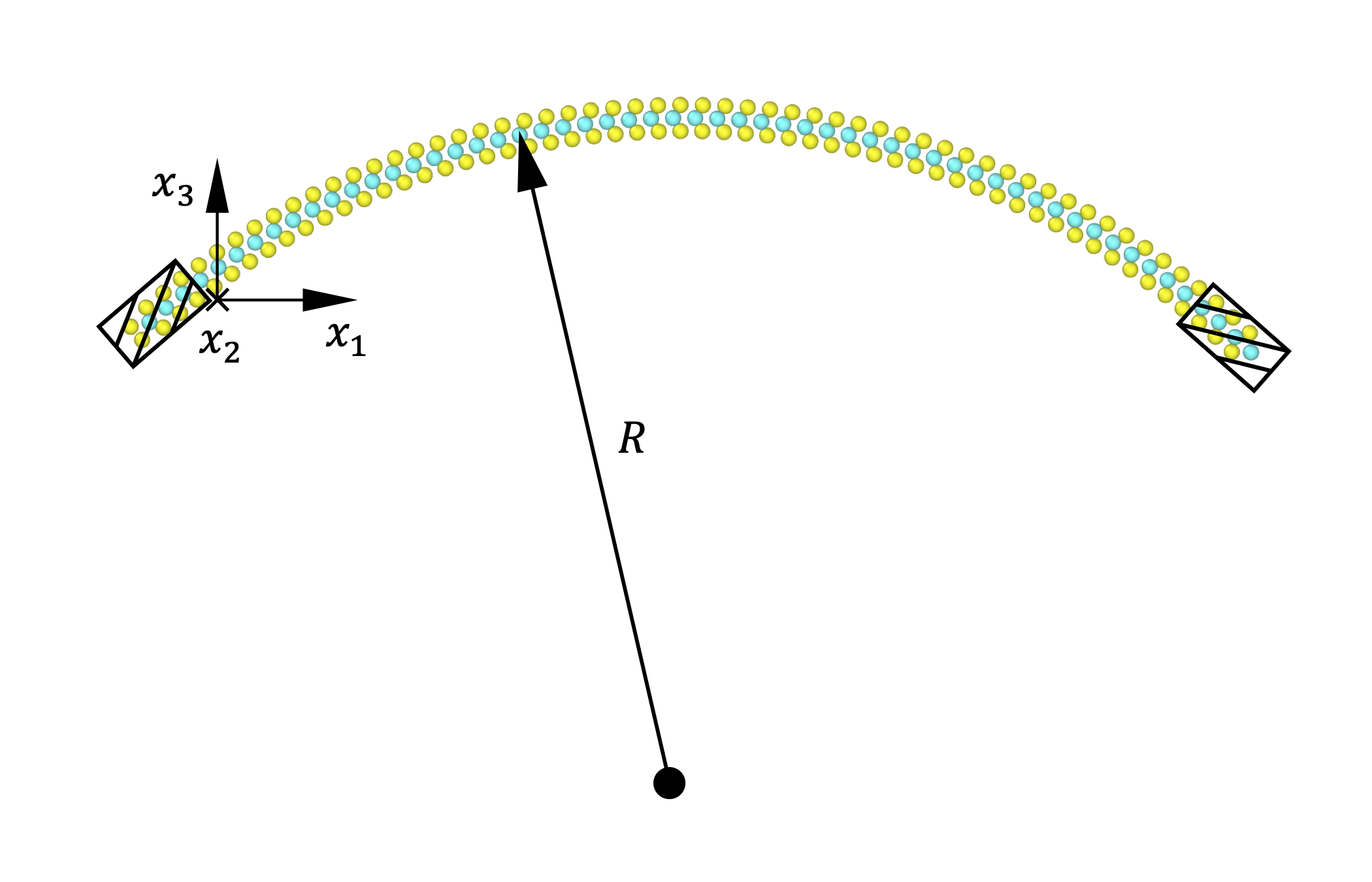}
\caption{Side view of a pure bending configuration of $\rm{MoS_2}$ with a radius of curvature $R$.}
\label{fig:MS_pure_bending}
\end{figure}

To isolate the effect of bending on $\mathbf{E}_\text{Mo}$, we perform an MS analysis of pure bending configurations. In beam theory, pure bending means that the deformed configuration of the beam is a circular arc involving no axial strain, and the extension of this to plate theory is that the deformed configuration of the plate is a section of a circular cylinder.  To model this using MS, these kinematics are prescribed at the edges of a specimen, as shown in Figure \ref{fig:MS_pure_bending}. 
We arranged the positions of atoms of an initially-flat $\rm{MoS_2}$ monolayer (of dimensions $15.3 \times 1.5\ {\rm nm}^2$, with 840 atoms) into a pure bending configuration. In the pure bending configuration, the radius of curvature of Mo atoms is $R$, as shown in the diagram. The maximum principal curvature of this configuration is $\vert\kappa_1\vert = R^{-1}$.  Top and bottom S atoms are arranged to have curvature radii $R+d_{\rm{Mo-S}}$ and $R-d_{\rm{Mo-S}}$, respectively.  Periodic boundary conditions are applied in the 2-direction. The atoms within $1.0\,\rm{nm}$ of the right and left edges of the flat monolayer are constrained after bending to maintain the curvature.  After relaxation of all atoms excepts for the constrained atoms, $\mathbf{E}_\text{Mo}$ is calculated at the center of the specimen. We perform this test for several values of $R$, and find that the relationship between the imposed $\vert\kappa_1\vert^2$ and extracted $(\mathbf{E}_\text{Mo})_{11}$ is very nearly linear, as assumed by the model \eqref{eq:e-mo-from-cont}.  The slope of this curve is therefore taken as $C_\text{bend}$.

\subsection{Calibrated parameters}\label{sec:results}
We now collect the results of the calibration tests of Sections \ref{sec:tensile}, \ref{sec:bending}, and \ref{sec:pure_bending} in Table \ref{tab:shell-params}.  
\begin{table}[!htb]\centering
\begin{tabular}{c|c|c}
$E$ & REBO & SW\\
\hline
$E$ (kg\,nm$^{-1}$s$^{-2}$) & 300.058 & 241.355\\
\hline
$\nu$ & 0.2962 & 0.2686\\
\hline
$h_\text{th}$ (nm) & 0.4671 & 0.4014\\
\hline
$C_\text{bend}$ (nm$^2$) & 0.048 & 0.020 
\end{tabular}
\caption{Calibrated parameters for the Kirchhoff--Love shell model of an $\rm{MoS_2}$ monolayer.}
\label{tab:shell-params}
\end{table}

\begin{remark}
\label{rem:isotropy}
The structure of an $\rm{MoS_2}$ monolayer has two distinct directions, referred to as ``zigzag'' and ``armchair'' in the literature (see Figure~\ref{fig:mos2_structure}).  MS simulations of the tensile test for both the zigzag and armchair directions produce very similar results, justifying the assumption of isotropy.  Previous work has also found that bending is essentially isotropic \cite{Xiong2016BendingRO}.  The values reported in Table \ref{tab:shell-params} correspond to taking the beam's axis to be the armchair direction. If the axis is the zigzag direction, the shell model's parameters are $E = 313.112$\,kg\,nm$^{-1}$s$^{-2}$, $\nu = 0.2962$, and $h_\text{th} = 0.4472$\,nm for the REBO potential, and $E = 245.069$\,kg\,nm$^{-1}$s$^{-2}$, $\nu = 0.2682$, and $h_\text{th} = 0.3959$\,nm for the SW potential. The values for the zigzag direction are very close to the  armchair direction values, and lead to essentially equivalent solutions for the deflection of the $\rm{MoS_2}$ layer. 
\end{remark}

\begin{remark}
Some prior work has explored material nonlinearity \cite{Mortazavi2016,Li2017}, and nonlocal effects \cite{Li2017,Barretta2019,Shokrieh2015} in 2D materials and nano-scale materials.  We find that the stress--strain behavior of the MS results is indeed nonlinear at larger strains, and that taking smaller specimens can lead to different material properties (which implies a length scale associated with nonlocal effects).  However, at the strain levels and specimen sizes relevant to 2D materials interacting with holey substrates, material nonlinearity and nonlocality do not play significant roles.
\end{remark}

\section{Comparison of IGA and MS for layer--substrate interaction}\label{sec:validation}
 After calibrating the continuum model, we now compare it with atomistic simulations on a series of validation tests that more closely resemble the target system of $\rm{MoS_2}$ interacting with a holey substrate.  We test the accuracy of the continuum model in Section \ref{sec:accuracy} and compare computational costs of atomistic and continuum models in Section \ref{sec:comp-cost}.  The systems that are directly accessible to MS analysis for validation are still significantly smaller than the size of geometric features in commercially-available holey substrates (cf. Figure \ref{fig:afm_substrate}) that are of practical interest for strain engineering.  However, this discrepancy in scale actually makes the validation stronger, in the sense that our continuum and geometric approximations from Section \ref{sec:continuum} become more reasonable as the system size increases.  Smaller-scale continuum--atomistic comparisons therefore serve as a stress test of the methodology.

\subsection{Accuracy of the continuum model}\label{sec:accuracy}
This section establishes a series of validation tests where the substrate surface $z_\text{sub}$ is defined analytically, and compares the results of MS and Kirchhoff--Love shell theory.  We take deflection of the $\rm{MoS_2}$ layer as a quantity of interest in this validation study and consider an acceptable tolerance for modeling error to be: continuum--atomistic discrepancies on the same order as differences between the two independently calibrated interatomic potentials (REBO and SW) in the atomistic modeling.  In other words, the continuum model can be seen as a useful substitute for atomistic simulation if it is as close to two widely-used atomistic models as the atomistic models are to each other.  

\subsubsection{Peeling of $\rm{MoS_2}$ from a flat substrate}\label{sec:peel}

\begin{figure}[H]
\vspace{10pt} 
\centering
\includegraphics[trim = 0mm 0mm 0mm 120mm, clip=true,width=0.5\textwidth]{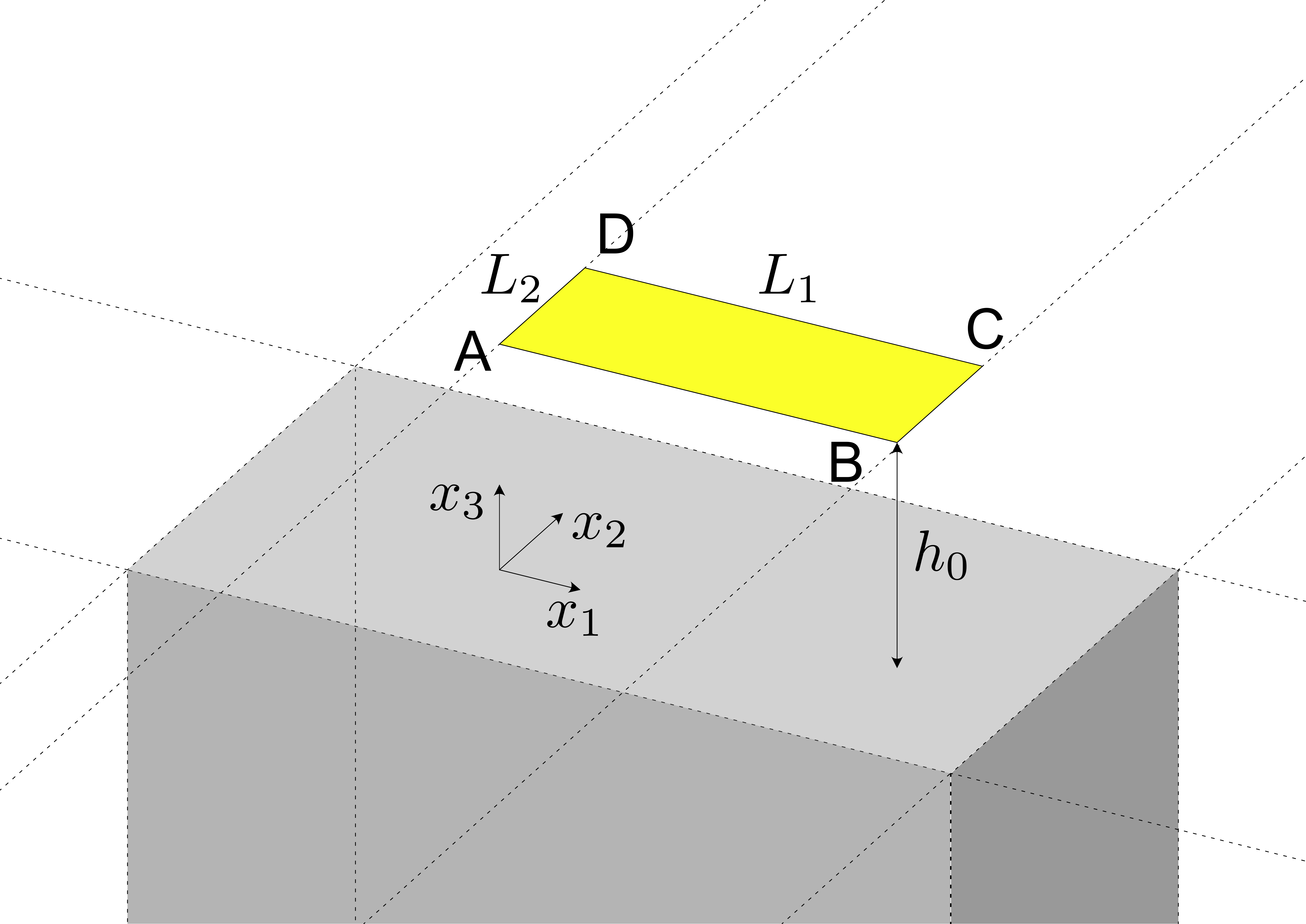}
\caption{Initial setup of an MoS$_2$ monolayer over a half-space flat substrate. The $1$-$2$ plane coincides with the top of the substrate, while $h_0$ is the initial height of the MoS$_2$ monolayer relative to the top of the substrate.}
\label{fig:peeling_test_setup}
\end{figure}

The first case we consider involves an MoS$_2$ monolayer over a flat substrate. The left end AD of the monolayer is pulled upward by imposing a fixed displacement. This causes the monloayer to peel away from the substrate, overcoming the vdW interactions. The use of a flat substrate isolates the approximation of $\rm{MoS_2}$ material behavior, since the geometric approximation of Section \ref{sec:geom-interact-approx} is not applied.  For consistency, in this case, the MS calculations also use the analytical half-space integral \eqref{eq:psi-a--b-approx} rather than interpolating between pre-computed energy values (as described in Section \ref{sec:ms-approx-lj}).

For both the MS and continuum simulations, we consider an $\rm{MoS_2}$ layer with dimensions
$L_1 = 33.94$\,nm and 33.48\,nm for the SW and REBO potentials, respectively, along the 1-direction, and $L_\text{2}=7.28$\,nm for the 2-direction.
Periodic boundary conditions are applied along the 2-direction.
Due to the imposed periodicity, $u_2=0$ and the solution is independent of $x_2$, which effectively reduces the setup to a 2D problem in the $1$-$3$ plane.
The top surface of the half-space substrate is $z_\text{sub}=0$\,nm, while the $\rm{MoS_2}$ layer reference position is $x_3=h_0=0.42$\,nm. In the continuum model, we fix $u_1 = 0$\,nm, on segment BC and impose impose $u_3=h_{p}-h_{0}$ on segment AD (see Figure \ref{fig:peeling_test_setup}). Consistent with this, in the MS simulation, Mo atoms are fixed to have $u_3=h_{p}-h_{0}$ on AD, and are fixed to have $x_1=0$ on BC. The S atoms on AD and BC are left unconstrained, so the structure is free to rotate about the $x_2$ axis.\footnote{In practice, some initial displacement must also be applied to S atoms on AD, to stably initialize the minimization procedure before relaxing to the unconstrained configuration.}
In the continuum simulation, the $\rm{MoS_2}$ layer is discretized using 100 $C^1$-continuous quadratic B-spline elements along the length in the $1-$direction. The MS simulations consist of a total of 8625 atoms.

\begin{figure}[H] 
\begin{center}
\subfloat[REBO potential.]{\includegraphics[trim = 30mm 90mm 30mm 90mm, clip=true,width=0.49\textwidth]{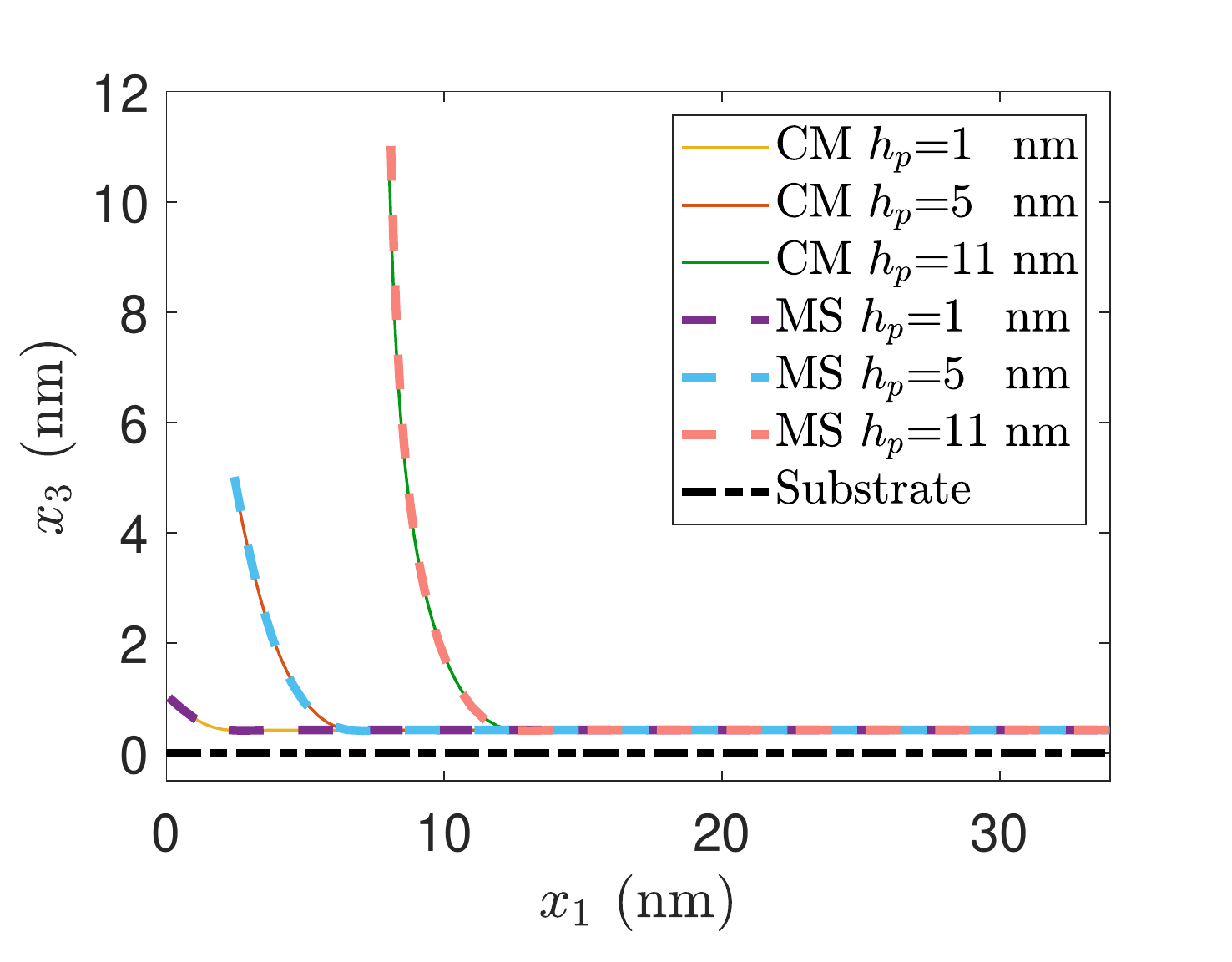}\label{peeling_test_results_REBO}}\hfill
\subfloat[SW potential.]{\includegraphics[trim = 30mm 90mm 30mm 90mm, clip=true,width=0.49\textwidth]{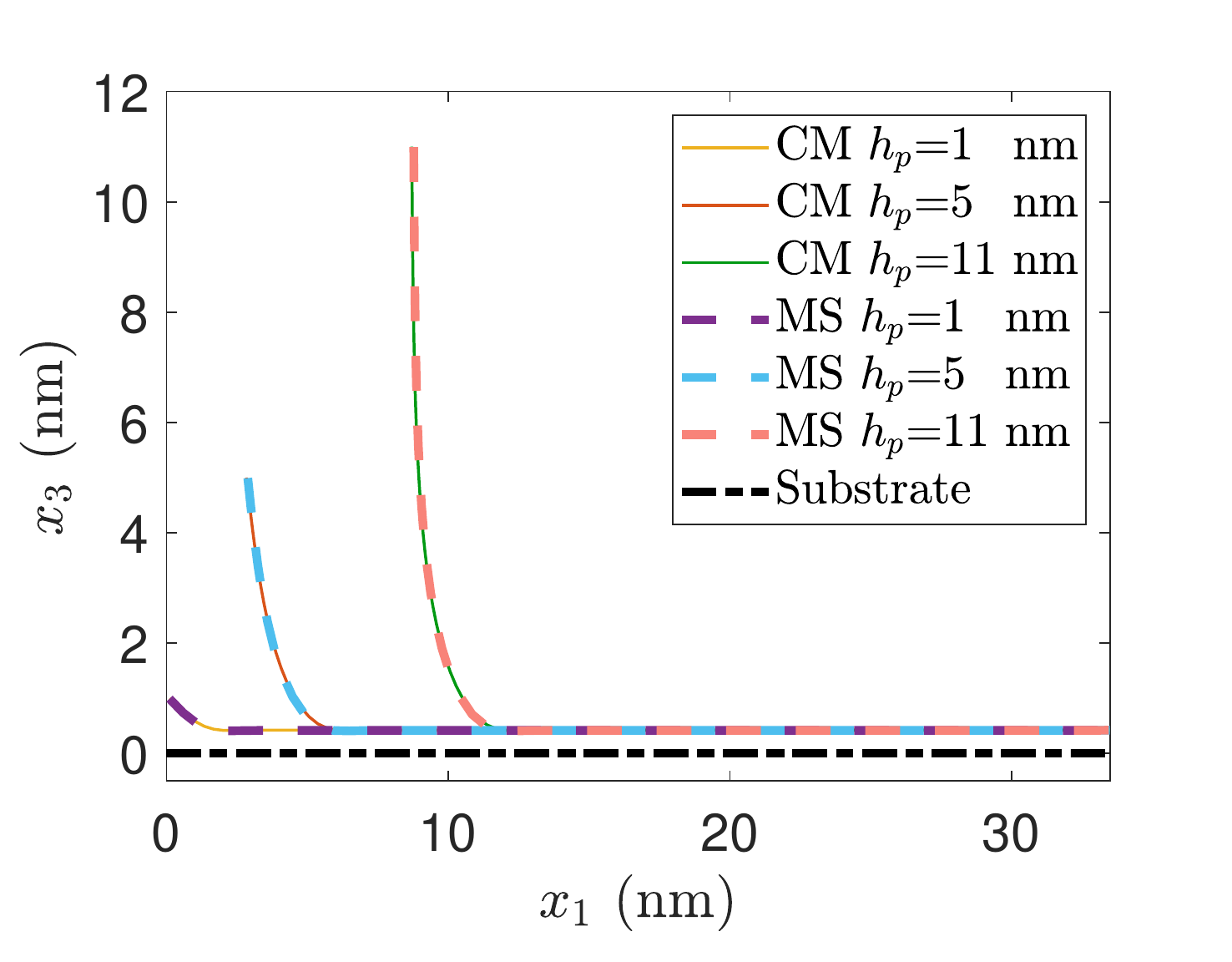}\label{peeling_test_results_SW}}\hfill  
\caption{Cross-sections of the peeled monolayer (solid and dashed lines) for several $h_p$ values, corresponding to continuum modeling (CM) and molecular statics (MS), for both the REBO and SW potentials.}
\label{peeling_test_results}
\end{center}
\end{figure}

The energy of the continuum and MS systems are minimized subject to the imposed constraints using a pseudo-time solution method and FIRE, respectively, as described above. The results are presented in Figure~\ref{peeling_test_results}, showing excellent agreement between the continuum and MS models for a range of end displaecements $h_p$ for both the SW and REBO potentials.  The bending near the peel-off point is induced by tension in the separated part of the monolayer, which undergoes a large rotation, of almost $90^\circ$ in the case of $h_p=11$\,nm.  Modeling this with Kirchhoff--Love shell theory requires the geometric nonlinearity of our formulation from Section \ref{sec:kl} in terms of the Green--Lagrange strain.  However, the linear material model \eqref{eq:svk} remains sufficient, even in the presence of large deflections.

\subsubsection{$\rm{MoS_2}$ monolayer over a trench}\label{sec:trench}

\begin{figure}[H]
\centering
\vspace{10pt}  
\includegraphics[trim = 10mm 60mm 10mm 60mm, clip=true,width=0.49\textwidth]{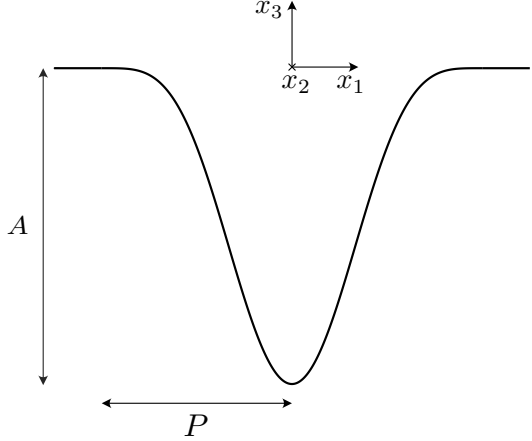}
\caption{Cross section representation in the $x_1$-$x_3$ plane of the trench substrate (see Eq. \ref{eq:trench_sub}). The structure is infinite in the $2-$direction.}
\label{fig:trench_cross_section}
\end{figure}

Next we consider an unloaded MoS$_2$ monolayer placed over a substrate containing a curved trench. Upon relaxation, the monolayer is pulled down into the trench due to the vdW interactions with the substrate. The non-flat substrate profile tests the geometric approximation in the continuum model. The substrate surface function is given by

\begin{equation}\label{eq:trench_sub}
    z_\text{sub} = \left\{\begin{array}{lcr}-\frac{A}{4}\left(\cos(\frac{\pi x_1}{P})+1\right)^2 &,&\vert x_1\vert < P~,\\ 0&,&\text{otherwise ,}\end{array}\right.
\end{equation}
where P modulates the width of the trench and A modulates its depth. This trench is considered to be part of a periodic array of trenches, with centers separated by distance $L = 40$\,nm in the $1$-direction. We take $A=16$\,nm and $P=16$\,nm, and consider a periodic $\rm{MoS_2}$ layer with dimensions
$L_\text{1}=40$\,nm, $L_\text{2}=2.7$\,nm, along the 1- and 2-directions, respectively, while its reference position is $x_3=0.42$\,nm. Periodic boundary conditions are applied in the 1- and 3-directions. As for the peeling problem discussed in Section \ref{sec:peel}, due to the problem setup and to the periodicity in the 2-direction, $u_2=0$ and the solution is independent of $x_2$.
In the continuum simulation the $\rm{MoS_2}$ layer is discretized using  100 $C^1$-continuous quadratic B-spline elements along the 1-direction. The MS simulations consist of a total of 9024 atoms.

\begin{figure}[H] 
\begin{center}
\subfloat[Results on a cross-section parallel to the $x_1$-$x_3$ plane.]{\includegraphics[trim = 10mm 60mm 10mm 60mm, clip=true,width=0.49\textwidth]{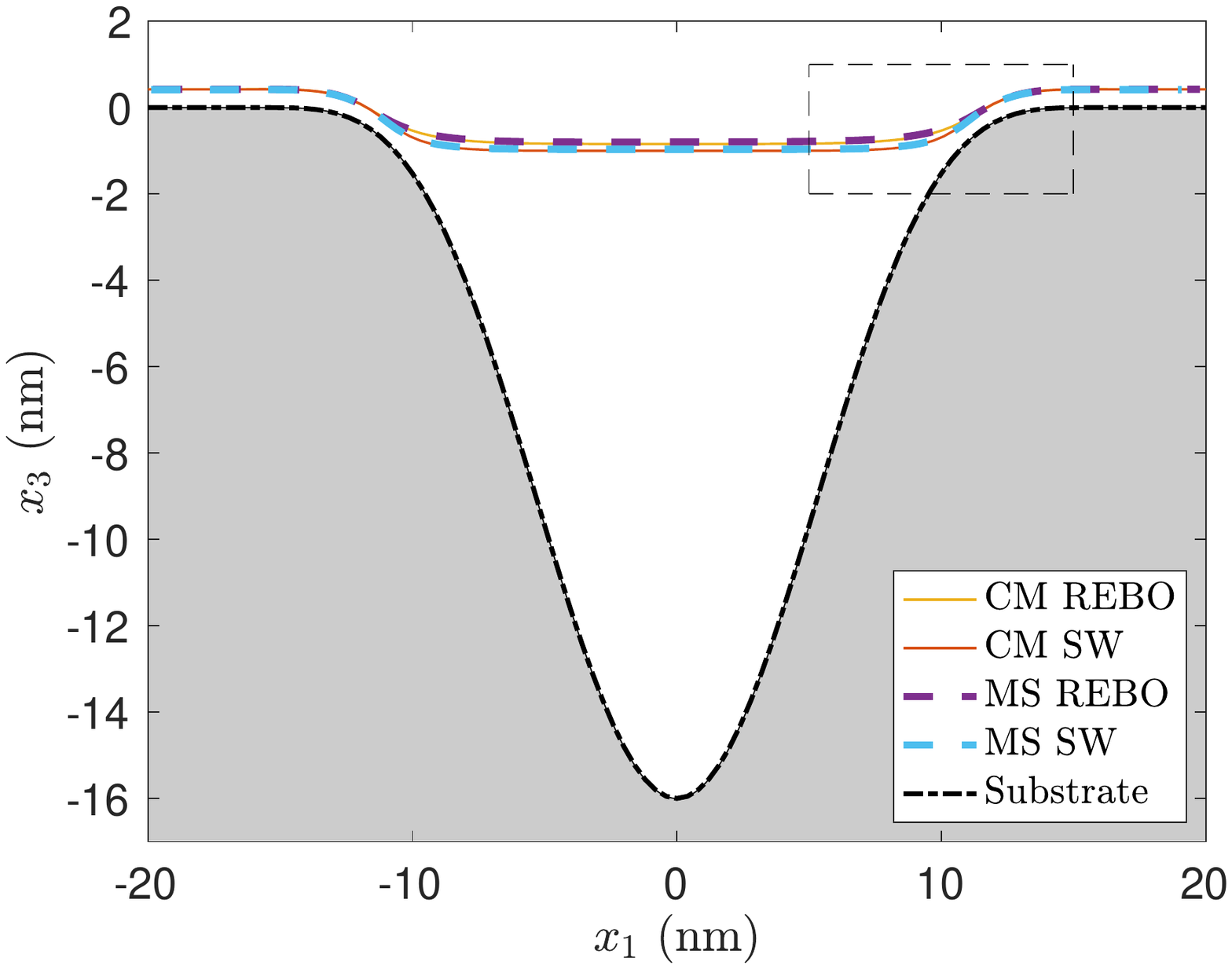}\label{trench_results}}\hfill
\subfloat[Zoomed-in view corresponding to the dashed rectangle in Figure \ref{trench_results}.]{\includegraphics[trim = 10mm 60mm 10mm 60mm, clip=true,width=0.49\textwidth]{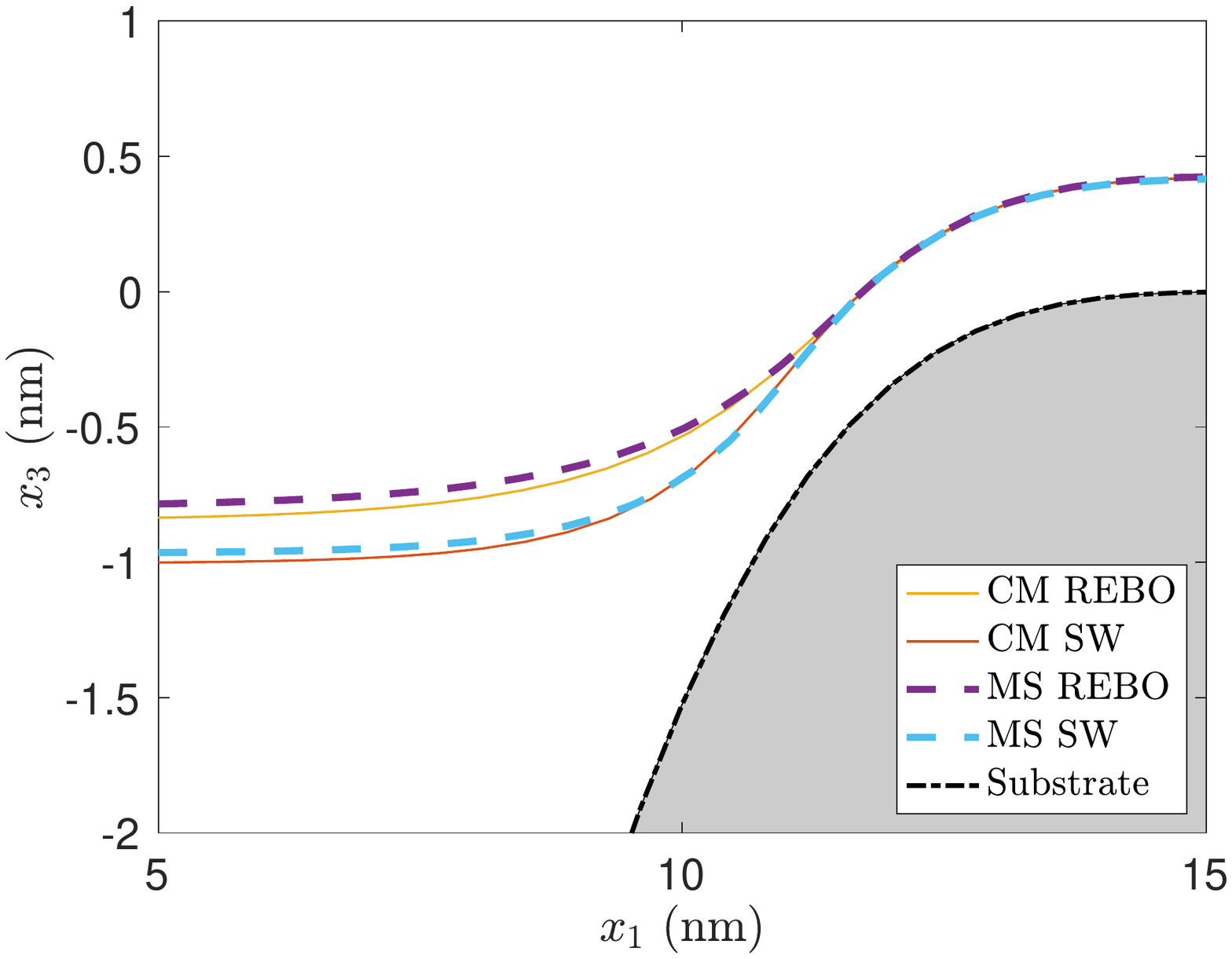}\label{trench_results_zoom}}\hfill  
\caption{Cross-sections of the deflected monolayer (solid and dashed lines) over the trench substrate (gray region), corresponding to continuum modeling (CM) and molecular statics (MS), for both the REBO and SW potentials.}
\label{trench_results_full}
\end{center}
\end{figure}

In this problem, the potential \eqref{eq:Omega_sub-int} is independent of $x_2$, so only a 2D grid is needed to interpolate precomputed energy values for the MS calculations.  (Integration over $\mathbb{R}$ with respect to $x_2$ is performed analytically.) 
Grid points have a spacing of 0.01\,nm in both 1- and 3-directions. The same solvers described in Section~\ref{sec:peel} are used.
Figure \ref{trench_results_full} shows the results obtained with both the continuum and MS approaches. The agreement is not as good as in Section~\ref{sec:peel}, due to the geometric approximation in the continuum model, but is still very good.

\subsubsection{$\rm{MoS_2}$ monolayer over a circular hole}\label{sec:hole}

\begin{figure} [ht]
\centering
\vspace{10pt}  
\includegraphics[trim = 0mm 0mm 0mm 0mm, clip=true,width=0.5\textwidth]{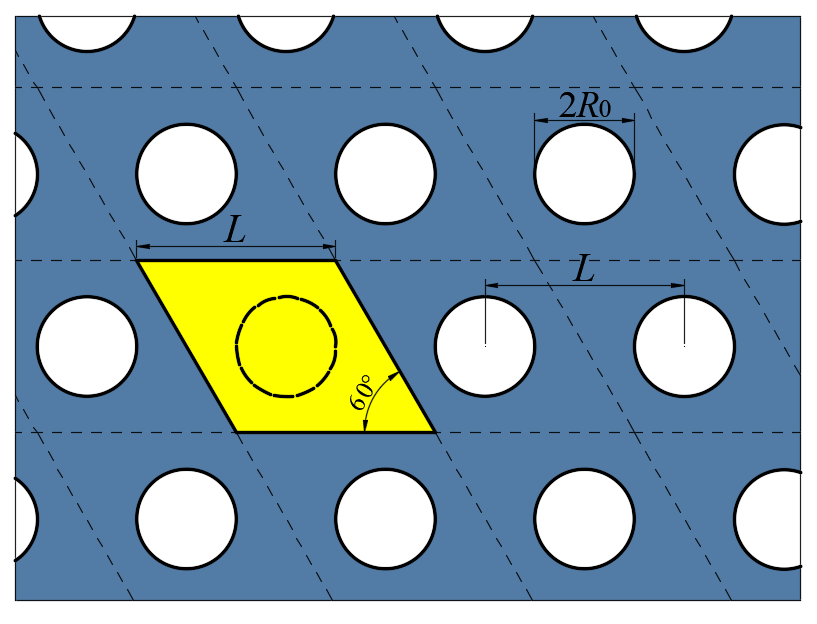}
\caption{Geometry of the substrate with a hexagonal pattern of holes. The geometry of the continuum shell is highlighted in yellow and it overlaps with a single periodic cell of the substrate. }
\label{fig:periodic-cell}
\end{figure}

As a final validation test, we consider a 3D problem involving an $\rm{MoS_2}$ monolayer suspended over a substrate with a hexagonal pattern of holes, which corresponds to the experimental system in \cite{holey_zhang}. Specifically, we consider a circular hole with center $\mathbf{x}_{00}=\left(x_{001},x_{002}\right)$ that is considered to be part of a periodic array of holes with centers $\mathbf{x}_{ij}$ given by 
\begin{equation}\label{substrate_circular_hole_centre_coordinates}
    \mathbf{x}_{ij}=\left(x_{ij1},x_{ij2}\right)=\left(x_{001}+iL-\frac{1}{2}jL~,~x_{002}+\frac{\sqrt{3}}{2}jL\right)\text{ .}
\end{equation}
The pattern is generated by a periodic parallelogram-shaped cell of length $L$ containing a single hole, as shown in Figure \ref{fig:periodic-cell}.  
As noted in Section~\ref{sec:introduction}, the substrate exhibits a periodic topography commensurate with the hole pattern. The substrate height profile within the cell centered at $\mathbf{x}_{00}$ is taken to be
\begin{equation}\label{substrate_circular_hole_continuum_sum}
    z_\text{sub}(\xi_1(x_1,x_2),\xi_2(x_1,x_2)) = \sum_{i=-p}^{p}\sum_{j=-q}^{q}-\mathbbm{1}_{A_{ij}}(\mathbf{x})\,\frac{A}{4}\left(\cos\left(\frac{\pi \lVert\mathbf{x}-\mathbf{x}_{ij}\rVert}{R}\right)+1\right)^{2},
\end{equation}
with
\begin{equation}\label{substrate_circular_hole_indicator_function_support}
    A_{ij}=\left\{\mathbf{x}\in\mathbb{R}^2:\lVert\mathbf{x}-\mathbf{x}_{ij}\rVert< R\right\}\quad\text{and}\quad\mathbbm{1}_{A_{ij}}(\mathbf{x}) = \left\{\begin{array}{lr}1&\mathbf{x}\in A_{ij}\\ 0&\text{otherwise}\end{array}\right.\text{ ,}
\end{equation}
where $R$ modulates the radius of the hole, and $A$ modulates the depth. As in Figure \ref{fig:afm_substrate}, the top of the substrate is at $z_\text{sub}=0$. The summation limits $p$ and $q$ can in principle be taken as infinite, although, for given $R$ and $L$, and focusing on a single periodic cell, the summation can be truncated, as, for all $\vert i\vert$ and $\vert j\vert$ beyond some finite value, $\mathbbm{1}_{A_{ij}}(\mathbf{x}) = 0$ for $\mathbf{x}$ inside the periodic cell centered at $\mathbf{x}_{00}$.

\begin{figure}[H] 
\begin{center}
\subfloat[Substrate periodic cell. Top view.]{\includegraphics[trim = 45mm 100mm 45mm 80mm, clip=true,width=0.34\textwidth]{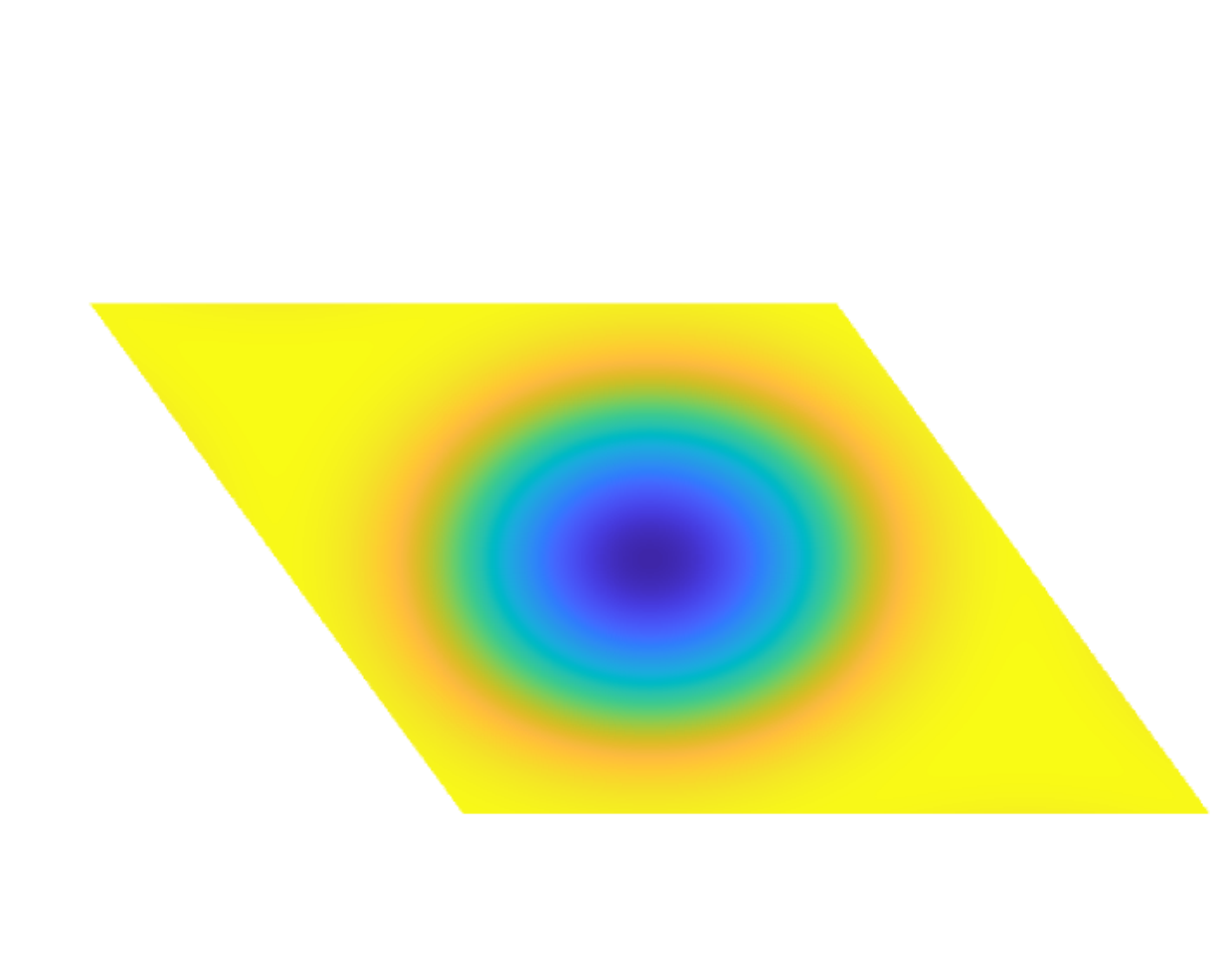}\label{substrate_circ_3D_top}}\hfill  
\subfloat[Substrate periodic cell. Side view.]{\includegraphics[trim = 50mm 100mm 45mm 80mm, clip=true,width=0.30\textwidth]{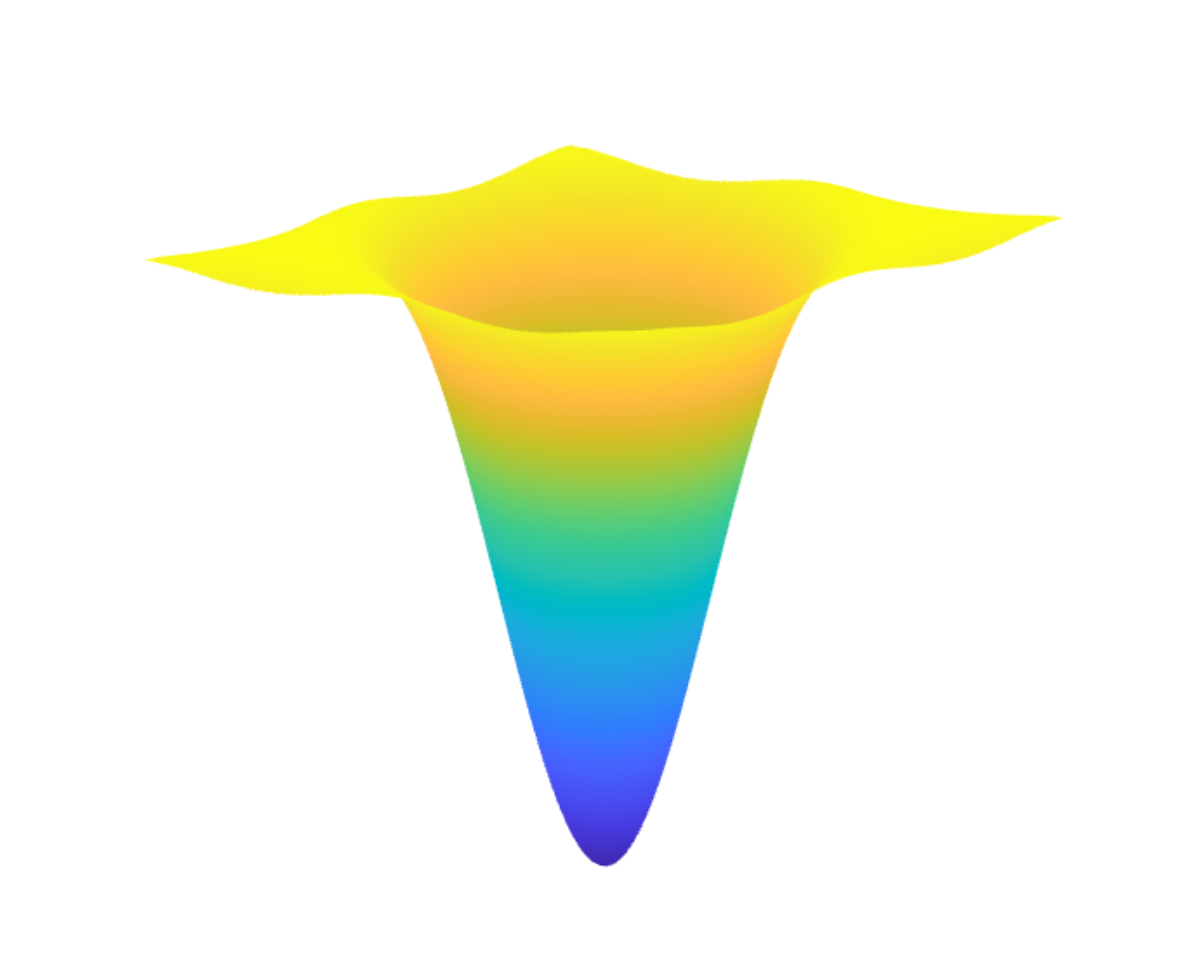}\label{substrate_circ_3D_side}}\hfill  
\subfloat[Multi-hole substrate.]{\includegraphics[trim = 45mm 100mm 45mm 80mm, clip=true,width=0.34\textwidth]{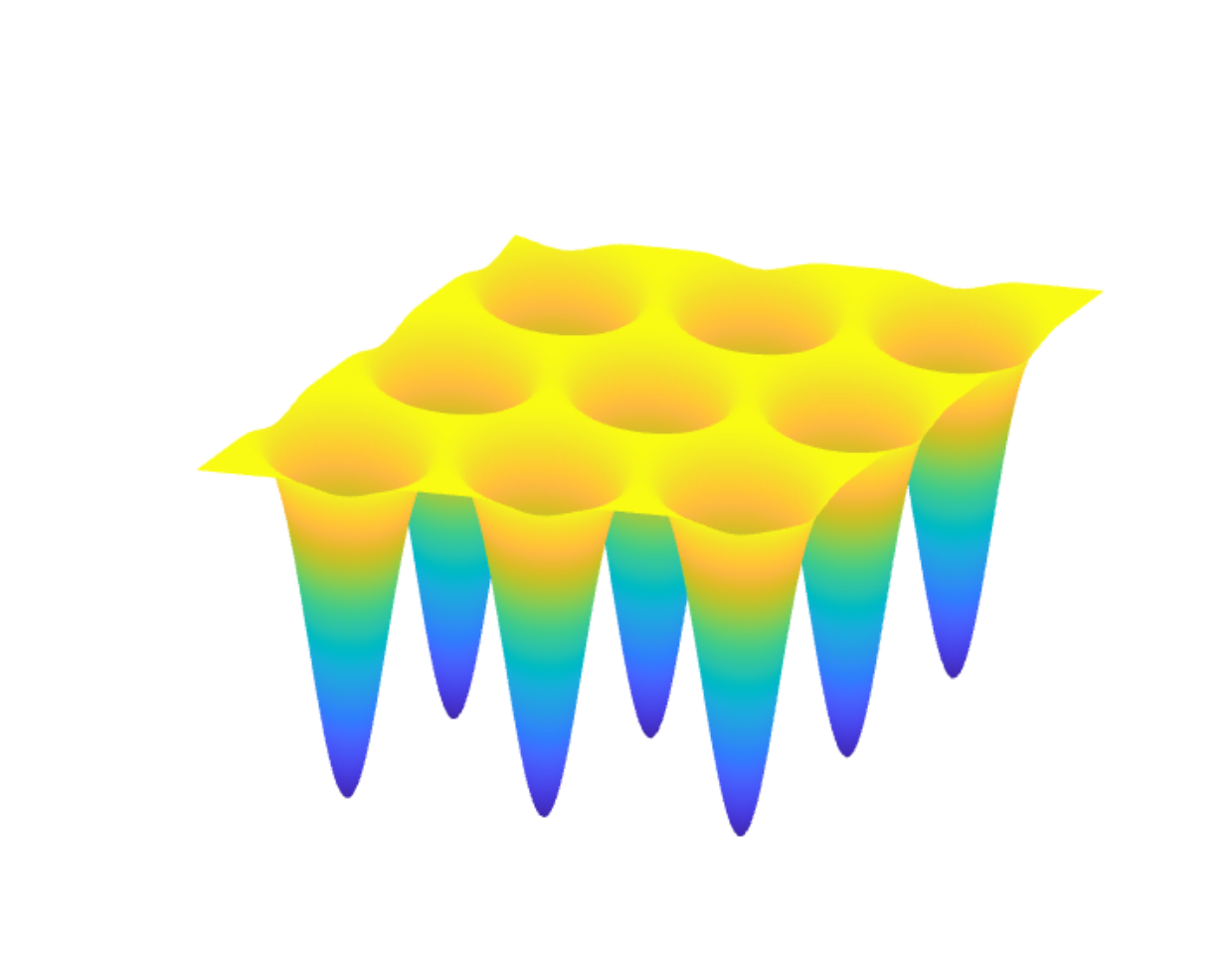}\label{substrate_circ_3D_multiple_side}}\hfill
\caption{Substrate as defined by Eq~(\ref{substrate_circular_hole_continuum_sum}) with $R = 240$\,nm, $A=240$\,nm, and $L=400$\,nm.}
\label{fig:substrate_circ_3D}
\end{center}
\end{figure}

\begin{figure}[H] 
\begin{center}
\subfloat[Comparison between AFM data and the continuum model (CM)'s substrate  height (defined by \eqref{substrate_circular_hole_continuum_sum}) along a cross-section in direction $\zeta$. The bottom figure shows the portion highlighted in red in the top figure.]{\includegraphics[trim = 0mm 0mm 0mm 0mm, clip=true,width=0.49\textwidth]{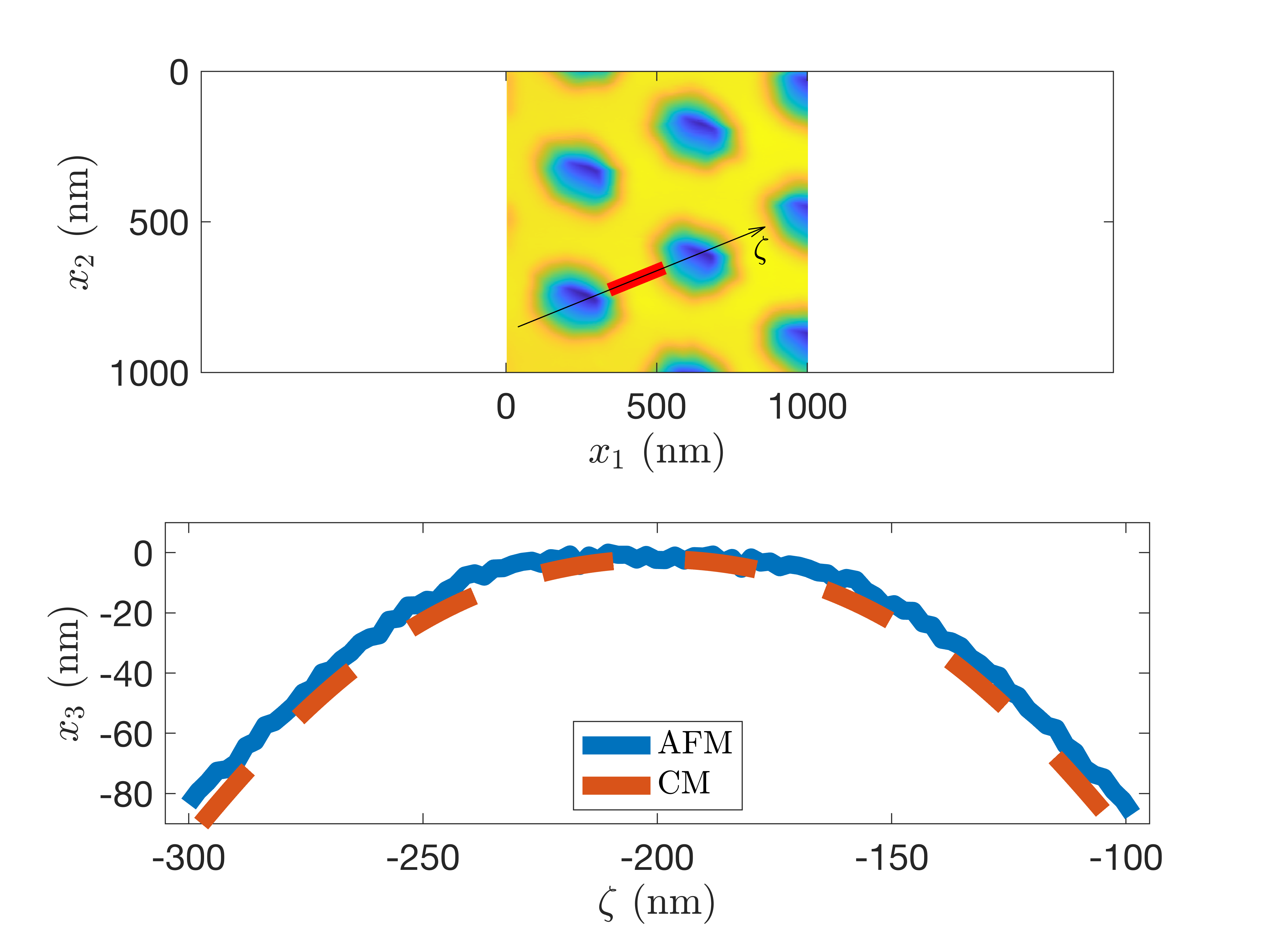}\label{fig:AFMvsCM_cross_section}}\hfill
\subfloat[3D view of measured substrate height (colored surface) and the surface defined by \eqref{substrate_circular_hole_continuum_sum} (black wireframe).]{\includegraphics[trim = 0mm -5mm 0mm 0mm, clip=true,width=0.49\textwidth]{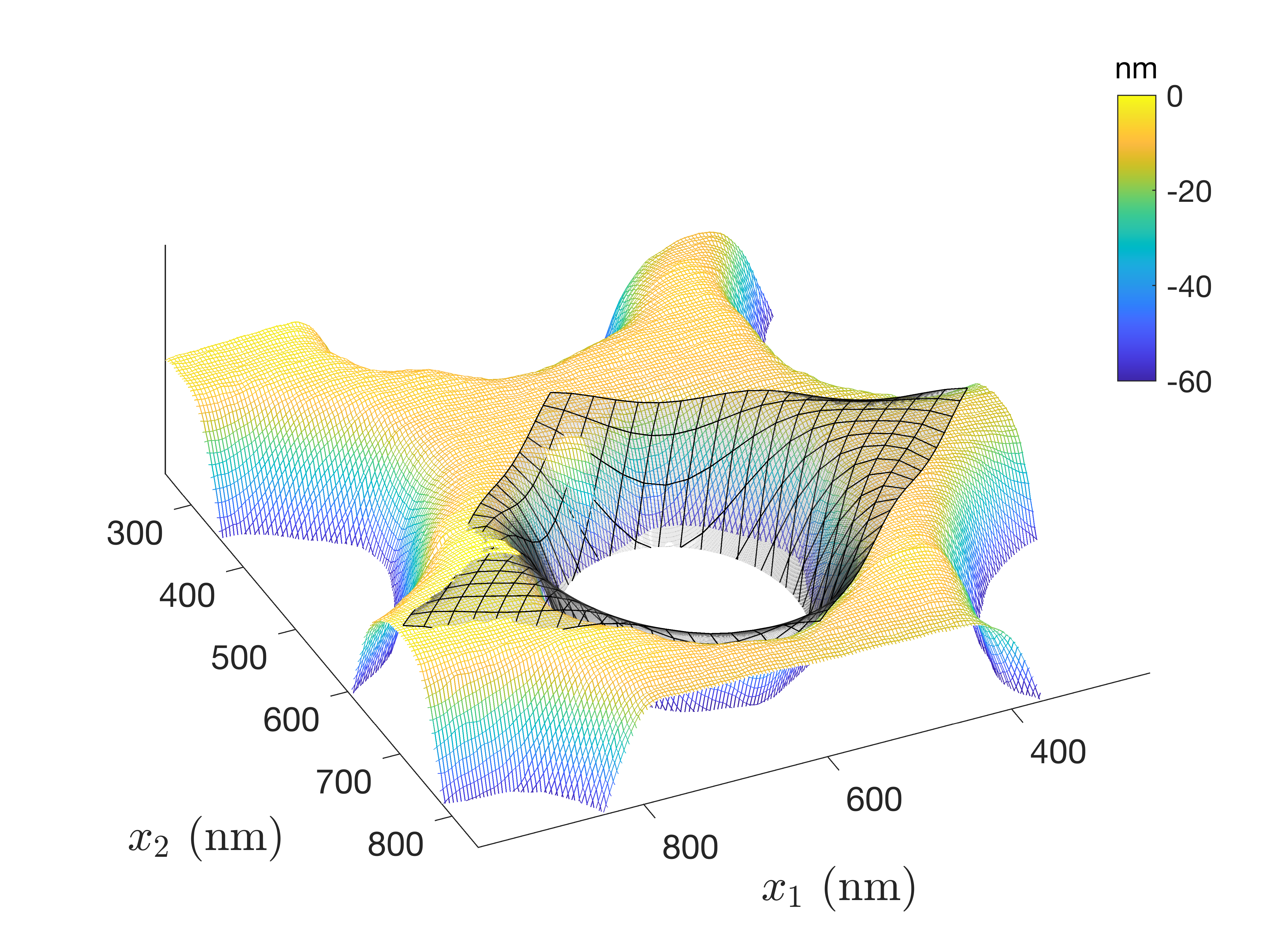}\label{fig:AFMvsCM_3D}}\hfill  
\caption{Comparison of \eqref{substrate_circular_hole_continuum_sum} using $R = 240$\,nm, $A=240$\,nm, and $L=400$\,nm with AFM data for a substrate with a nominal hole radius of $100$\,nm.  Both surfaces are cut off below $-60$\,nm, because material below this point has negligible interaction with the $\rm{MoS_2}$ layer and AFM data becomes unreliable.}
\label{fig:AFMvsCM}
\end{center}
\end{figure}

The functional form of the substrate surface in \eqref{substrate_circular_hole_continuum_sum} is motivated by a qualitative comparison with AFM measurements of  $\rm{Si_3N_4}$ holey substrates reported in \cite{holey_zhang}, such as that plotted in Figure \ref{fig:afm_substrate}.  Figure \ref{fig:substrate_circ_3D} shows a surface of the form \eqref{substrate_circular_hole_continuum_sum}.  The parameters $R$, $A$, and $L$ can be calibrated to provide reasonable quantitative agreement with the measured surface of a substrate, as illustrated in Figure \ref{fig:AFMvsCM}.  We note, however, that, after calibration, the parameter $R$ in \eqref{substrate_circular_hole_continuum_sum} may differ significantly from the nominal hole radius stated by the substrate manufacturer, i.e., $R_0$ in Figure \ref{fig:periodic-cell}, due to ambiguity of what constitutes the ``edge'' of a hole when the real substrate height varies continuously.  

For $R/L$ sufficiently small, we can take $p=q=0$, so that interactions within a periodic cell are limited to the topography within that cell. This is the case that we consider for the validation test, because it allows for a simplified calculation of the Lennard-Jones potential in the MS analysis.  Specifically, if $R$ is significantly larger than the Lennard-Jones length scales $\{\sigma_{\text{A--B}}\}$, it is reasonable to ignore the influence of neighboring holes during the calculation of the Lennard-Jones potential, i.e., calculate it for a single hole in an otherwise flat infinite substrate.  This assumption makes the integral \eqref{eq:Omega_sub-int} independent of azimuthal angle about the $x_3$ axis. Azimuthal integration can then be performed analytically, such that a 2D Cartesian grid along the radius and $x_3$ can be used to interpolate energy values; our MS analysis of this case uses increments of 0.01\,nm for both the radial direction and the 3-direction.   

\begin{figure}[H] 
\begin{center}
\subfloat[Results on a cross-section of the hole passing through its center.]{\includegraphics[trim = 10mm 60mm 10mm 60mm, clip=true,width=0.49\textwidth]{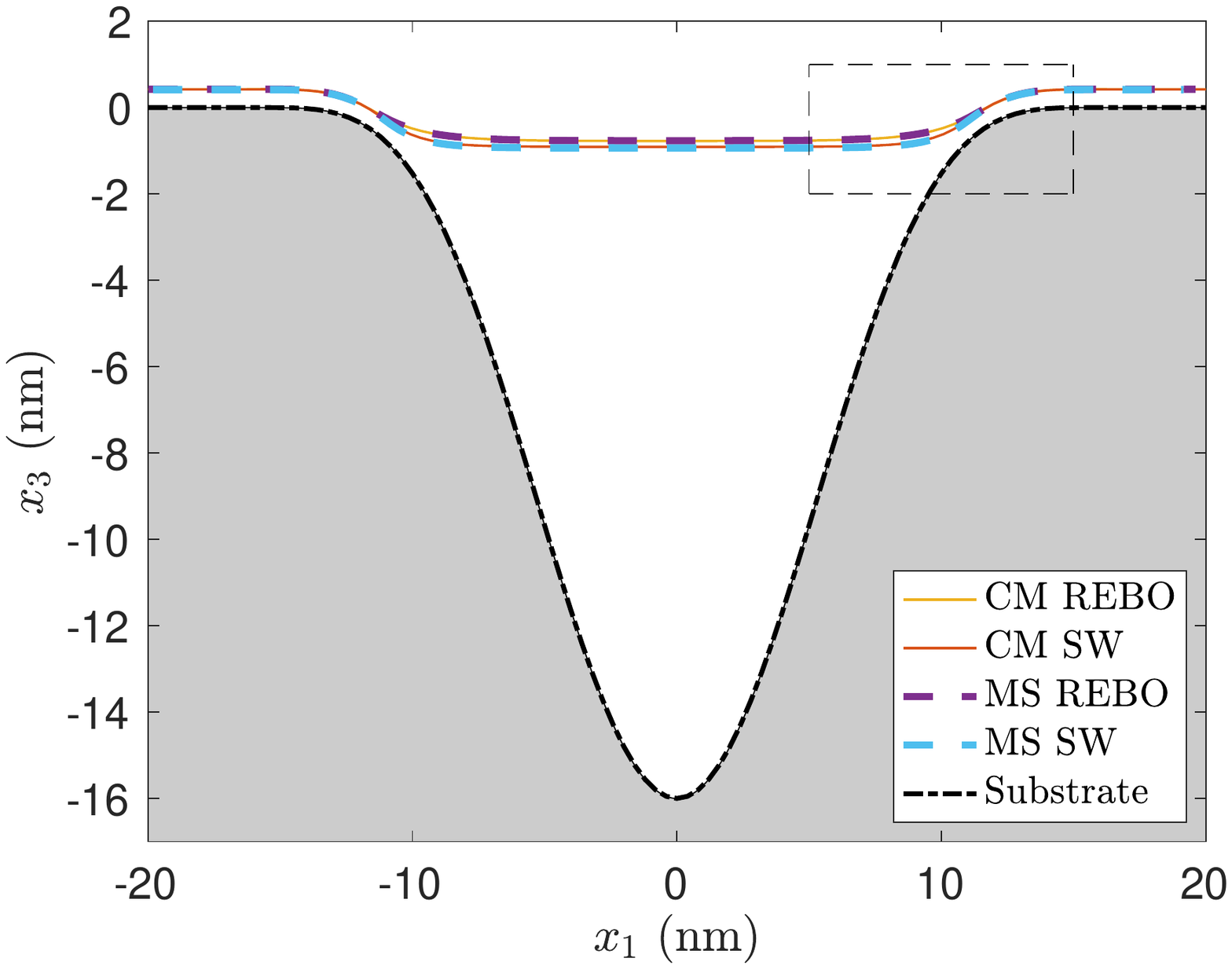}\label{cos_sub_results}}\hfill
\subfloat[Zoomed-in view corresponding to the dashed rectangle in Figure \ref{cos_sub_results}.]{\includegraphics[trim = 10mm 60mm 10mm 60mm, clip=true,width=0.49\textwidth]{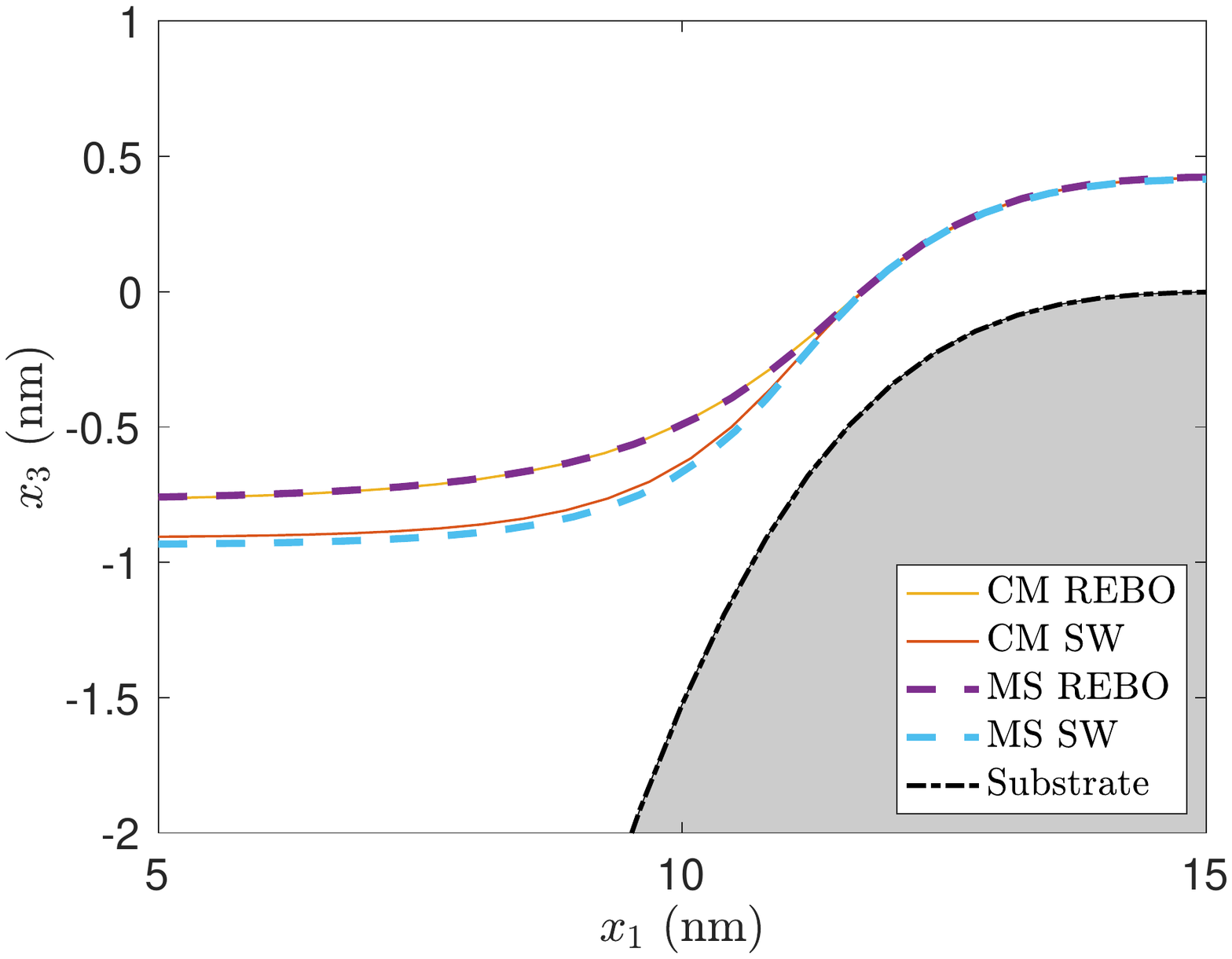}\label{cos_sub_results_zoom}}\hfill  
\caption{Cross-sections of the deflected monolayer (solid and dashed lines)  over the circular substrate (gray region), corresponding to continuum modeling (CM) and molecular statics (MS), for both the REBO and SW potentials.}
\label{cos_sub_results_full}
\end{center}
\end{figure}

For our validation test, we take $L=40$\,nm, $A=16$\,nm and $R=16$\,nm, which permits $p=q=0$, as discussed above.  
In the continuum model, we use a structured grid of $100\times 100$ $C^1$-continuous quadratic B-spline elements on the repeating cell and apply periodic boundary conditions.  The monolayer's deflections on cross-sections passing through the center of the hole are shown in Figure \ref{cos_sub_results_full}, for all combinations of continuum/MS and REBO/SW.  We see that deflections of the monolayer are predicted to within a much smaller error than the difference between the REBO and SW potentials, implying that the continuum approximation is a valid method for predicting deflections.  

\begin{figure}[H] 
\begin{center}
\subfloat[MS, REBO.]{\includegraphics[trim = 0mm 0mm 0mm 0mm, clip=true,width=0.49\textwidth]{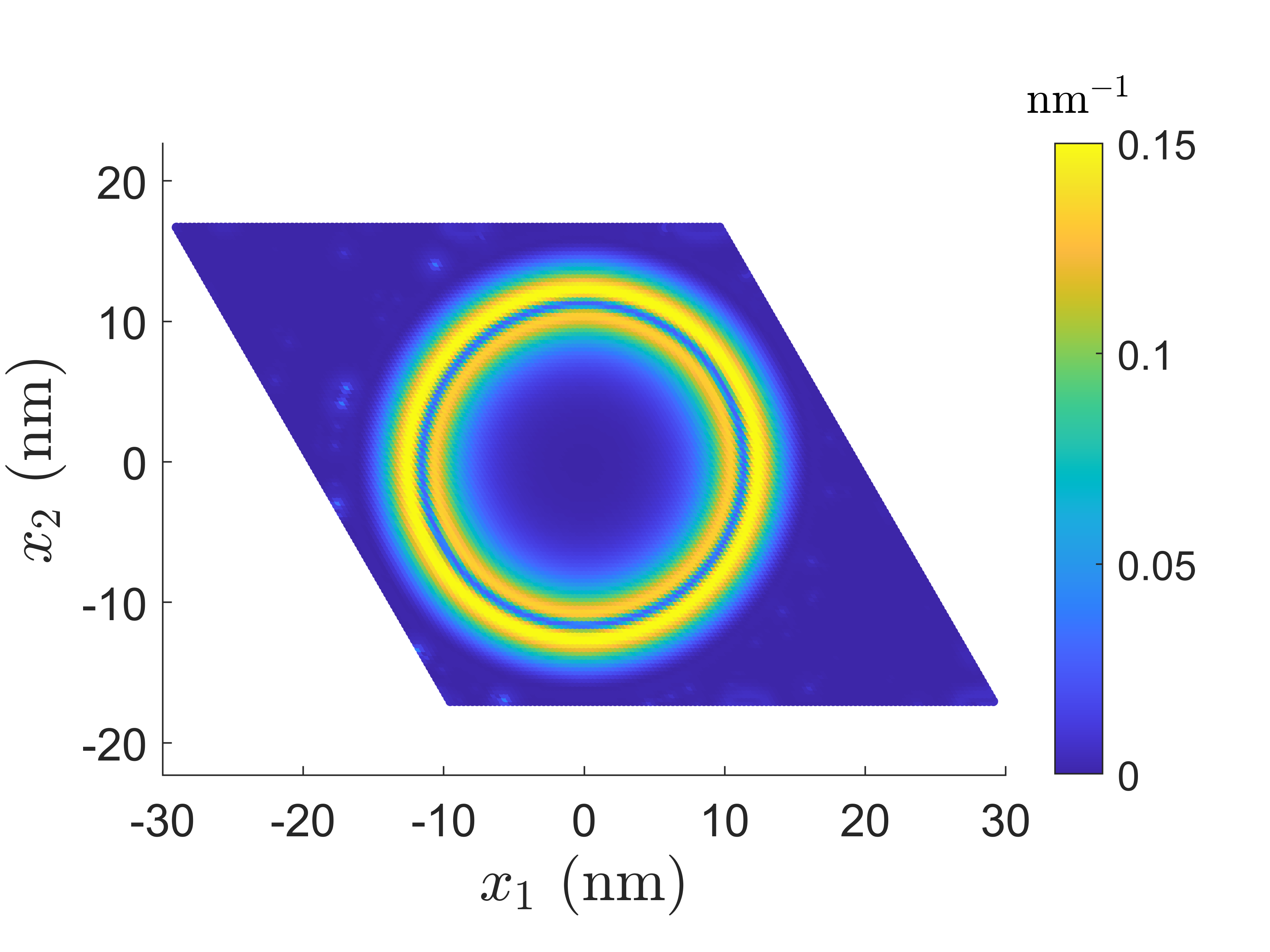}\label{fig:cos_L40_curv_MS_REBO}}\hfill
\subfloat[CM, REBO.]{\includegraphics[trim =  0mm 0mm 0mm 0mm, clip=true,width=0.49\textwidth]{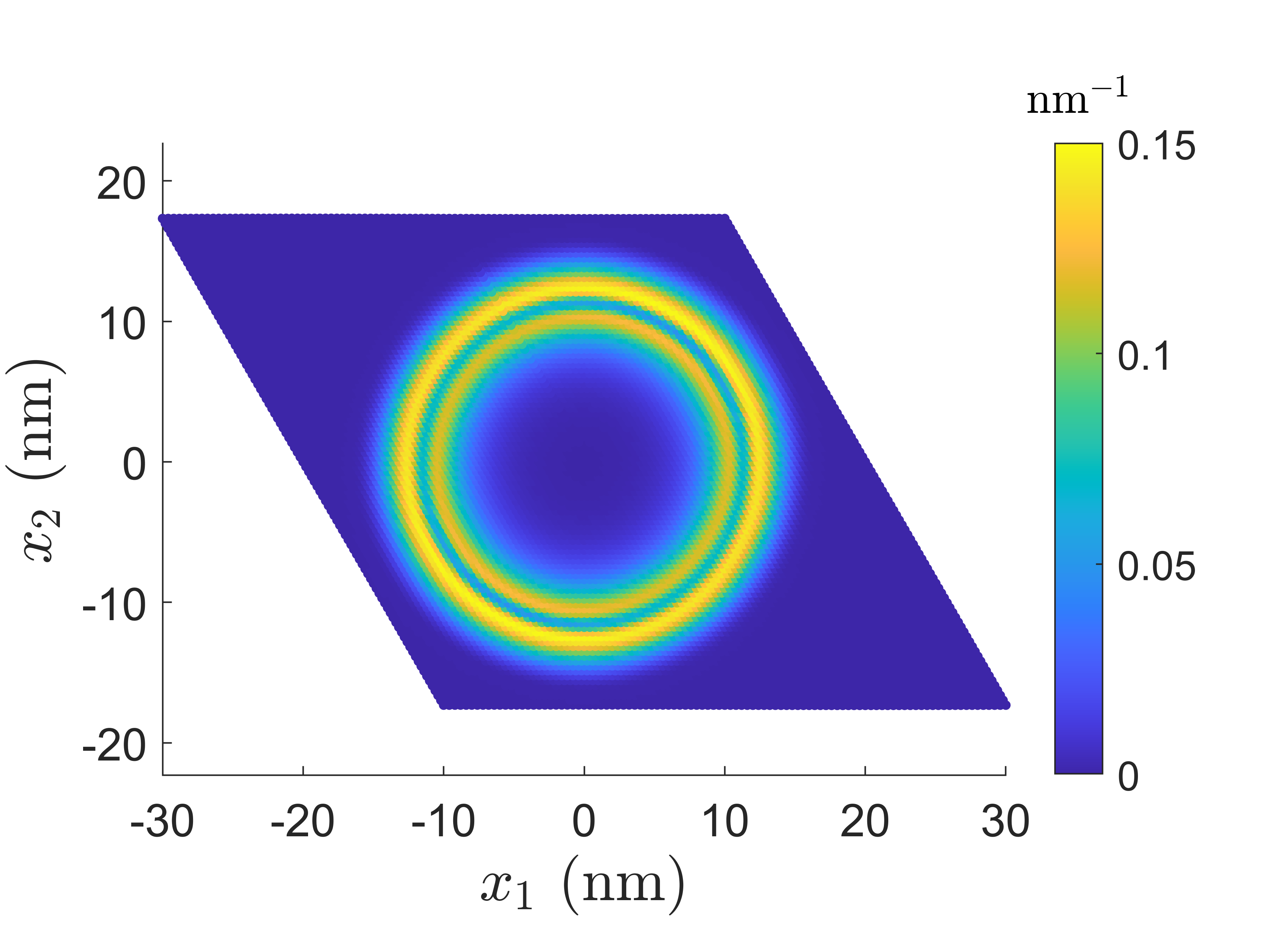}\label{fig:cos_L40_curv_CM_REBO}}\hfill  
\subfloat[MS, SW.]{\includegraphics[trim = 0mm 0mm 0mm 0mm, clip=true,width=0.49\textwidth]{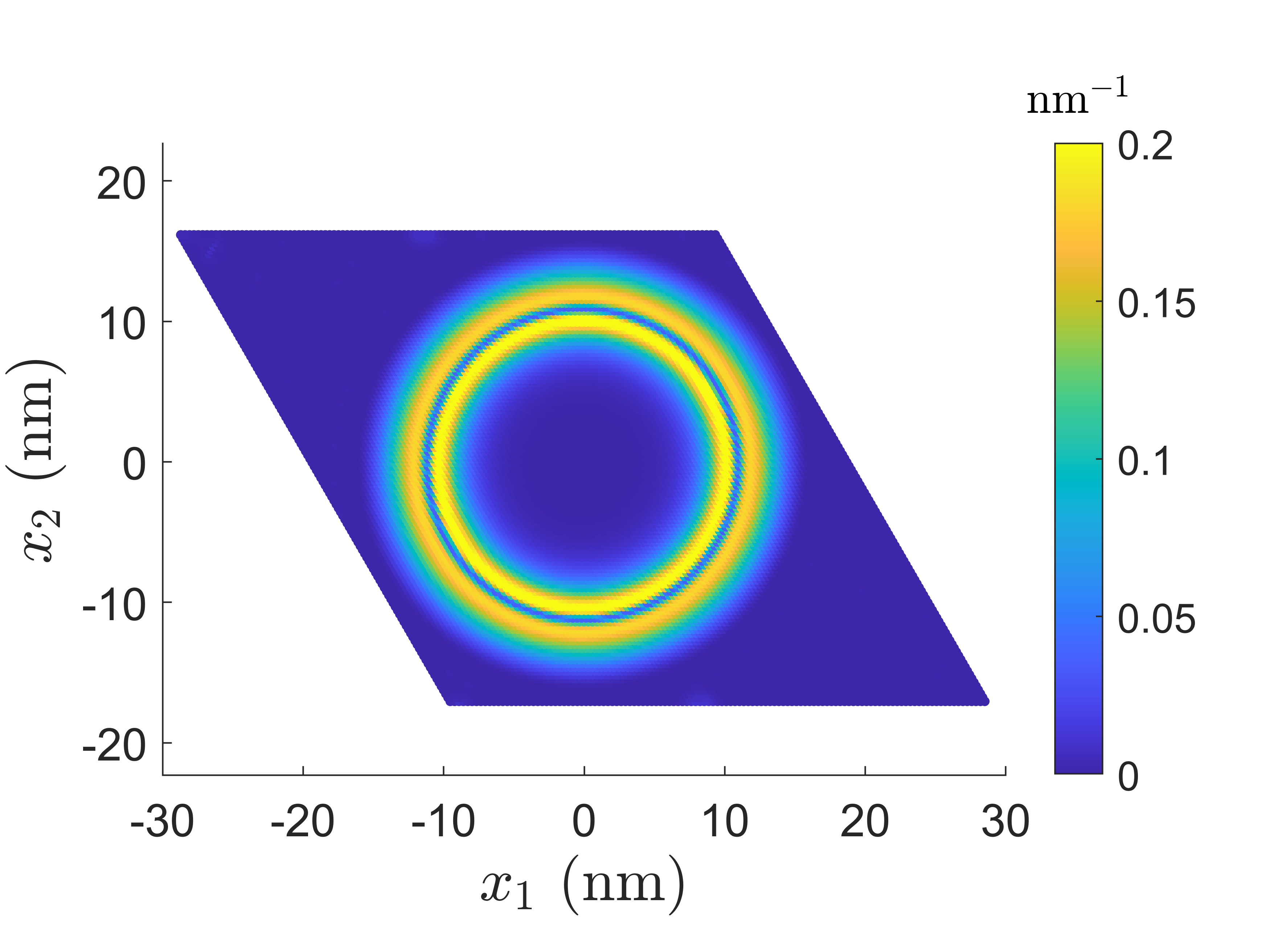}\label{fig:cos_L40_curv_MS_SW}}\hfill
\subfloat[CM, SW.]{\includegraphics[trim =  0mm 0mm 0mm 0mm, clip=true,width=0.49\textwidth]{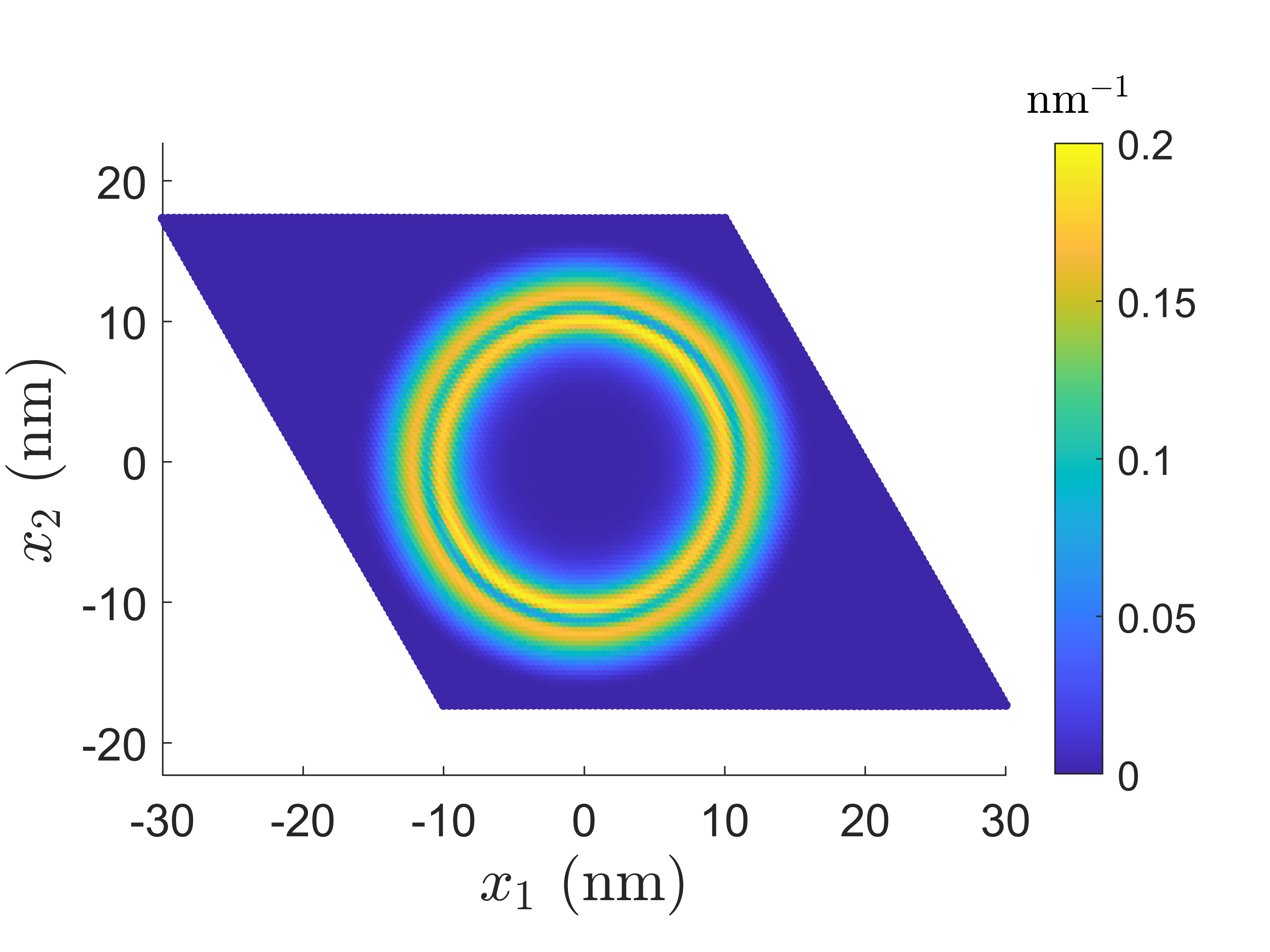}\label{fig:cos_L40_curv_CM_SW}}\hfill  
\caption{2D contour plots for the spectral radius of the bending strain tensor $\bm{\kappa}$ of the Mo layer and of the midsurface for the molecular statics (MS) and the continuum model (CM), respectively. Both REBO and SW potentials are considered.}
\label{fig:cos_curv_L40_REBO}
\end{center}
\end{figure}

\begin{figure}[H] 
\begin{center}
\subfloat[MS, REBO.]{\includegraphics[trim = 0mm 0mm 0mm 0mm, clip=true,width=0.49\textwidth]{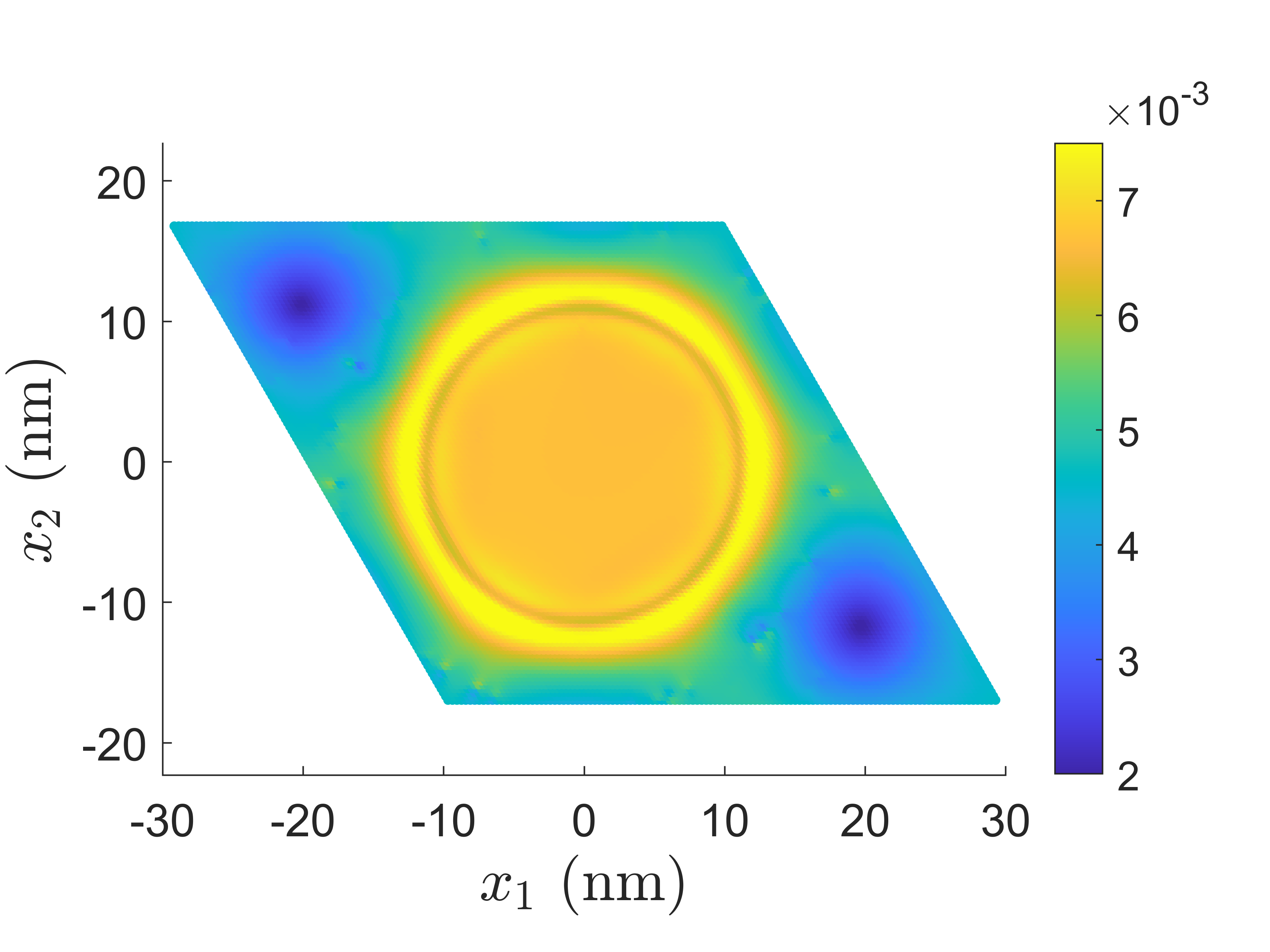}\label{fig:cos_L40_strain_MS_REBO}}\hfill
\subfloat[CM, REBO.]{\includegraphics[trim =  0mm 0mm 0mm 0mm, clip=true,width=0.49\textwidth]{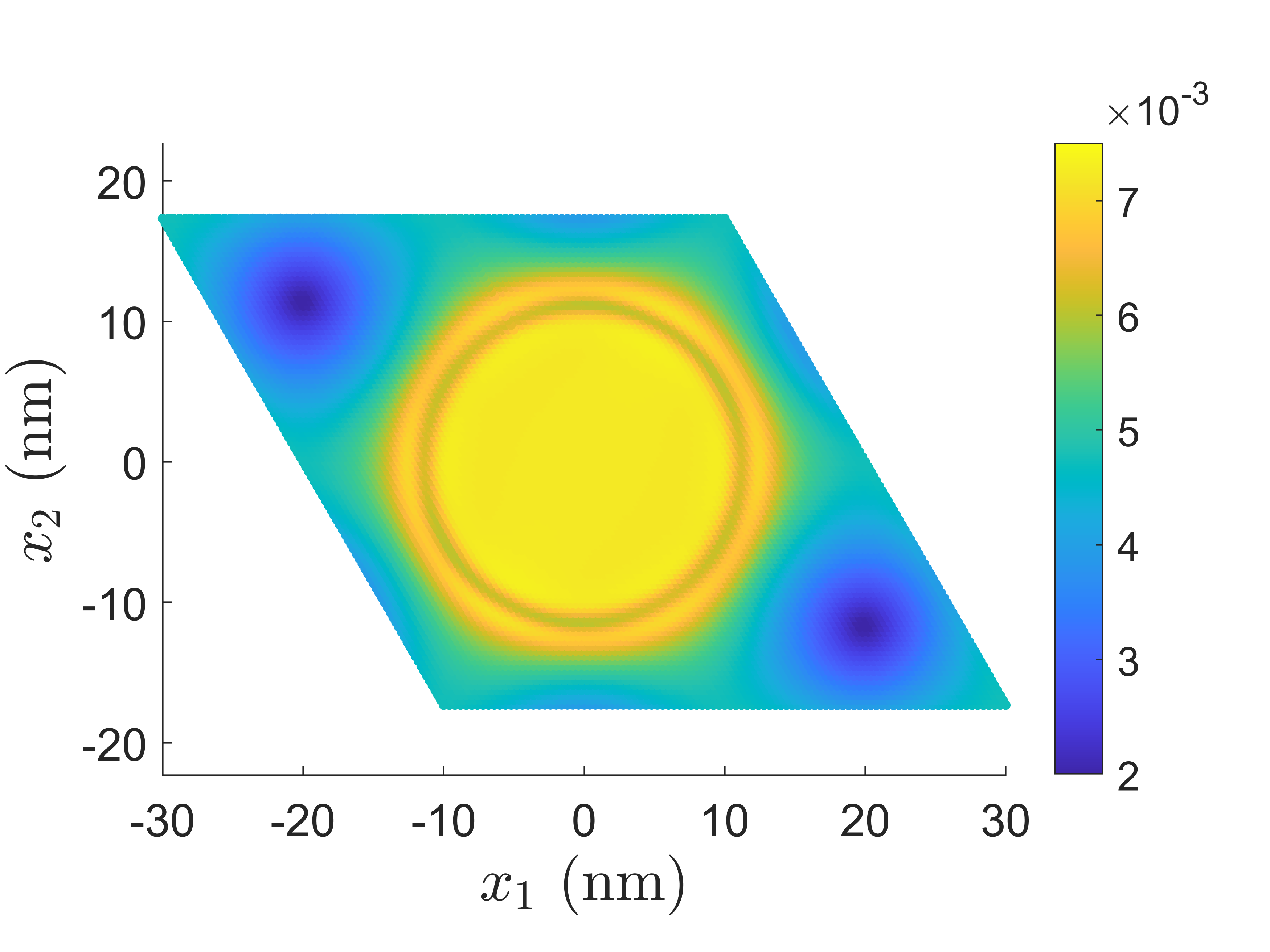}\label{fig:cos_L40_strain_CM_REBO}}\hfill  
\subfloat[MS, SW.]{\includegraphics[trim = 0mm 0mm 0mm 0mm, clip=true,width=0.49\textwidth]{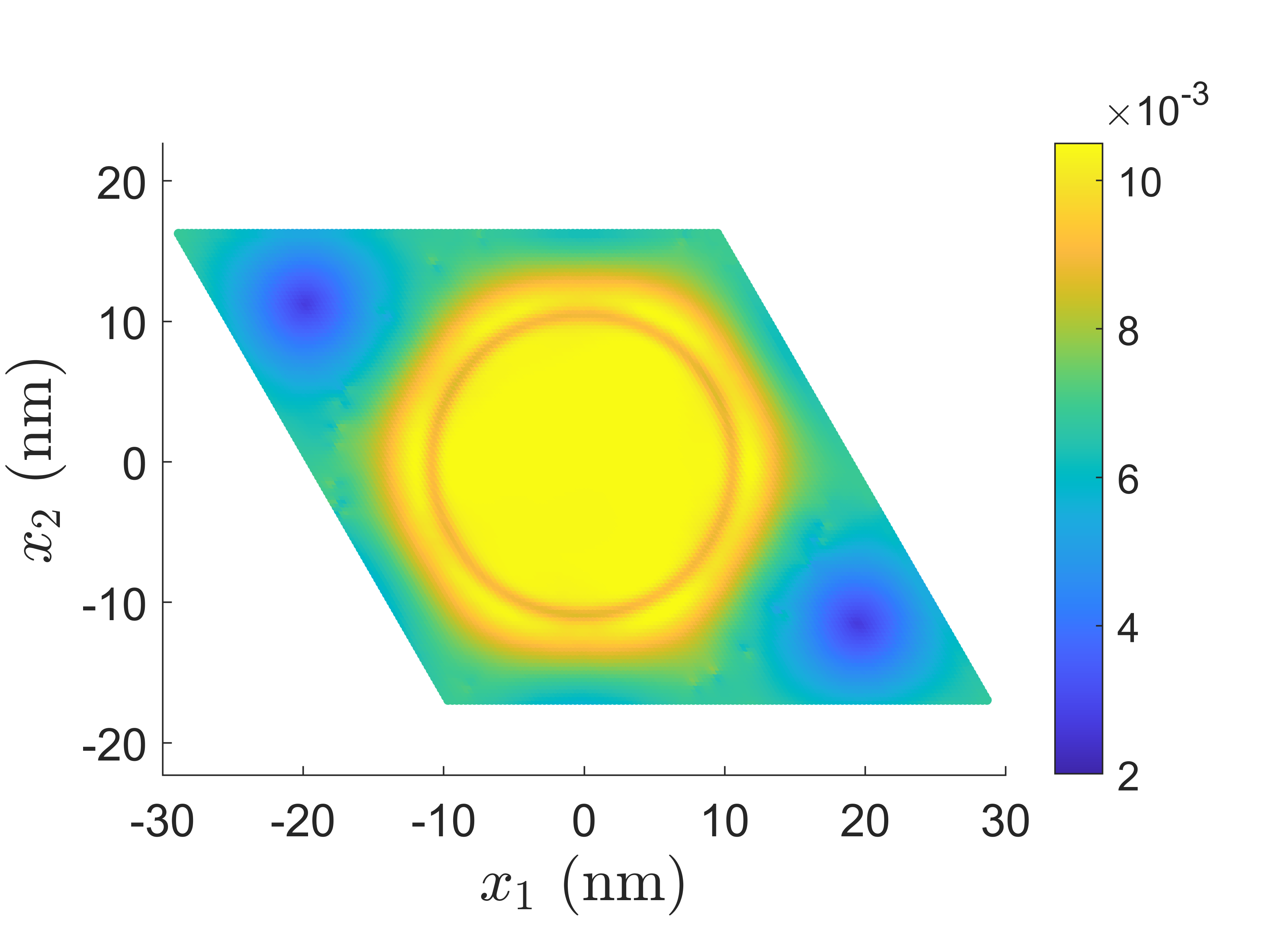}\label{fig:cos_L40_strain_MS_SW}}\hfill
\subfloat[CM, SW.]{\includegraphics[trim =  0mm 0mm 0mm 0mm, clip=true,width=0.49\textwidth]{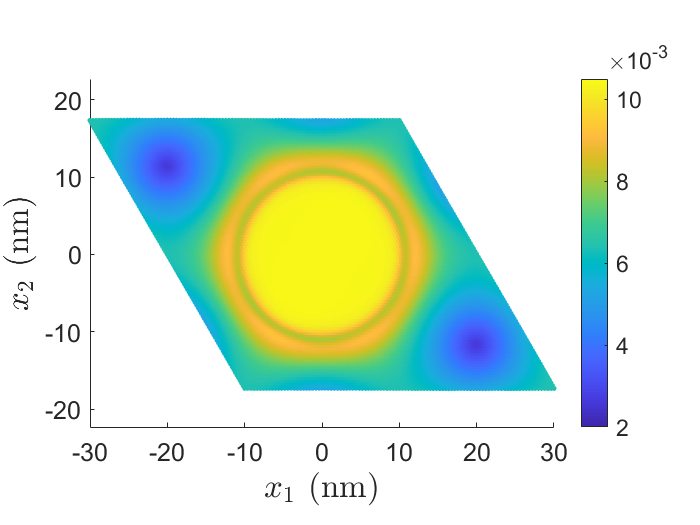}\label{fig:cos_L40_strain_CM_SW}}\hfill  
\caption{2D strain fields for both the REBO and SW potentials. Figures for the molecular statics (MS) show the maximum eigenvalue of $\mathbf{E}_\text{Mo}$. Figures for the continuum model (CM) show the maximum eigenvalue of $\mathbf{E}_\text{Mo}^\text{CM}$.}
\label{fig:cos_strain_L40}
\end{center}
\end{figure}

\begin{figure}[H] 
\centering
\subfloat[Spectral radius of $\bm{\kappa}$: cross-sections of Figure \ref{fig:cos_curv_L40_REBO}.]{\includegraphics[trim = 0mm 0mm 0mm 0mm, clip=true,width=0.49\textwidth]{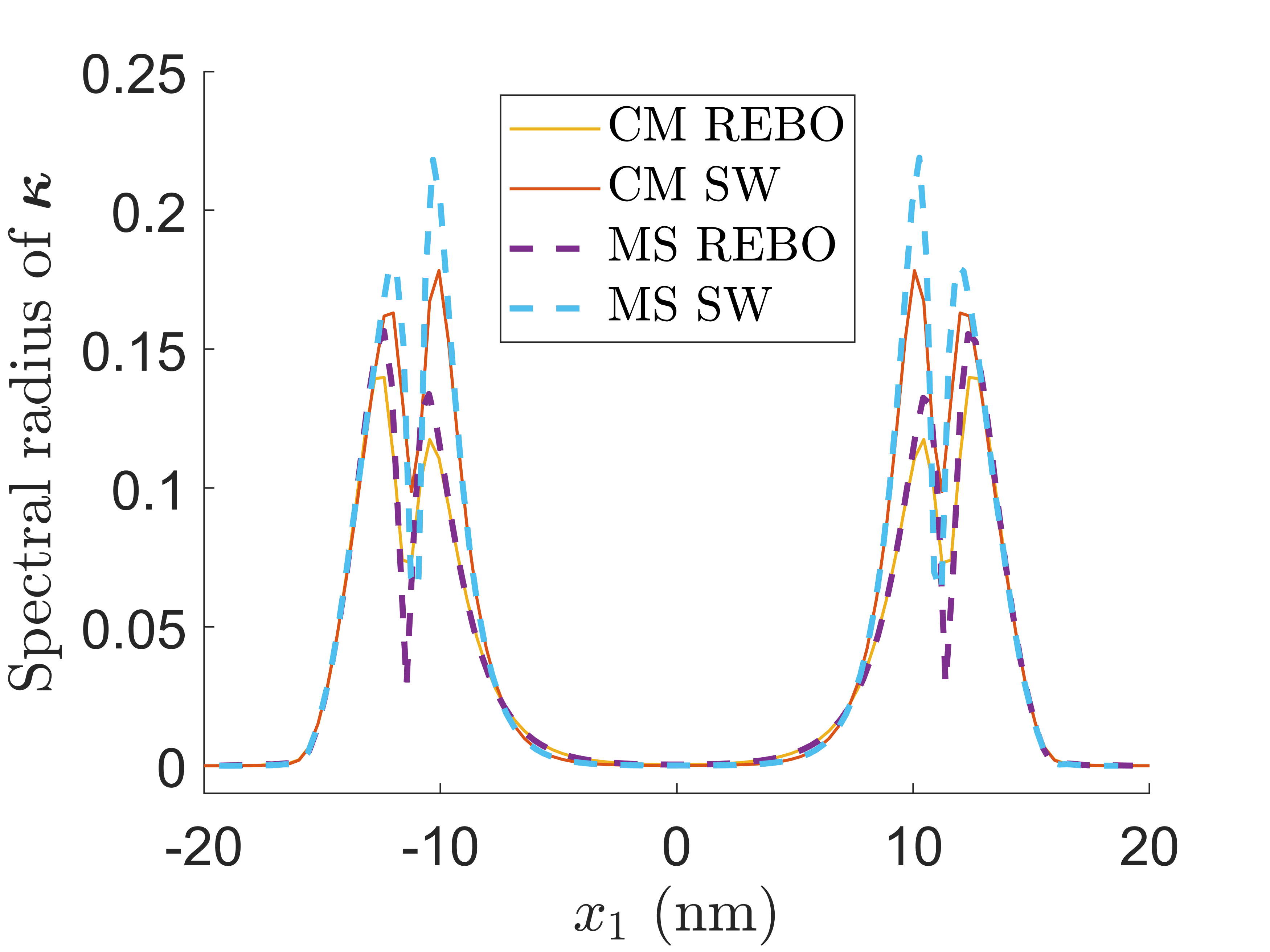}\label{fig:cos_L40_curvature_CMvsMS}}\hfill
\subfloat[Strain: cross-sections of Figure \ref{fig:cos_strain_L40}.]{\includegraphics[trim = 0mm 0mm 0mm 0mm, clip=true,width=0.49\textwidth]{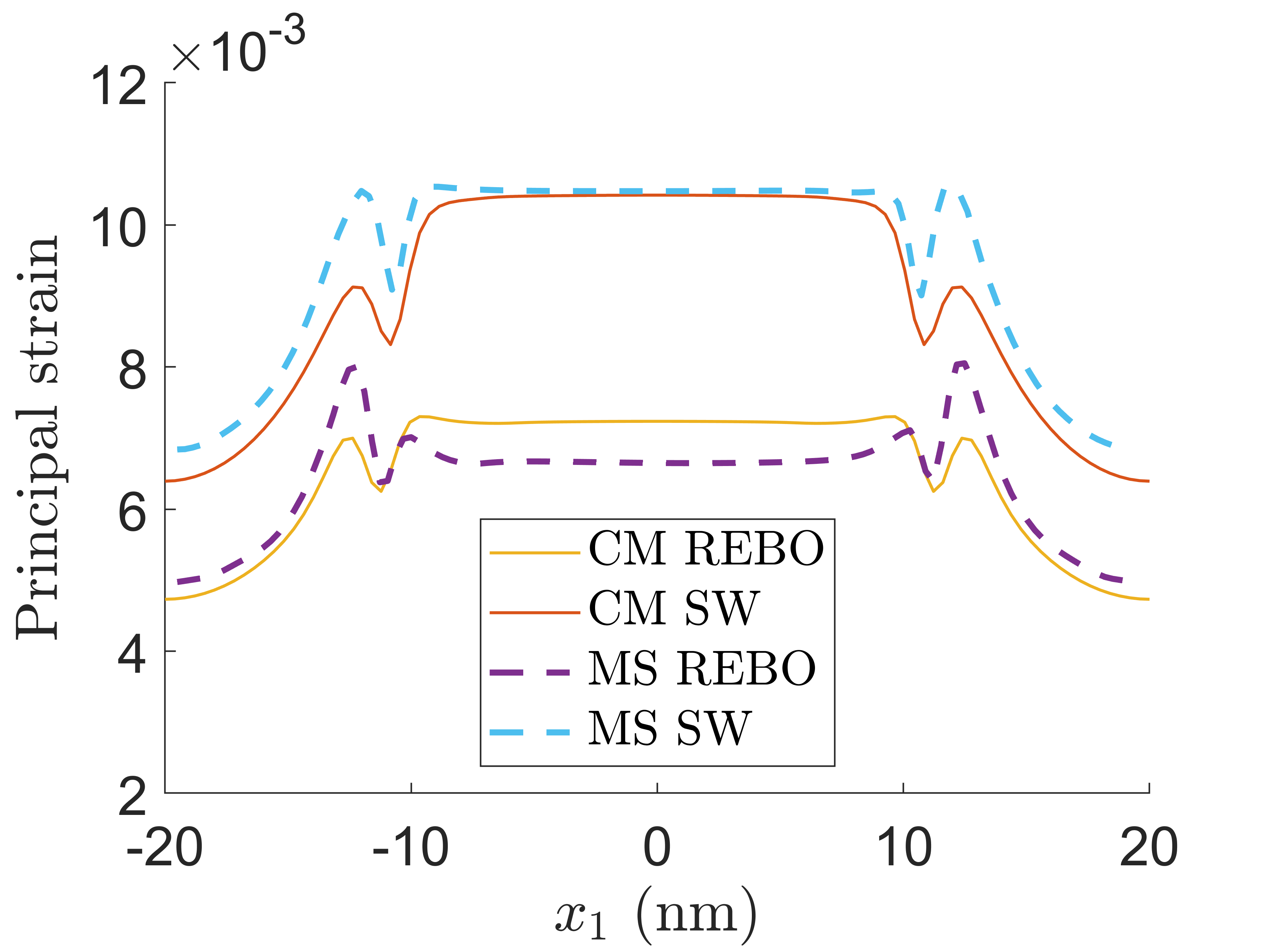}\label{fig:cos_L40_strain_CMvsMS}}\hfill
\caption{Bending and Mo sub-layer strains (solid and dashed lines) over the substrate with circular holes, corresponding to continuum modeling (CM) and molecular statics (MS) for both the REBO and SW potentials, at a cross-section passing through the center of symmetry of the $\rm{MoS_2}$ monolayer.}
\label{fig:strain_CMvsMS}
\end{figure}

We next consider the strain quantities $\bm{\kappa}$ and $\mathbf{E}_\text{Mo}$ discussed in Section \ref{sec:ms-strain}. Figure \ref{fig:cos_curv_L40_REBO} shows the spectral radius of $\bm{\kappa}$, i.e., the largest absolute value of the eigenvalues of $\bm{\kappa}$, computed from both continuum and MS analyses (for both the REBO and SW interatomic potentials), while Figure~\ref{fig:cos_strain_L40} shows the maximum eigenvalue of $\mathbf{E}_\text{Mo}$.  Comparison of these quantities along a cross section passing through the center of the hole are also given in Figure ~\ref{fig:strain_CMvsMS}. The results show close agreement of both bending and Mo sub-layer strains, again defining ``close'' relative to the substantially-larger difference between strains predicted by the SW and REBO potentials.  Note that the use of \eqref{eq:e-mo-from-cont} with $C_\text{bend} > 0$ is essential to capturing the rings of high strain in the Mo sub-layer around the edges of the hole.

\subsection{Computational cost comparison}\label{sec:comp-cost}
For the circular hole geometry studied in Section~\ref{sec:hole}, we now make a comparison of the computational cost of shell analysis and MS, for holes of varying size, with
\begin{equation}
    R = \alpha R_0,\quad L = \alpha L_0,\quad\text{and}\quad A = \alpha A_0\text{ ,}
\end{equation}
where $R_0 = 4$\,nm, $L_0 = 10$\,nm, $A_0 = 4$\,nm, and $\alpha$ is a dimensionless scalar we use to vary overall size.  The continuum model can use the same discretization (of $100\times 100$ elements) for different values of $\alpha$, whereas the number of degrees of freedom in the MS computation necessarily increases with $\alpha$.  Overall trends in computational performance are not affected by the choice of SW or REBO as an interatomic potential, so the comparison of this section uses only REBO, for simplicity.

    Due to differences in computer hardware and some arbitrary choices of solver settings, we caution against reading too much into a direct quantitative comparison of run times.  The shell analysis computations were performed in serial on a laptop with a 2.6\,GHz Intel Core i9-11950H processor and 32\,GB of RAM.  MS analysis was performed using a computing cluster with 24 2.6\,GHz AMD Opteron 6344 processor cores on each node. The precomputation of the Lennard-Jones potential on the radial-$x_3$ grid was performed using MATLAB on a single node with a 24 cores for parallelization, while the energy minimization was performed in parallel using 10 nodes (for a total of 240 cores). Note that the MS precalculation increases in cost with $\alpha$ since the grid spacing remains constant while the domain increases in size.  The shell analysis used a direct solver (MUMPS \cite{MUMPS:1,MUMPS:2}) for each step of Newton iteration within each step of the implicit pseudo-time marching scheme.  Algebraic residuals of the inner Newton loop and displacement magnitude within steps of the outer pseudo-time marching scheme were converged to within predefined relative tolerances.  Energy gradients in the MS energy minimizations were also converged to within a predefined tolerance.  The time reported for the shell analysis does not include just-in-time compilation of element kernels performed by FEniCS, because this step does not need to be repeated for different problem instances.  Due to differences in solution algorithms and semi-arbitrary choices of discretization and solver parameters, the timing data here should be taken as merely representative of the overall magnitude of computational cost for different methods, supporting the following broad conclusions:
\begin{itemize}
    \item There is a large disparity in computational cost between continuum shell structural analysis and MS analysis.
    \item The cost of the MS analysis becomes unmanageable as problem size increases, while shell analysis remains tractable. 
\end{itemize}
Performance testing was not sufficiently controlled to draw more precise conclusions (e.g., estimating an asymptotic relationship between problem size and wall clock time that could be reliably extrapolated to larger problems) and one could expect to change running times by moderate constant factors by changing convergence criteria, solver parameters, etc.  Keeping these caveats in mind, we refer the reader to Table \ref{tab:comp-cost} for a comparison of the wall clock times for the shell and MS analysis performed under the above-stated conditions.  Although the number of degrees of freedom in the continuum model remains constant, the energy functional becomes moderately more difficult to minimize as the ratio of Lennard-Jones length scales to geometric length scales decreases, leading to an increase in computation time as $\alpha$ increases.  In the case of the MS analysis, this effect is compounded by an increase in the number of degrees of freedom, which grows quadratically with $\alpha$. The number of atoms for $\alpha=1, 2, 4 ,8$ are 3072, 11907, 47628, and 190512, respectively. 
\begin{table}[H]\centering
\begin{tabular}{c|c|c|c|c}
$\alpha$ & {\bf Kirchhoff--Love shell} & MS minimization & MS precomputation & {\bf MS total}\\
\hline
1 & 566 & 423 & 12189 & 12612\\
\hline
2 & 654 & 3468 & 32165 & 35633\\
\hline
4 & 765 & 13150 & 72120 & 85270\\
\hline
8 & 982 & 62512 & 164252 & 226764
\end{tabular}
\caption{Computational cost (wall time in seconds) of continuum and atomistic simulations.}
\label{tab:comp-cost}
\end{table}

\section{Application to strain engineering}\label{sec:application}

Having validated the accuracy of the continuum model, we proceed to larger-scale problems that would require extreme computational resources to analyze using atomistic MS modeling.  As a demonstration of the potential usage of our continuum modeling framework in strain engineering, we consider a parametric study of spacing between circular holes, using a realistic radius corresponding to the holey substrate imaged in Figure \ref{fig:afm_substrate}.  This substrate has a nominal hole radius of $R_0=200$\,nm, which we approximate with $R = 240$\,nm and $A = 240$\,nm in the functional form of \eqref{substrate_circular_hole_continuum_sum}, as shown in Figure \ref{fig:AFMvsCM} for $L=400$\,nm.\footnote{Recall our discussion from Section \ref{sec:hole} on the distinction between the parameter $R$ and the nominal hole size $R_0$.} The spacing between holes is then controlled by the parameter $L$, while holding $R$ and $A$ fixed.  Our choice of $L$ as a variable parameter is based on the fact that it is easiest to control from a manufacturing standpoint, whereas variables controlling hole size and shape would be more constrained by the details of the lithographic process used.  However, the purpose of the present study is mainly to demonstrate the scale of analysis enabled by our continuum modeling approach, rather than to draw detailed scientific conclusions.  Again, we only consider a REBO monolayer in this section for simplicity.

\begin{figure}[H] 
\begin{center}
\subfloat[$L = 400$\,nm.]{\includegraphics[trim = 0mm 0mm 0mm 0mm, clip=true,width=0.49\textwidth]{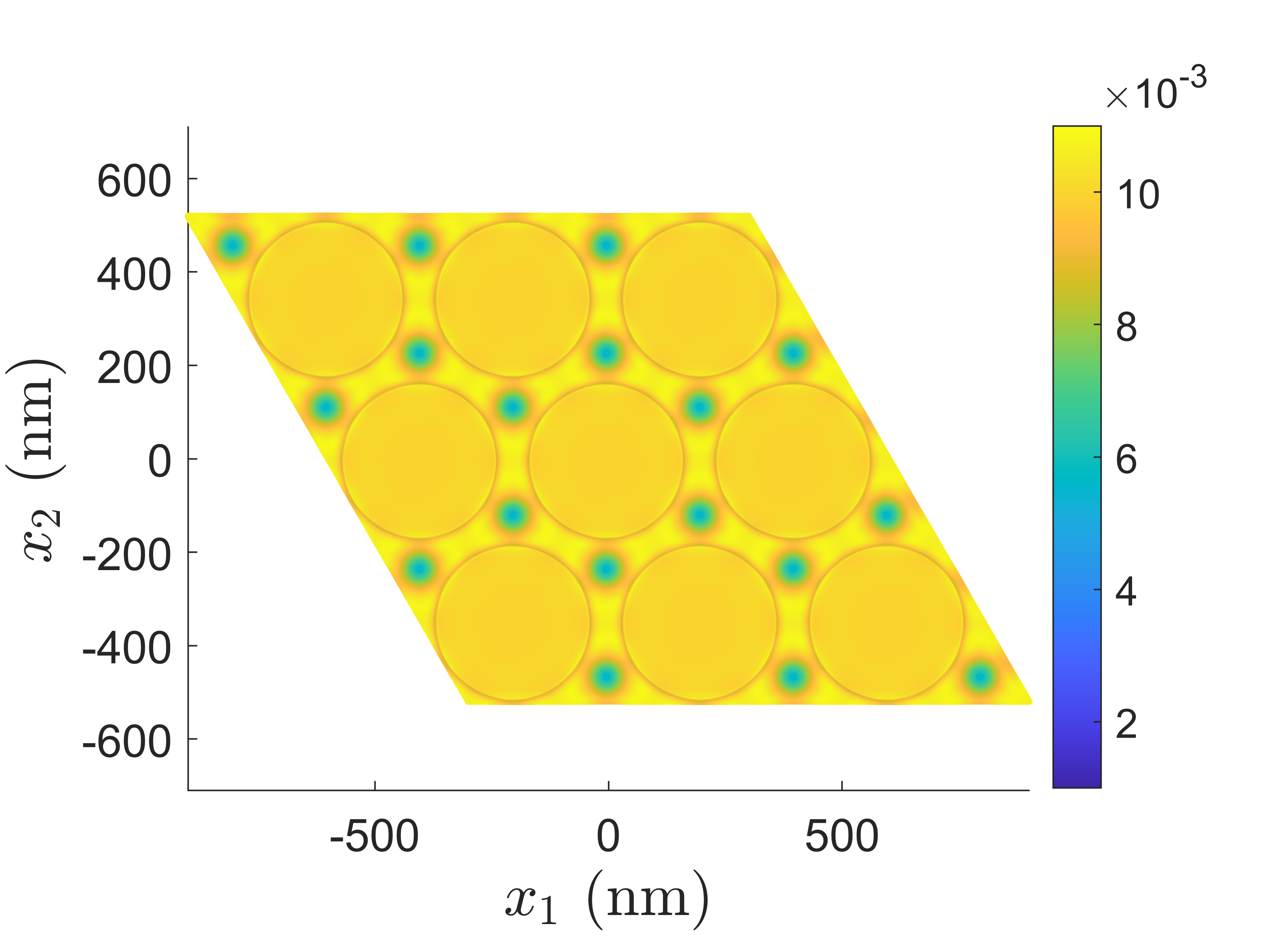}\label{paraview_cosSub_L400}}\hfill
\subfloat[$L = 440$\,nm.]{\includegraphics[trim =0mm 0mm 0mm 0mm, clip=true,width=0.49\textwidth]{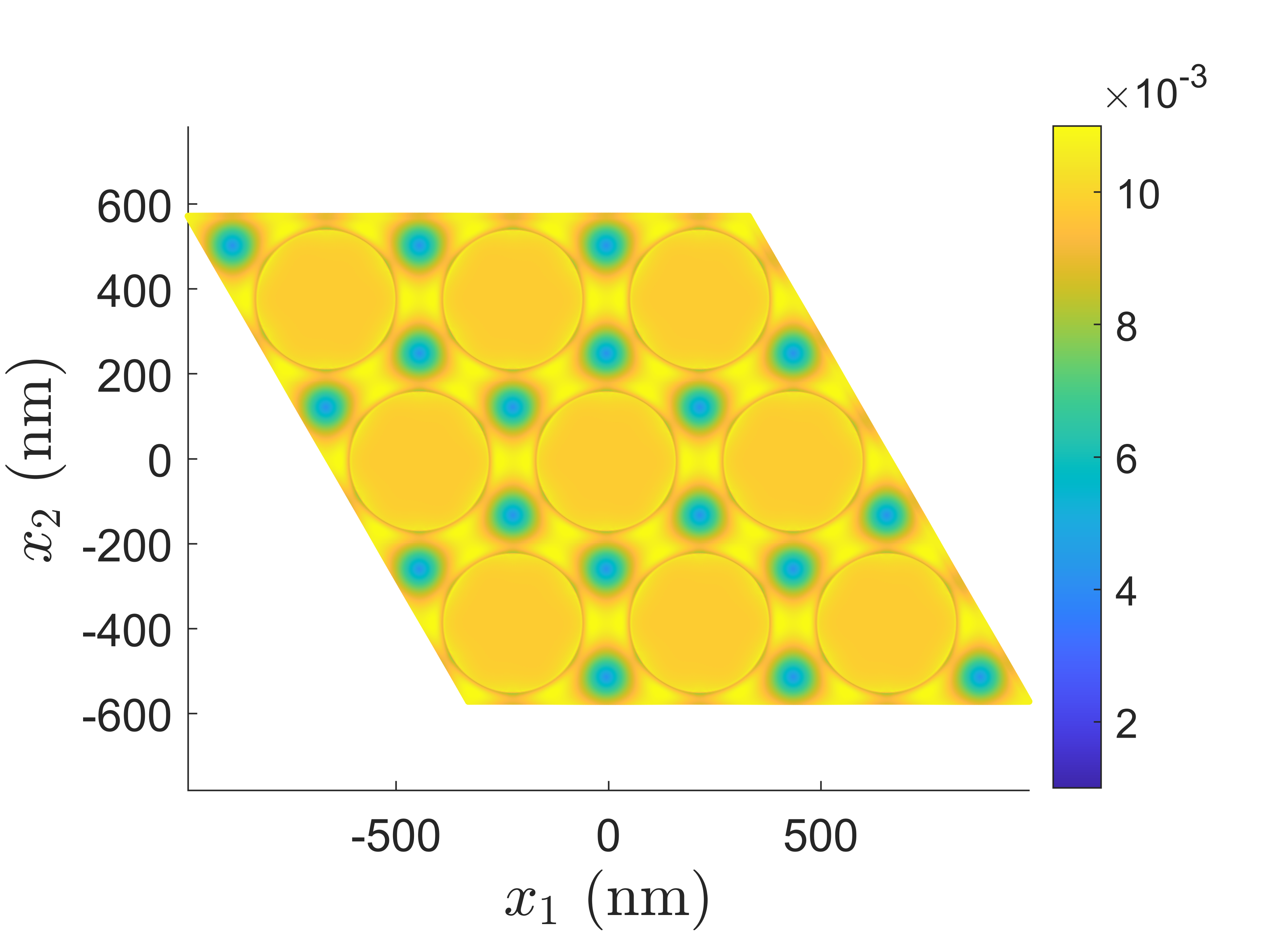}\label{paraview_cosSub_L440}}\hfill
\subfloat[$L = 480$\,nm.]{\includegraphics[trim = 0mm 0mm 0mm 0mm, clip=true,width=0.49\textwidth]{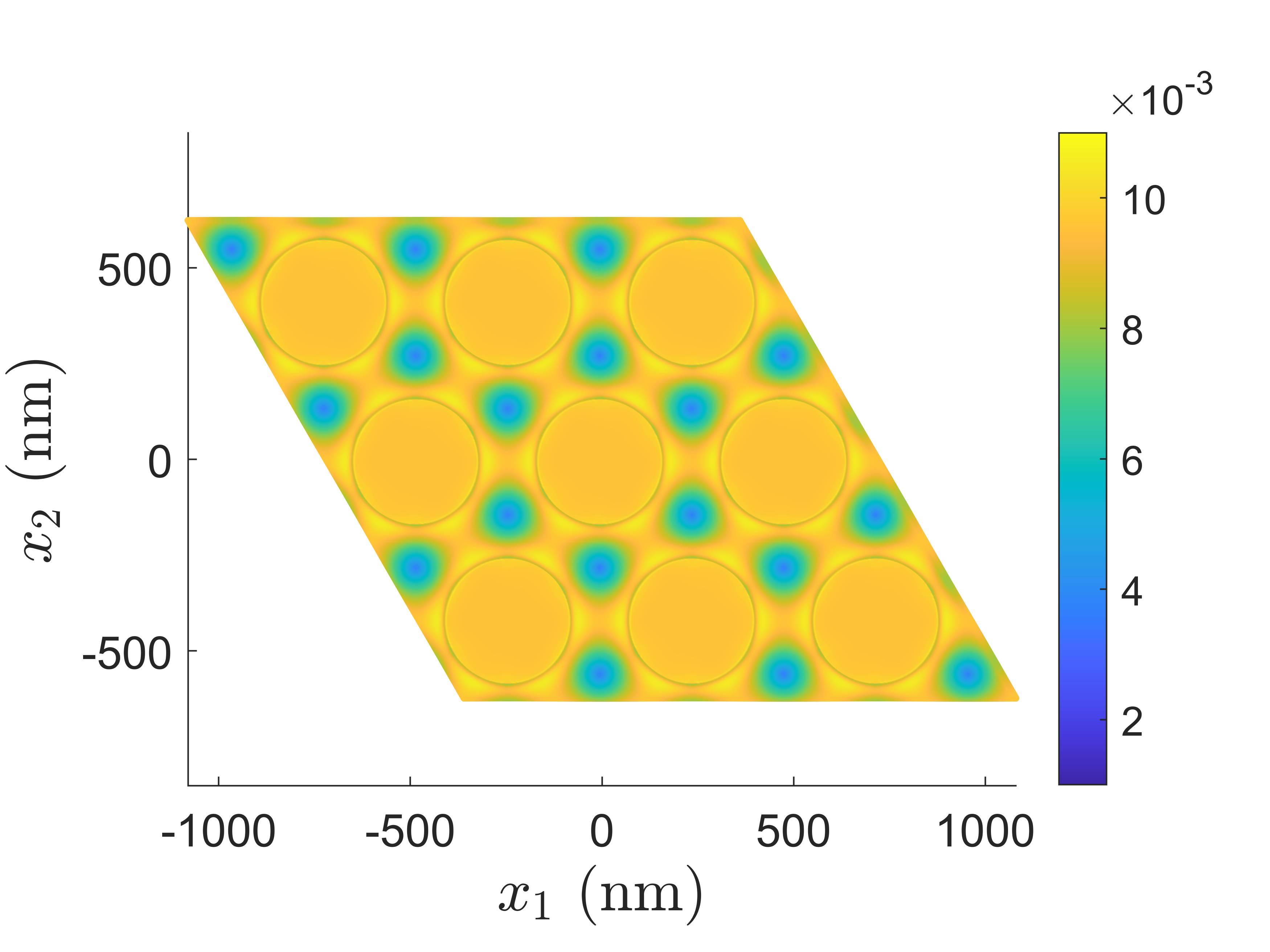}\label{paraview_cosSub_L480}}\hfill
\subfloat[$L = 520$\,nm.]{\includegraphics[trim =0mm 0mm 0mm 0mm, clip=true,width=0.49\textwidth]{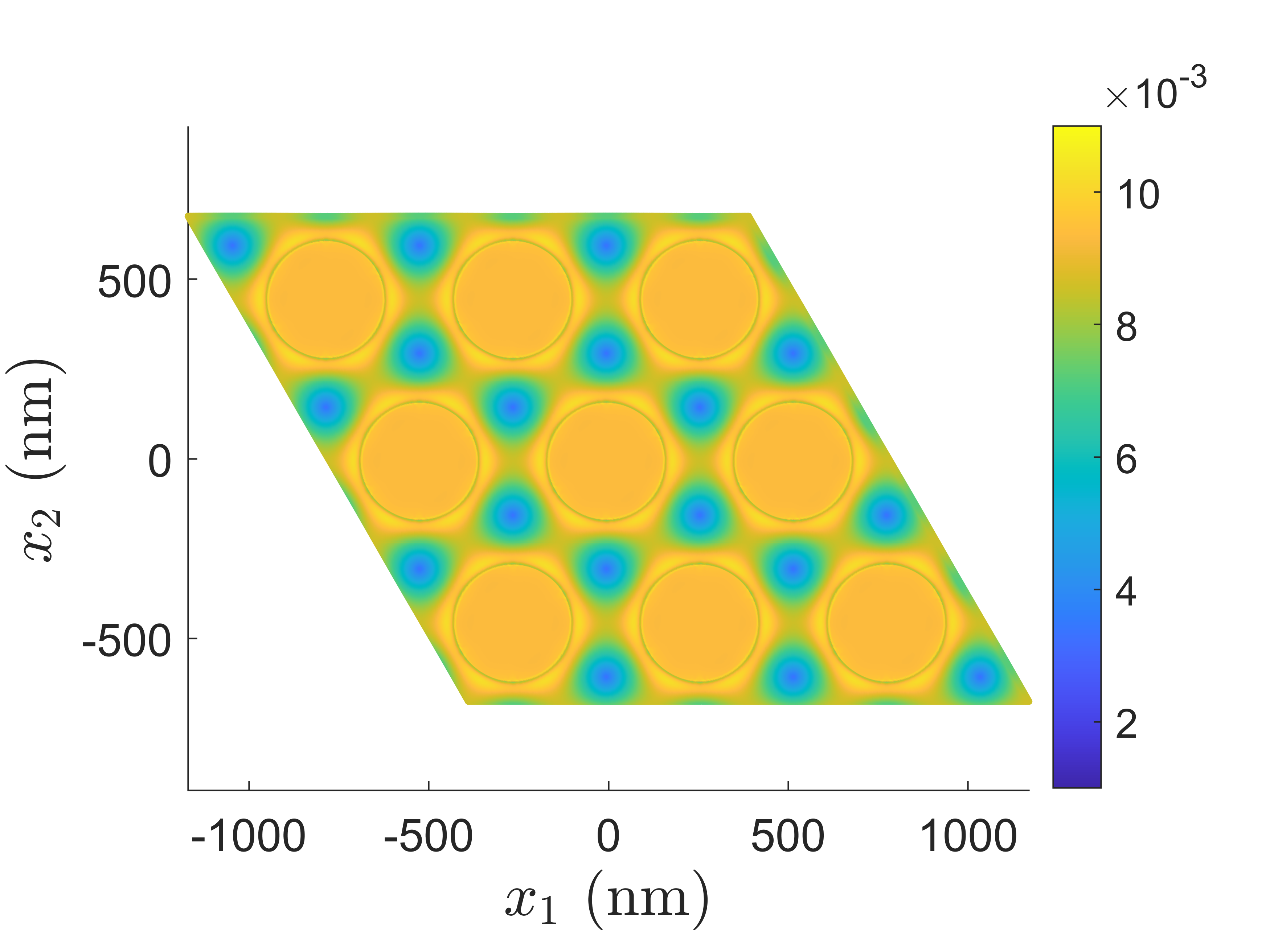}\label{paraview_cosSub_L520}}\hfill
\caption{Maximum eigenvalue of $\mathbf{E}_\text{Mo}^\text{CM}$ of a monolayer for various values of $L$.}
\label{paraview_cosSub_large}
\end{center}
\end{figure}

\begin{figure}[H] 
\begin{center}
\subfloat[$L = 400$\,nm.]{\includegraphics[trim = 0mm 0mm 0mm 0mm, clip=true,width=0.49\textwidth]{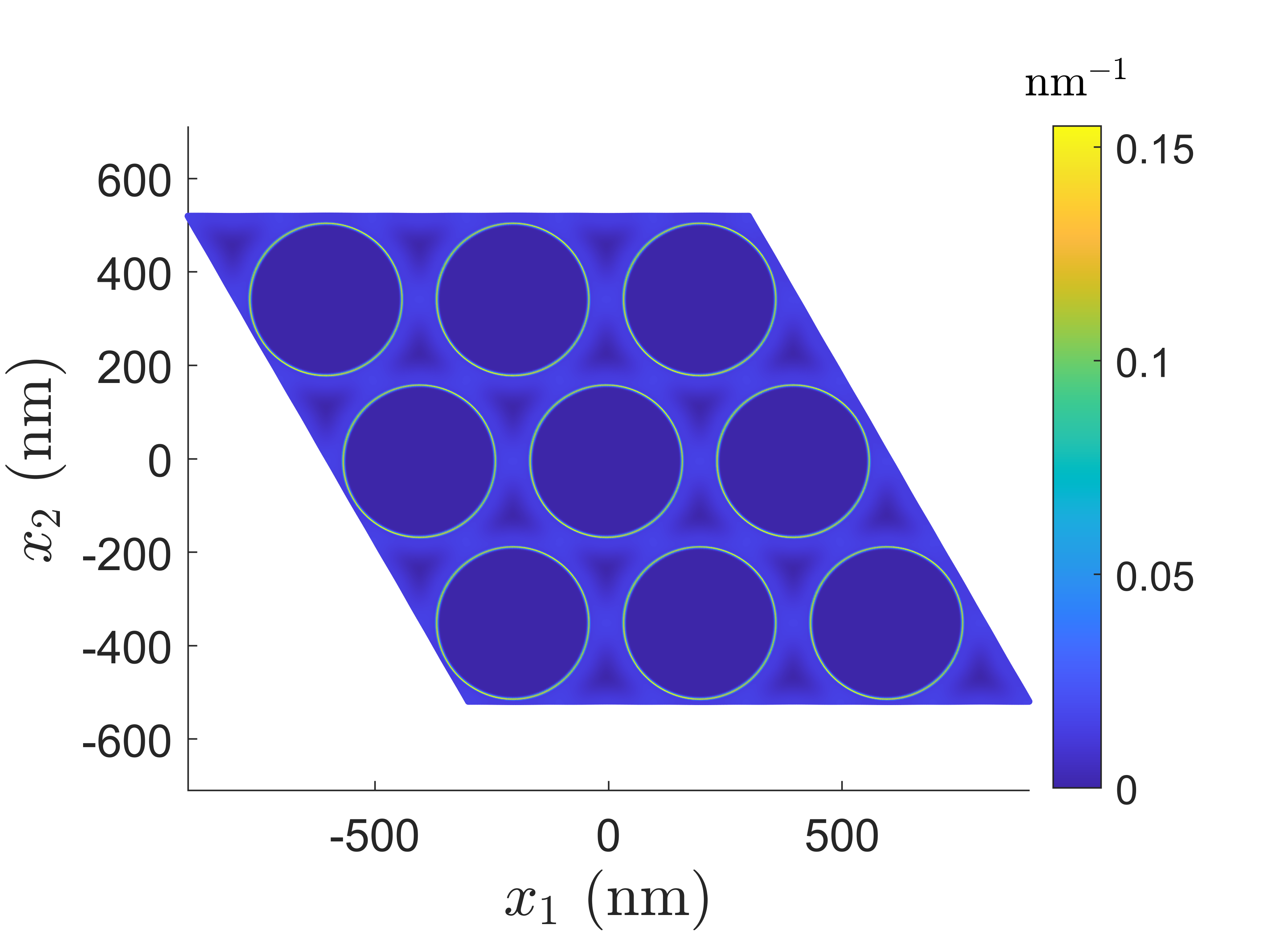}\label{cosSub_L400_curvature}}\hfill
\subfloat[$L = 440$\,nm.]{\includegraphics[trim = 0mm 0mm 0mm 0mm, clip=true,width=0.49\textwidth]{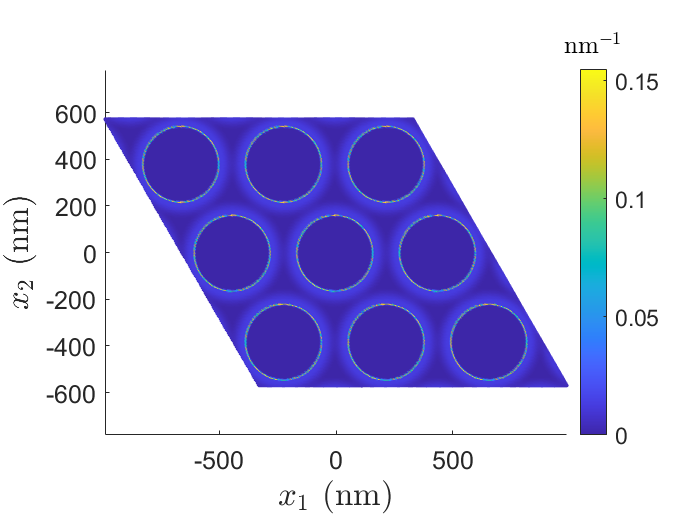}\label{cosSub_L440_curvature}}\hfill
\subfloat[$L = 480$\,nm.]{\includegraphics[trim = 0mm 0mm 0mm 0mm, clip=true,width=0.49\textwidth]{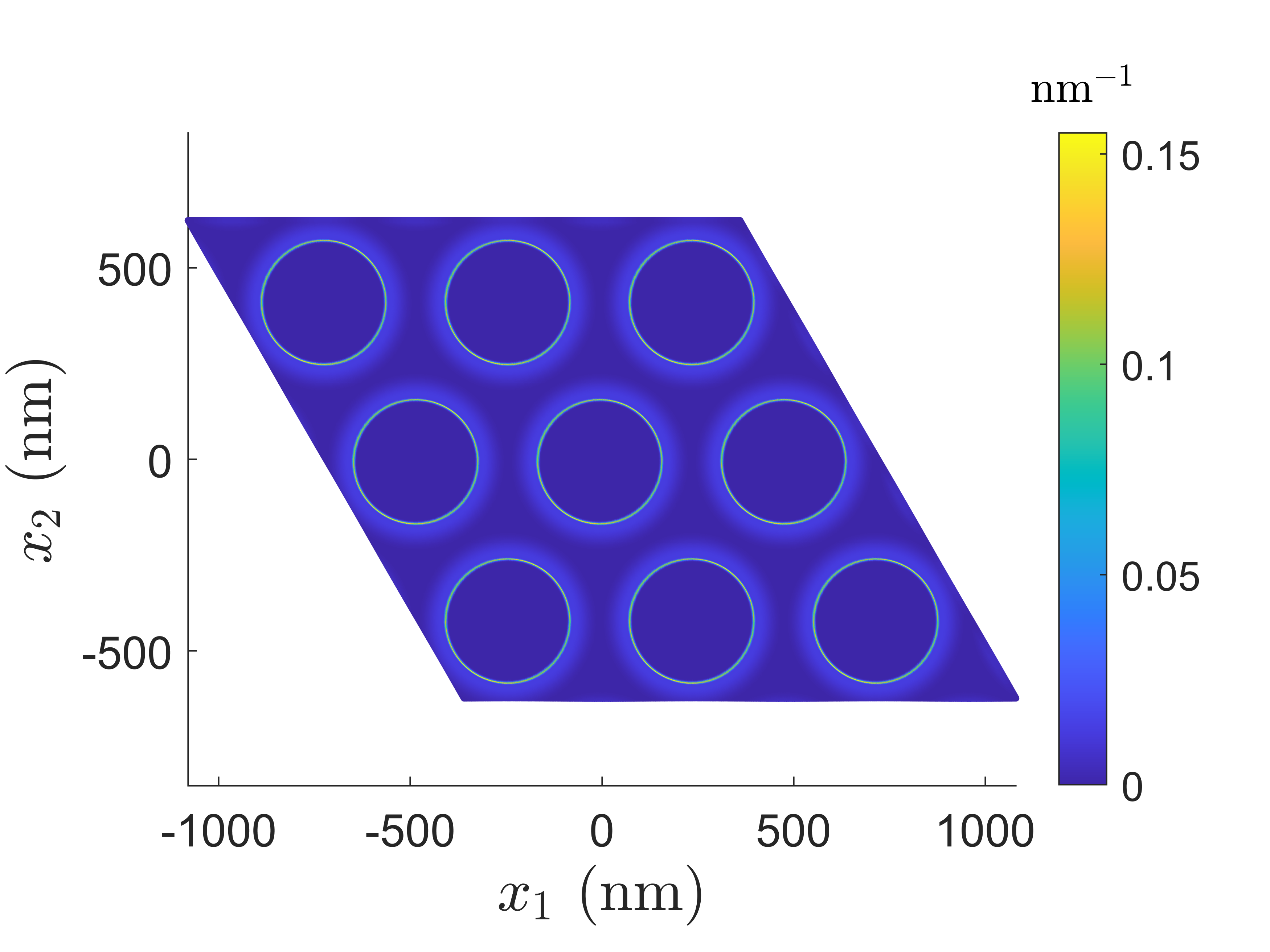}\label{cosSub_L480_curvature}}\hfill
\subfloat[$L = 520$\,nm.]{\includegraphics[trim = 0mm 0mm 0mm 0mm, clip=true,width=0.49\textwidth]{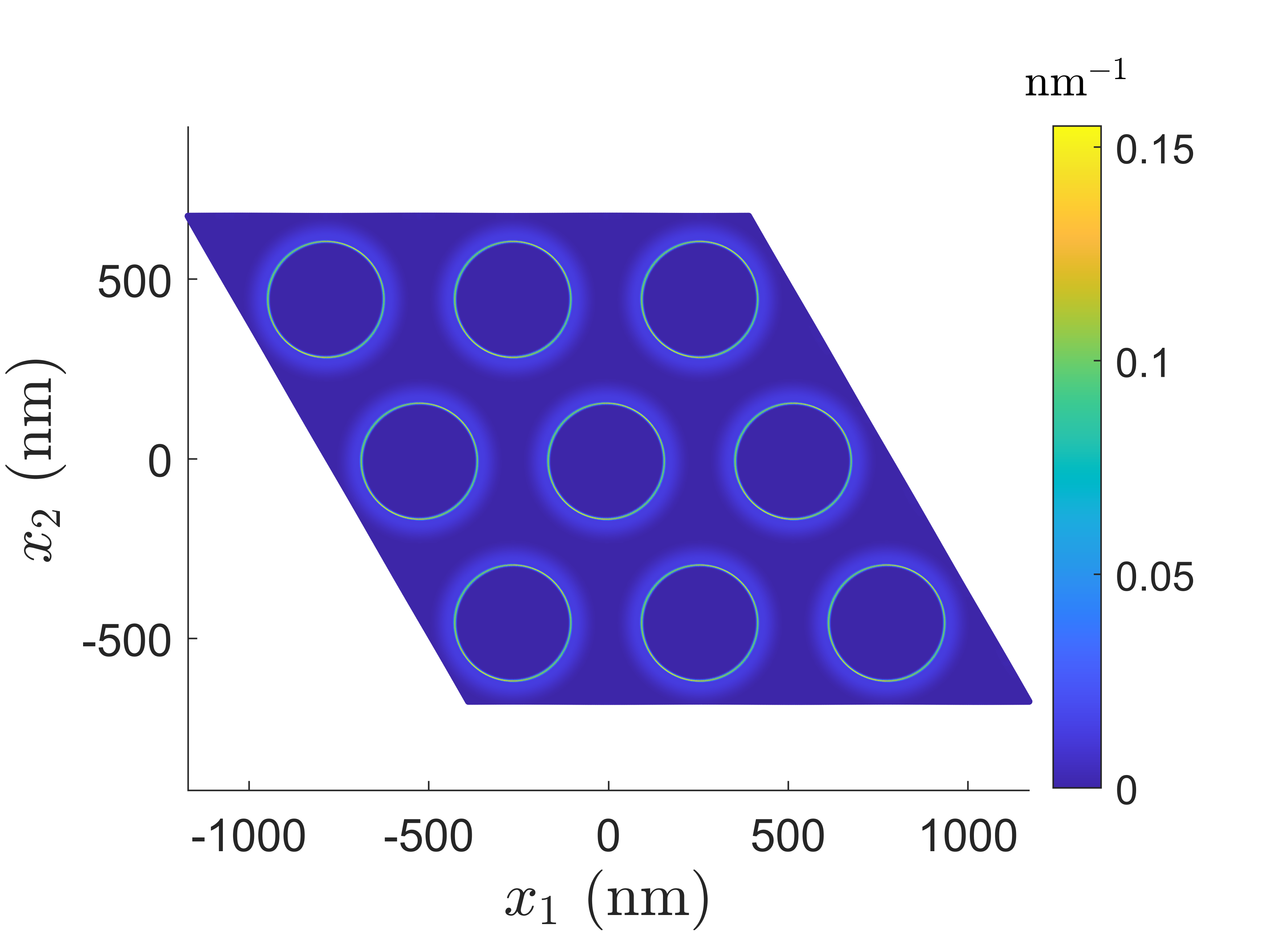}\label{cosSub_L520_curvature}}\hfill
\caption{Spectral radius of $\bm{\kappa}$ at the midsurface of the monolayer for various values of $L$.}
\label{fig:cos_large_curvature}
\end{center}
\end{figure}

Figure \ref{paraview_cosSub_large} shows how varying $L$ within the range of 400--520\,nm produces Mo sub-layer strain distributions that are similar in broad qualitative terms, with minima occurring at vertices of the hexagonal Voronoi cells about hole centers.  The overall magnitude of the strains are within measurement uncertainties of those observed experimentally in similar scenarios \cite[Table S2]{holey_zhang}.  Comparing the bending strain plotted in Figure \ref{fig:cos_large_curvature} with results from Section \ref{sec:hole}, we see that bending strain in larger specimens remains concentrated where the monolayer peels away from the substrate. This is because the curvature of the substrate geometry itself decreases as it is scaled-up, but the length scale over which peel-off occurs is controlled by the vdW length scales, which are independent of substrate geometry.

\begin{figure}[H] 
\begin{center}
\subfloat[Maximum eigenvalue of $\mathbf{E}_\text{Mo}^\text{CM}$.]{\includegraphics[trim = 0mm 0mm 0mm 0mm, clip=true,width=0.49\textwidth]{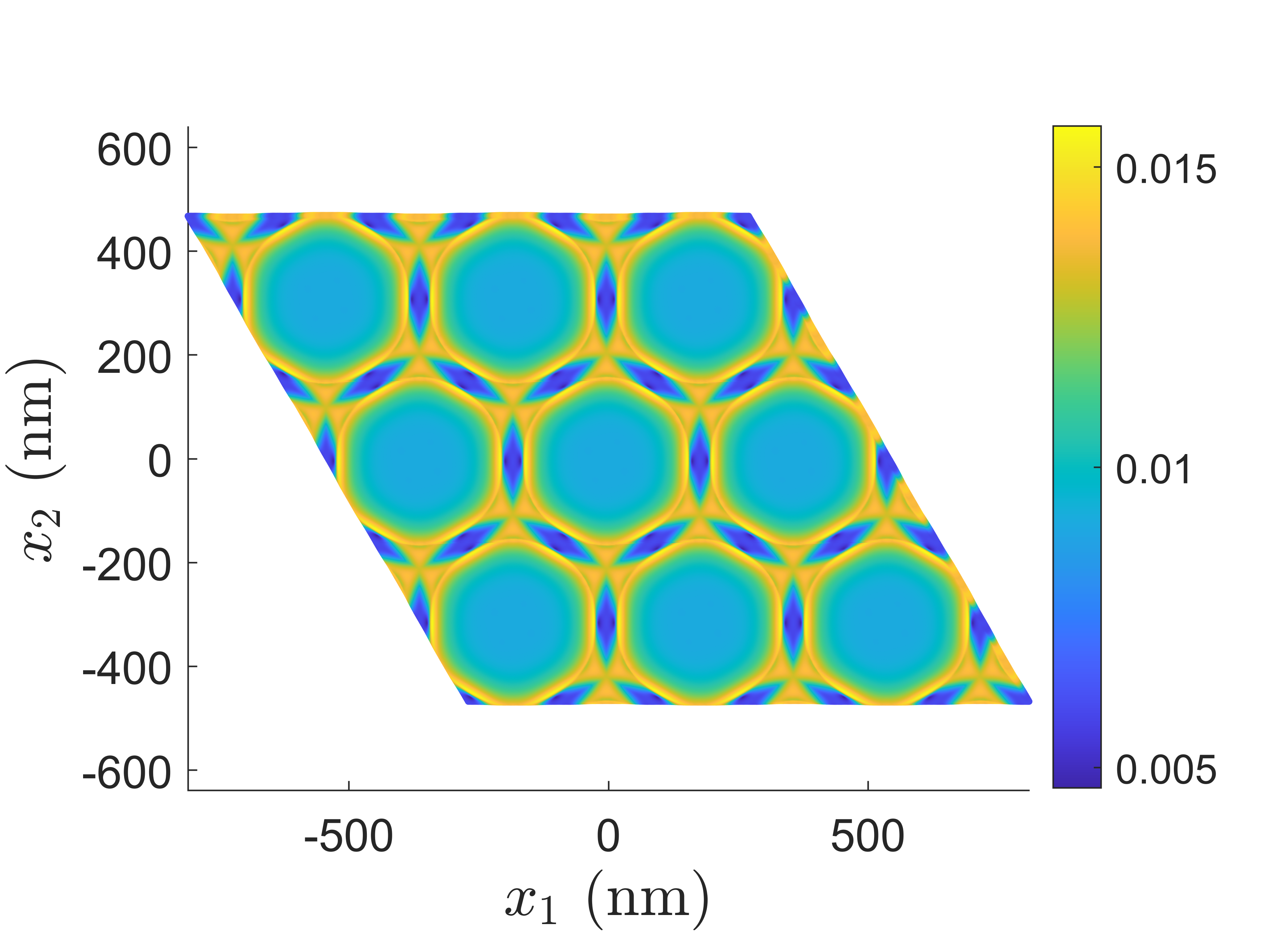}} 
\subfloat[Spectral radius of $\bm{\kappa}$.]{\includegraphics[trim = 0mm 0mm 0mm 0mm, clip=true,width=0.49\textwidth]{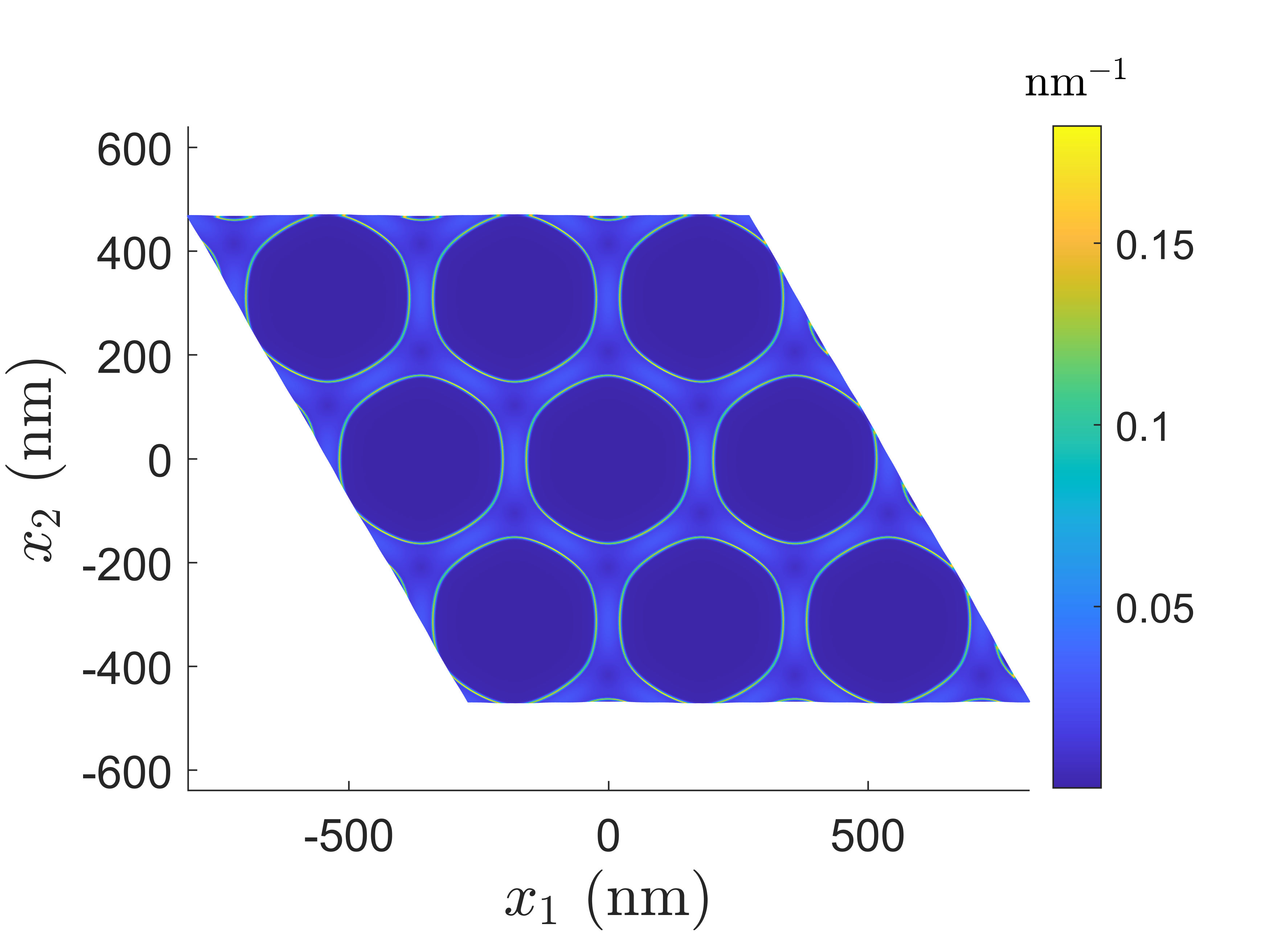}\label{cosSub_L360_curvature}}\hfill
\caption{Bending and Mo sub-layer strain fields for $L = 360$\,nm. Note that the range on the color bar differs from Figure \ref{paraview_cosSub_large} and Figure \ref{fig:cos_large_curvature}.}
\label{paraview_cosSub_large_0.9}
\end{center}
\end{figure}

Decreasing $L$ to 360\,nm, we see a qualitative phase change in the strain distribution, where strain minima migrate to the faces of the Voronoi cells and larger peaks appear where the monolayer peels away from the substrate, as shown in Figure \ref{paraview_cosSub_large_0.9}. The qualitative distribution of bending strain remains similar, i.e., concentrated around the edges of the holes.  The precise implications of these results for strain- and curvature-dependent electronic properties are outside the scope of this paper, however we note that the properties such as the band gap of MoS$_2$ are strongly affected by local strain and curvature, as reviewed in Section \ref{sec:introduction}. 
The parametric study of this section demonstrates the ability of continuum modeling to make unanticipated predictions that are out of reach for MS analysis.

\section{Conclusions}\label{sec:conclusions}
We have developed an atomistically-informed Kirchhoff--Love shell model of a 2D layer of MoS$_2$ and its interaction with a  Si$_3$N$_4$ substrate, and discretized it using IGA.  The deflections, strains, and curvatures computed in this model for a several validation problems involving substrates of different goemetries (flat, trench, hole) compare favorably with those found by reference fully-atomistic MS simulations.  In particular, differences between shell analysis and MS are much smaller than differences between MS simulations using the REBO and SW potentials, indicating that the error introduced by homogenizing the atomistic problem into a continuum model is lower than the error inherent in common interatomic potentials used for MS.    The agreement between continuum and MS modeling hinges crucially on {\em geometric} nonlinearity in the Kirchhoff--Love shell model.  However, we find that material nonlinearity in the stress--strain model is of lesser significance, and the aforementioned continuum--MS agreement is obtained using only a simple linear isotropic St.\,Venant--Kirchhoff model, in contrast to more elaborate 2D material models like those of \cite{Shirazian2018,Ghaffari2018,Mokhalingam2020} for graphene.

A major benefit of the continuum approach is that the computational cost of the isogeometric shell analysis is much lower than that of MS, especially for larger problem sizes. This is demonstrated by studying the dependence of strain and curvature in MoS$_2$ suspended over a substrate with a hexagonal pattern of holes as a function of the hole spacing on scales inaccessible to atomistic calculations. The results show an unexpected qualitative change in the deformation pattern below a critical hole separation, transitioning from a pattern of strain minima at hexagonal Voronoi cell vertices to minima at the Voronoi cell faces with larger strain peaks.

In addition to strain engineering, electronic properties of TMDs like MoS$_2$ can also be tuned by stacking them within multi-layer structures.  An exciting future research direction is therefore to extend the framework proposed in this paper to multiple layers of MoS$_2$, incorporating vdW interactions between layers.  However, it remains to be determined whether the layer--substrate interaction approximation of Section \ref{sec:tangent-plane} could be adapted to layer--layer interaction.  Another potential use for this continuum modeling framework would be to provide initial guesses for MS energy minimizers like FIRE, although the details of deriving suitable atomic positions from the continuum solution are beyond the scope of the present work.  

\section*{Acknowledgements}
DK and MP were supported by start-up funding from the University of California San Diego. ET and MC were supported in part by the National Science Foundation through the University of Minnesota MRSEC under Award Number DMR-2011401. The AFM data plotted in Figure \ref{fig:afm_substrate} was provided by Yichao Zhang (University of Minnesota).

\bibliographystyle{unsrt}
\bibliography{main}

\end{document}